\documentclass{article}
\usepackage{latexsym}

\def\be{\begin{eqnarray}}
\def\ee{\end{eqnarray}}
\def\nn{\nonumber}

\def\i{i}  %imaginary unit
\def\vy{y_\lambda} 
\def\vmu{\mu^\lambda}

\hoffset= - 1.0in         % switch off for draft style
\usepackage{ifthen}
\usepackage{graphicx}
\usepackage{amsfonts,amsmath,amssymb}

\begin{document}

\hfill OCU-PHYS 284

\hfill ITEP/TH-58/07

\bigskip

\centerline{\Large{Boundary Ring or a }}
\centerline{\Large{Way to Construct Approximate NG Solutions}}
\centerline{\Large{with Polygon Boundary Conditions}}
\centerline{\Large{II. Polygons $\bar\Pi$, which admit an
inscribed circle
}}

\bigskip

\centerline{\it H.Itoyama  and  A.Morozov}

\bigskip

\centerline{Osaka City University, Japan}
\centerline{ITEP, Moscow, Russia}

\bigskip

\centerline{ABSTRACT}

\bigskip

{\footnotesize
We further develop the formalism of arXiv:0712.0159
for approximate solution of Nambu-Goto (NG) equations with
polygon conditions in AdS backgrounds,
needed in modern studies of the string/gauge duality.
Inscribed circle condition is preserved,
which leaves only one unknown function $y_0(y_1,y_2)$
to solve for, what considerably simplifies our presentation.
The problem is to find a delicate balance
-- if not exact match --
between two different structures: NG equation
-- a non-linear deformation of Laplace equation
with solutions non-linearly deviating from holomorphic
functions, --
and the boundary ring, associated with polygons made from
null segments in Minkovski space.
We provide more details about the theory of these structures
and suggest an extended class of functions to be used at
the next stage of Alday-Maldacena program:
evaluation of regularized NG actions.
}

\bigskip

\bigskip

\tableofcontents

\newpage

\section{Introduction}
\setcounter{equation}{0}

In this paper we begin consideration of %consider
the next class of
approximate solutions to Nambu-Goto (NG) equations with
null-polygon boundary conditions
by the method suggested in \cite{malda3}.
This problem is important for the study
of the string/gauge (AdS/CFT) duality \cite{Pol,AdS},
reformulated recently \cite{am1}-\cite{amlast}
as an identity between regularized minimal areas
in $AdS_5$ and BDS/DHKS/BHT \cite{BDS,AMoth3,BHT,CS1}
amplitudes for gluon scattering in $N=4$ SUSY YM.
Unfortunately, even after this ground-breaking
reformulation \cite{am1},
explicit check of duality is escaping,
even in the leading order of the strong-coupling
expansion -- as usual because of the technical
difficulties on the stringy side.
In this particular case the first hard problem is
explicit solution to a special version of
Plateau minimal-surface problem \cite{Pla}:
to Nambu-Goto equations in $AdS_5$ geometry with
the boundary conditions at the $AdS$ boundary,
represented by a polygon $\Pi$
with $n$ light-like (null) segments.
We refer to \cite{am1} for explanation of how this
polygon emerges in the problem after a sequence
of transformations,
$$
{\rm NG\ model} \rightarrow \sigma-{\rm model}
\ \ \stackrel{T-{\rm duality}\ a\ la\ \cite{KT}}
{\longrightarrow}\ \
\sigma-{\rm model} \rightarrow {\rm NG\ model},
$$
and to \cite{mmt1,mmt2} for additional comments and
notations. Irrespective of these motivations, the
current formulation of the gauge/string duality is now
made pure geometric, at least in the leading order:
\be
{\rm regularized}\ \Big({\rm area\ of\ a\ minimal\ surface\
in}\ AdS_5,\ {\rm bounded\ by}\ \Pi\Big)
= {\rm regularized}
\left(\oint_\Pi\oint_\Pi \frac{dy^\mu dy'_\mu}{(y-y')^2}\right)
\ee
and the {\bf first problem} is to find what the minimal
surface is (with problems of regularization and higher-order
corrections arising at the next stage).
As surveyed in \cite{malda3}, explicit solution to the
{\bf first problem} is currently available only for
$n\leq 4$ \cite{Kru,am1} and the maximally symmetric case
($\Pi = S^1$)  at $n=\infty$. As usual with Plateau problem,
even approximate methods are not immediately available
beyond this exactly-solvable sector. In \cite{malda3} an
line-of-attack was suggested and the first approximate
results obtained for the simplest $Z_n$ configurations.
The present paper describes the next step in the same direction,
generalizing the results of \cite{malda3} to the next
non-trivial case: of polygons which do not have symmetry,
but still have a restricted geometry, identified as
``$\bar\Pi$ possesses an inscribing circle'' in \cite{malda3}.
In this case the boundary conditions and thus the solution
are lying in $AdS_3$ subspace of $AdS_5$, and
the problem is reduced to finding a single non-trivial
function, say $y_0(y_1,y_2)$, while the other two
are expressed through the $AdS_3$ constraints \cite{am1},
\be
\begin{array}{ccc}
y_3 = 0 & \ \ &(Y_3=0)  \\
y_0^2 + 1 = y_1^2 + y_2^2 + r^2&& (Y_4=0)
\end{array}
\label{ads3}
\ee

%\newpage

\section{Approach to approximate solution:
a summary of \cite{malda3}}
\setcounter{equation}{0}

The strategy, suggested in \cite{malda3} was to:

$\bullet$ First, represent $y_0$ as a power series,
\be
y_0 = \sum_{i,j\geq 0} a_{ij} y_1^iy_2^j
\label{y0ser}
\ee
and rewrite NG equations in the form of recurrence
relations for $a_{ij}$, with recursion relating the
two adjacent "levels" $k=i+j$ and leaving a number
of free parameters.
If the structure of the {\it boundary ring} is explicitly
known, then
expansion (\ref{y0ser}) can be modified in order to
take boundary conditions into account from the very
beginning, though this can cause additional convergency
problems for the series.

$\bullet$ Second, truncate the series at some level $N$
and specify the remaining free parameters which
match  the boundary conditions in the best possible way
at given truncation level.
Increasing $N$ provides better and better fit to both
the NG equations and boundary conditions.

$\bullet$ Fitting criteria and thus the resulting
approximations can be different, depending on the further
application. As explained in \cite{malda3}, one can
improve either local or global approximation to boundary
conditions or instead try to better match the behavior at
the angles of the polygon, which is responsible for the
main IR divergence of the regularized area of the minimal
surface.

\subsection{Recurrent relations and free parameters}

The first recurrence relations were already found
in \cite{malda3}:

There are no relations at levels zero and two:
all the corresponding coefficients,
$a_{00}$ and $a_{10},a_{01}$ are free parameters,
i.e. there are $\nu_0=1$ and $\nu_1=0$ of them.

At level two NG equations impose a single relation:\footnote{
As clear already from this formula the choice of
$a_{k0}$ and $a_{0k}$ (instead of, say, $a_{k0}$ and
$a_{k,k-1}$) makes the limit $a_{10},a_{01} \rightarrow 0$
singular. Note that original solution of \cite{am1} is
exactly of this kind: $a_{ij} = \delta_{i1}\delta_{j1}$
and singularities are easily resolved for it.}
\be
a_{11} = -\frac{a_{02}(1+a_{00}^2-a_{10}^2) +
a_{20}(1+a_{00}^2-a_{01}^2)}
{a_{01}a_{10}} = -\frac{a_{02}A_{10}+a_{20}A_{01}}{a_{01}a_{10}}
\label{11thr0220}
\ee
where $A_{01} = 1+a_{00}^2-a_{01}^2$ and
$A_{10} = 1+a_{00}^2-a_{10}^2$.
Next formulas involve a generalization of these quantities:
\be
A_{kl} = 1+a_{00}^2-ka_{10}^2-la_{01}^2
\ee

At level three we get two relations:
%$$
\be
a_{12} =
\frac{6a_{03}a_{01}^3a_{10}^2A_{10} -
3a_{30}a_{01}^2a_{10}A_{01}^2 -
a_{01}^2a_{10}^2(a_{11}^2-4a_{02}a_{20})A_{03}
+a_{00}a_{01}^2\big(a_{02}(A_{10}A_{21}+4a_{01}^2a_{10}^2)
+a_{20}A_{01}A_{41}\big)}
{a_{01}^2a_{10}(A_{01}A_{10}-4a_{01}^2a_{10}^2)},
%$$
\nn \\
\ \ \ \ \ \ \ \nn \\
%$$
a_{21} = \frac{6a_{30}a_{10}^3a_{01}^2A_{01} -
3a_{03}a_{10}^2a_{01}A_{10}^2 -
a_{01}^2a_{10}^2(a_{11}^2-4a_{02}a_{20})A_{30}
+a_{00}a_{10}^2\big(a_{20}(A_{01}A_{12}+4a_{01}^2a_{10}^2)
+a_{02}A_{10}A_{14}\big)}
{a_{01}a_{10}^2(A_{01}A_{10}-4a_{01}^2a_{10}^2)}
\nn \label{12thr0330}
\ee
%$$
and $\nu_3=2$ free parameters $a_{03}$ and $a_{30}$.

Similarly, at level $k$ there will be $k-1$ relations
imposed by NG equations, and $\nu_k=2$ out of $k+1$
coefficients $a_{i,k-i}$ at this level will remain
free. We always choose $a_{k0}$ and $a_{0k}$ for these
free parameters. They can be associated with two arbitrary
functions -- of $y_1$ and $y_2$ respectively, and this
freedom resembles  the general solution of the
archetypical equation
$\frac{\partial^2 Y}{\partial y_1\partial y_2} = 0$,
given by $Y(y_1,y_2) = f(y_1) + g(y_2)$
with two arbitrary functions $f$ and $g$.
We shall see in s.\ref{comprec} below that
even more relevant can be analogy
with the ordinary Laplace equation, solved by arbitrary
holomorphic and antiholomorphic functions.

If series (\ref{y0ser}) is truncated at level $N$,
it contains $(N+1)^2$ different coefficients $a_{ij}$,
of which $2N+1$ remain free parameters, unconstrained
by NG equations.

\subsection{Boundary conditions and the boundary ring}

According to \cite{malda3} the boundary conditions can
be formulated in terms of the boundary ring ${\cal R}_\Pi$,
which consists of all polynomials of $y$-variables
that vanish on the boundary polygon $\Pi$.\footnote{
By definition, solutions of our problem belong to the
intersection of the space of $r=0$ asymptotes of NG
solutions with the completion of the boundary ring.
In still other words, anzatze that we substitute into
NG equations should be taken from completion of the
boundary ring. Completion here means first, that,
say, $r^2$ rather than $r$ itself belongs to ${\cal R}_\Pi$
according to (\ref{ads3}), i.e. $r$ belongs to the
{\it algebraic} completion of the ring. Second, our
anzatze can be looked for among formal series made out
of elements of ${\cal R}_\Pi$.}
Since $y_3=0$ this ${\cal R}_\Pi$ includes $y_3$ as
a generator and we can actually restrict considerations
to polynomials, depending on only three variables
$y_0,y_1,y_2$.

\begin{figure}
\begin{center}
{\includegraphics[width=200pt,height=150pt]
{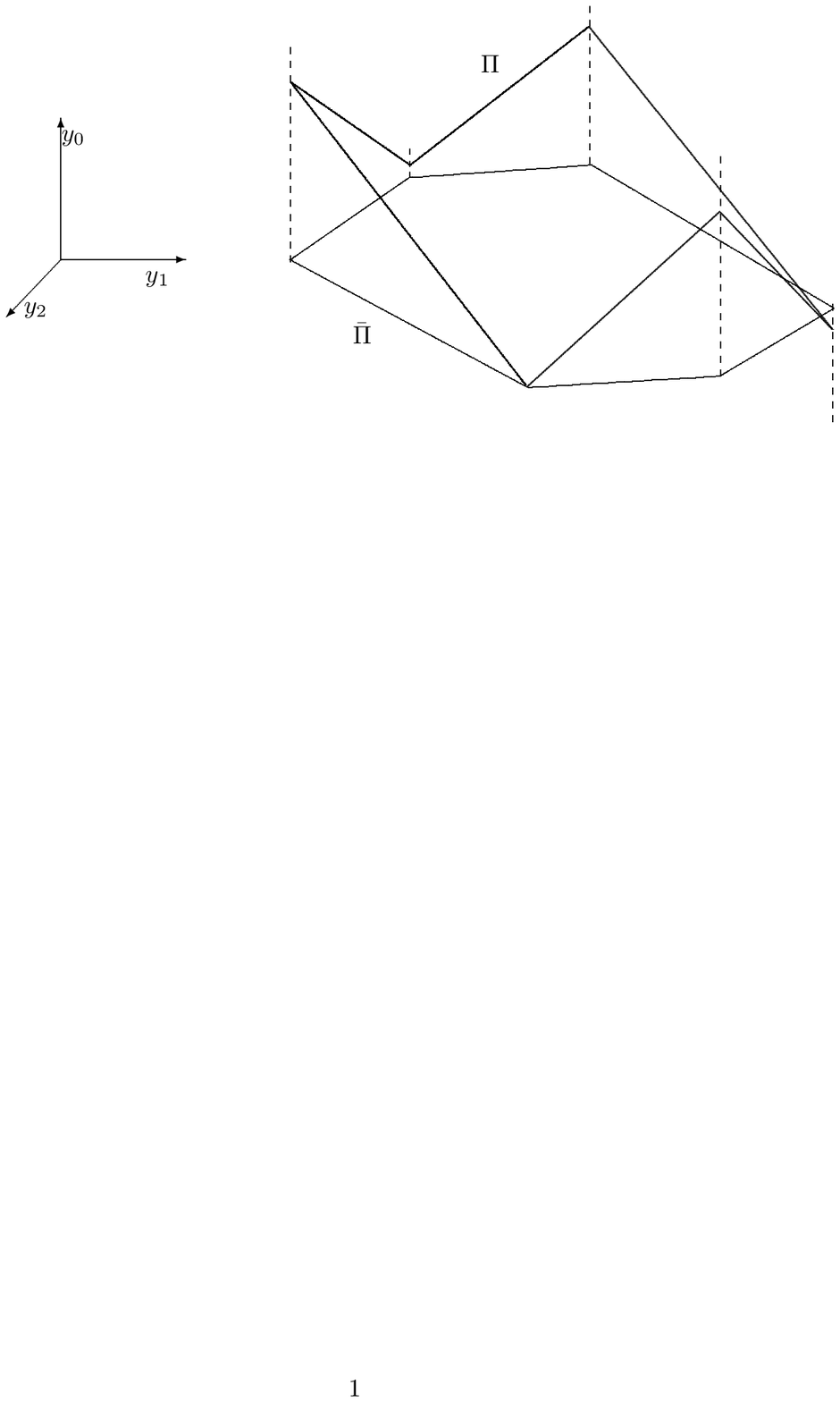}}
{\includegraphics[width=150pt,height=150pt]
{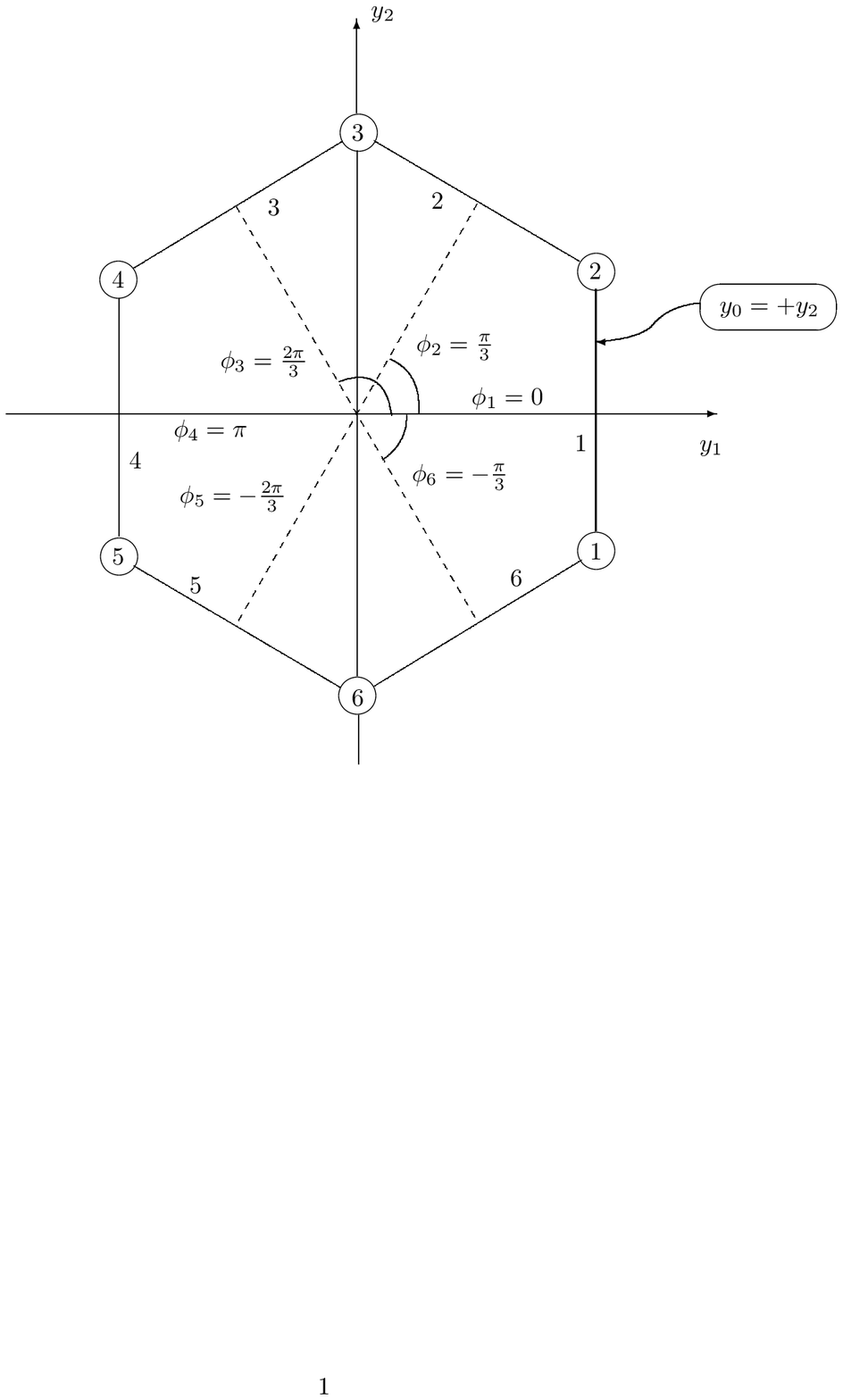}}
%{\includegraphics[width=200pt,height=200pt]
%{./pics/rho.jpg}}
%\input{./pics/FigZnpol.tex}
\caption{{\footnotesize
Convention for labeling sides and vertices of
the $Z_n$-symmetric polygon $\bar\Pi$ (right).
Its corresponding $\Pi$ is shown in the left picture.
}}
\label{Znpol}
\end{center}
\end{figure}

As further explained in \cite{malda3},
if $n$ edges of $\Pi$ are defined by the equations:
\be
\left\{
\begin{array}{c}
c_ay_1 + s_ay_2 = h_a,  \\
y_0 = y_{0a} + \sigma_a(-s_a y_1+c_ay_2),
\end{array}\right.
\label{sidea}
\ee
with $c_a = \cos \phi_a$, $s_a=\sin\phi_a$ and
$\sigma_a=\pm 1$, see Fig.\ref{Znpol}, then

$\bullet$
the condition that $\Pi$
closes in $y_0$ direction is
\be
\sum_{a=1}^n \sigma_a l_a = 0,
\ee
where $l_a$ are the lengths of $\bar\Pi$, which
is projection of $\Pi$ onto the $(y_1,y_2)$ plane,
and

$\bullet$
the following three polynomials are
the obvious elements of ${\cal R}_\Pi$:
\be
P_\Pi(y_1,y_2) = \prod_{a=1}^n \Big(c_a y_1+s_a y_2-h_a\Big),\nn \\
\tilde P_\Pi(y_0,y_1) =
\prod_{a=1}^n
\Big( y_1 + (-)^{a+1}s_a(y_0-y_{0a}) - c_ah_a\Big),\nn\\
\widetilde{\tilde P}_\Pi(y_0,y_2) =
\prod_{a=1}^n \Big( y_2 + (-)^{a}c_a(y_0-y_{0a}) - s_ah_a\Big)
\label{Ppolsdef}
\ee

\bigskip
\bigskip

So far we imposed only one of the constraints (\ref{ads3}),
$y_3=0$. The second constraint can be imposed
only if all $h_a=1$, and this is what we assume below in
the present paper.
As already stated, this condition implies the existence
of a unit circle, inscribed into $\bar\Pi$.
If additionally $n$ is even and $\sigma_a = (-)^a$, then
also all the $n$ parameters $y_{0a}$ coincide and we can
put $y_{0a}=0$ by a constant shift of $y_0$:
this choice corresponds to $y_0$ vanishing
at all points where the sides of $\bar\Pi$ touch the
inscribing circle.

\subsection{Boundary conditions as sum rules}

A possible way to describe above boundary conditions
is also to write them down for each particular side of $\Pi$.
On (\ref{sidea}) we have (with $h_a=1$ and $y_{0a}=0$):
\be
y_1 = c_a-s_at_a, \nn \\
y_2 = s_a + c_at_a, \nn \\
y_0 = \sigma_at_a
\ee
i.e.
\be
z=y_1+iy_2 = (c_a+is_a)(1+it_a)=e^{i\phi_a}(1+i\sigma_ay_0),
\label{segmcomp}
\ee
where $t_a$ is a parameter along the corresponding straight
line. Along its segment, which is the side of $\Pi$
it changes within some region $t_a\in [-t_{a1},+t_{a2}]$.
Then boundary conditions imply that
\be
\sum_{i,j\geq 0} a_{ij}(c_a-s_at_a)^i(s_a+c_at_a)^j = \sigma_a t_a,
\ \ \ \ t_a\in [-t_{a1},+t_{a2}], \ \ \ a=1,\ldots,n
\ee
A set of sum rules arise is we consider these equalities
as term-by-term identities for series in powers of $t_a$.
For example, if coordinate system is rotated to put
$c_1=1$, $s_1=0$, we get an infinite
set of relations
\be
\sum_{j\geq 0} a_{ij} = \sigma_1 \delta_{i1}
\ee
-- to be supplemented by $(n-1)$ more similar sets, associated with
other sides of $\Pi$.
The free parameters $a_{k0}$ and $a_{0k}$ are defined
by boundary conditions.

Among other things, this consideration seems to imply %implies
that $y_0$ should satisfy (\ref{sidea}) along entire straight line,
not only within the segment.
This is consistent with the known property of solutions to
Plateau problem in the flat Euclidean space \cite{Oss}.

\subsection{Approximate methods}

Unfortunately no way is known at the moment to solve above
relations exactly, except for in a few simple situations,
listed in s.2 of \cite{malda3}.
In order to proceed one is naturally turned to approximate
considerations.
However, there are no ready methods to address this kind
of problems and one needs to practice the {\it trial and error}
approach.

Usually approximation starts from making the best thinkable
anzatz, explicitly taking into account all the already known
properties of the problem (symmetries, to begin with) with
remaining infinite-parameter freedom contained in
adequately defined formal series.
Then this anzatz is substituted into original equation, and
-- if the formal series was introduced in an adequate way
(what is more a matter of art or lack than of a rigorous theory),
-- the equation turns into a {\it recurrent} relation for
coefficients of the series.
So far everything was exact, even if not fully deductive,
approximation comes at the next stage:
when infinite series is {\it truncated} at some level $N$.
Success of the method depends on the choice of
"original knowledge", of particular anzatz,
including a point to expand around and particular
expansion parameters, and -- not the least -- on the properties
of the problem, i.e. the very existence of effective truncations,
producing reasonable approximation at sufficiently low $N$.

In \cite{malda3} various attempts were described to
apply this procedure, and some of them seem relatively
successful.
The main problem appears to be a balance between
reasonable introduction of formal series in {\it local}
parameters (say, coordinates $y$) consistent with
differential nature of NG equations, and adequate
imposition of {\it global} boundary conditions,
relatively far from expansion point.
It turns out that, somewhat unusually, the balance
should better be shifted towards the boundary conditions.

\subsection{The goal of this paper
\label{goal}}

Success of \cite{malda3} was due to construction of
specific polynomials, named ${\cal P}_n(y_0,y_1,y_2)$,
which had four important properties:

$\bullet$
${\cal P}_\Pi \in {\cal R}_\Pi$ was an element
of the boundary ring, thus an anzatz
\be
{\cal P}_\Pi = 0
\label{calPanza}
\ee
satisfies boundary conditions exactly
and it can be further generalized (perturbed) to
\be
{\cal P}_\Pi = P_2 {\cal B}
\label{calPanzapert}
\ee
which continue to satisfy boundary conditions
with any perturbation function
${\cal B}(y_0;y_1,y_2)$.\footnote{
Additional $Z_n$ symmetry, assumed in \cite{malda3},
allowed to put  ${\cal B}(y_0;y_1,y_2) = y_0B(y_1,y_2)$,
but this is not the case generically.}

$\bullet$
${\cal P}_n$ was linear in $y_0$,
\be
{\cal P}_\Pi = y_0Q_\Pi(y_1,y_2) - K_\Pi(y_1,y_2)
\label{calPy0linear}
\ee
what allowed to resolve (\ref{calPanza}) and treat it
as an anzatz for a {\it single-valued} function
\be
y_0^{(0)}(y_1,y_2) = \frac{K_\Pi(y_1,y_2)}{Q_\Pi(y_1,y_2)}
\label{mainanza0}
\ee
After that (\ref{calPanzapert}) could be solved
iteratively, {\it a la} \cite{MS}, and provides
%while iterative resolution of (\ref.{calPanzapert})
%made it into
a formal series perturbation of this function.

$\bullet$
The polynomial $Q_\Pi$ in (\ref{calPy0linear})
did not have zeroes {\it inside} $\bar\Pi$,
in particular, it did not vanish at the origin,
\be
Q_\Pi(y_1,y_2) = 1 + O(y_1,y_2),
\label{linearity1}
\ee
what made the function (\ref{mainanza0}) free of singularities,
and this property was inherited by all perturbative
corrections implied by (\ref{calPanzapert}).\footnote{
A little care is needed at this point if one wishes to include
the $y_0$-linear terms from ${\cal B}(y_0,y_1,y_2)$
into {\it denominators} of perturbation series,
i.e. sum up the corresponding parts of the series exactly,
what can always be done.
}

$\bullet$
The polynomial $K_\Pi(y_1,y_2)$ satisfied NG equations
in the first approximation, i.e. application of NG operator
provided only terms with higher powers of $y_1,y_2$
than were present in $K_\Pi$.
This property was easy to formulate in \cite{malda3}
because $Z_n$-symmetric $K_\Pi$ considered there were
homogeneous polynomials (of degree $n/2$), but it becomes
a subtler concept in generic situation.
Still it is this property that allows to honestly treat
${\cal B}$ as a {\it perturbation}, needed to {\it correct}
(\ref{mainanza0}) in order to make it satisfying the NG
equations.\footnote{
Note that this approach is somewhat unusual, because it
shifts emphasize from differential equations to boundary
conditions.
Ref.\cite{malda3} describes in length how this shift of
accents occurs, here we use this modified view from the
very beginning.
Still, it deserves reminding that one of the reasons for it
is that the modern opinion is that Plateau problem arises
in string/gauge duality in a special context:
we need minimal surfaces in AdS space with boundaries
lying at its boarder
(infinity or the origin, depending on parametrization of AdS),
so that their areas are diverging near the boundary.
What we need are regularized areas, but regularization
requires {\it exact} knowledge of behavior at the boundary,
i.e. of allowed {\it type} of asymptotics
-- in order to define physical quantities, which are
independent of the {\it coefficients} in front of these
asymptotical terms.
This is what makes care about the boundary conditions the
first priority.
If they are taken into account in exact way, then one
can always deal with equations {\it a posteriori},
by minimizing the resulting regularized area over remaining
free parameters, which could otherwise be fixed {\it a priori}
by exactly solving the original equations.
As explained in \cite{mmt1}, this approach can be much simpler
and more practical.
}

\bigskip

Thus in this paper our primary goal is to search for an analogue
of the polynomials ${\cal P}_\Pi$ in the case of generic
$\Pi$ with $n$ angles and inscribed circle in $\bar\Pi$.
If they are constructed, then we can look at approximations
to minimal surfaces provided by (\ref{mainanza0}) and
consider the actual role of corrections, which are obligatory
non-vanishing, since ${\cal B}=0$ is inconsistent with
NG equations.
Actually, the present paper is only a step in this direction.
We begin by constructing the theory from the very beginning,
but leave many important branches of possible development
only mentioned, what finally prevents us from providing
an exhaustive answer.
Thus {\it de facto} the goal is to describe the context, what
opens a lot of room for improvements and for getting better
and wider results.

\subsection{Plan of the paper
\label{plan}}

Our first subject in s.\ref{comprec} is conversion of
NG equations into recurrence relations.
Such conversion can be made over different "backgrounds",
the $c$- and $b$-series of \cite{malda3} being particular
examples.
In s.\ref{comprec} we concentrate on the "basic" example,
with background zero, so that all other sets of
recurrence relations
can be considered as subalgebras of this main one.
Our main interest here is deviation from harmonic functions
due to the difference between non-linear NG operator and linear
Laplace in one complex dimension -- on the $(y_1,y_2)$ plane.

The next s.\ref{andi} addresses the problem of sharp angles
-- an important issue for applications in Alday-Maldacena
program, because angles are the sources of most important
quadratic divergencies of regularized actions.
We explain how sharp-angle conditions can be formulated
analytically.
Of course, elements of the polygon boundary rings satisfy
these conditions, but they are of course violated by
generic solutions to NG equations, exact or approximate,
before boundary conditions are imposed.
Moreover, if boundary conditions are matched approximately,
not exactly (like some options considered in \cite{malda3}),
one still has an opportunity to require that angles are
sharp (not smoothened) -- and it is here that
these analytical formulas are especially useful.

In s.\ref{quadrila} we address the problem of NG solutions
for generic quadrilaterals.
Despite it is solved in \cite{mmt1,mmt2}, solution is not
found in the form of explicit function $y_0(y_1,y_2)$.
A way to bring it to such form is provided by technique of
non-linear algebra \cite{NOLAL,nolal2}.
We demonstrate that
%In this paper we do not complete this tedious calculation,
%but explain that
at $n=4$ this $y_0(y_1,y_2)$ is always a solution to
an explicit {\it quadratic} equation, like it turned out
to be in the particular case of rhombi \cite{malda3}.

The following subject in s.\ref{bori} is boundary rings for
polygons $\Pi$.
The main puzzle here is the structure behind the polynomials
$K_\Pi$ in eq.(\ref{calPy0linear}).
In \cite{malda3} they were obtained from somewhat mysterious
manipulations with $P$'s from (\ref{Ppolsdef}) and were
found to have a form, which is very similar to (\ref{Ppolsdef})
in $Z_n$-symmetric situation with even $n$:
\be
K_{n/2} \sim  \prod_{a=1}^{n/2} \Big(s_a y_1+c_a y_2\Big)
\label{Kn2def}
\ee
The problem is that this time
the product at the r.h.s. is only over
a half of segments and thus can not be immediately
generalized to asymmetric cases
(going from even to odd $n$ introduces additional problem:
the simplest choice of $\sigma_a = (-)^{a-1}$ can not be made).
We demonstrate in s.\ref{bori} how such polynomials can
actually be constructed -- though they probably do not play
the same role as they did in \cite{malda3}.
The reason is that already in the first non-trivial
asymmetric configuration -- at $n=4$ -- exact solution
is associated with the boundary ring element, which is
not {\it linear}, but {\it quadratic} in $y_0$,
see s.2.6 of \cite{malda3}.
This is the first signal that the proper analogue
of ${\cal P}_\Pi$ in asymmetric case should not be linear.
At the same time, s.\ref{quadrila} demonstrates
that {\it quadratic} can be enough, at least at $n=4$ it is
the case.
It is still unclear what the situation is going to be
beyond for $n>4$, where explicit solutions of NG equations
are yet unknown.
A promising option is to look for the adequate anzatze
among the boundary ring elements of order $n/2$ in $y_0$.
%It would be interesting if the relevant order of equation
%is always $n/2$.
According to the strategy, outlined in \cite{mmt1} and
\cite{malda3} we suggest to parameterize potentially relevant
elements of the boundary rings by {\it a few} parameters,
and treat them as if they were {\it moduli} of NG solutions,
i.e. evaluate the regularized action and minimize it w.r.t.
these parameters.
This approach can finally turn simpler then direct solution of
NG equations by methods, considered in s.\ref{comprec}.

Appendix at the end of the paper contains some remarks about
sophisticated notations used throughout the text.

\newpage

\section{NG equations as recurrence relations
\label{comprec}}
\setcounter{equation}{0}

The first recurrence relations were already found
in \cite{malda3}.
It will be more convenient to switch to the complex
coordinates $z=y_1+{\rm i}y_2$, $\bar z=y_1-{\rm i}y_2$
in the $(y_1,y_2)$ plane and
write instead of (\ref{y0ser})
\be
y_0 = \sum_{k,j\geq 0}\left(\alpha_{kj} z^k +
\bar\alpha_{kj} \bar z^k\right)(z\bar z)^j =
\sum_{k,j\geq 0}{\rm Re}\left(\alpha_{kj} z^k\right)(z\bar z)^j
\label{y0serC}
\ee

\subsection{Reduced NG action}

Recurrent relations result from substitution of a formal series
representation for $y_0(y_1,y_2)$ into NG equations, which for
$y_3=0$ have the form
\be
\frac{\partial}{\partial y_1}
\left(\frac{\partial y_0}{\partial y_1}
\frac{H_{22}}{r^2L_{NG}}\right) +
\frac{\partial}{\partial y_2}
\left(\frac{\partial y_0}{\partial y_2}
\frac{H_{11}}{r^2L_{NG}}\right) -
\frac{\partial}{\partial y_1}
\left(\frac{\partial y_0}{\partial y_2}
\frac{H_{12}}{r^2L_{NG}}\right) -
\frac{\partial}{\partial y_2}
\left(\frac{\partial y_0}{\partial y_1}
\frac{H_{12}}{r^2L_{NG}}\right) = 0, \nn\\
\frac{\partial}{\partial y_1}
\left(\frac{\partial r}{\partial y_1}
\frac{H_{22}}{r^2L_{NG}}\right) +
\frac{\partial}{\partial y_2}
\left(\frac{\partial r}{\partial y_2}
\frac{H_{11}}{r^2L_{NG}}\right) -
\frac{\partial}{\partial y_1}
\left(\frac{\partial r}{\partial y_2}
\frac{H_{12}}{r^2L_{NG}}\right) -
\frac{\partial}{\partial y_2}
\left(\frac{\partial r}{\partial y_1}
\frac{H_{12}}{r^2L_{NG}}\right) + \frac{2L_{NG}}{r} = 0
\label{NGeq}
\ee
where
\be
H_{ij} = \frac{ -\frac{\partial y_0}{\partial y_i}
\frac{\partial y_0}{\partial y_j}
+ \frac{\partial r}{\partial y_i}\frac{\partial r}{\partial y_j}
+ \delta_{ij} }{r^2}
\label{NGH}
\ee
and
\be
L_{NG} = \sqrt{ \det_{ij}^{\phantom .} H_{ij} }
= \sqrt{H_{11}H_{22}-H_{12}^2}
\ee
After substitution of (\ref{ads3}) the two equations become
dependent and we can consider any one of them.
Even more convenient is to make the substitution (\ref{ads3})
directly in NG action, then it depends on a single
function $y_0(y_1,y_2)$ and looks like
\cite{malda3}
\be
\int L_{NG}dy_1dy_2 =
%\frac{dy_1dy_2}{r^3}
\int \sqrt{\frac{(y_i\partial_i y_0 - y_0)^2
- (\partial_i y_0)^2 + 1}{\left(1+y_0^2-y_1^2-y_2^2\right)^3}}
\,dy_1dy_2
\label{LNGred}
\ee

\subsection{Linear approximation to NG equation and its
generic solution}

Equations (\ref{NGH}) are highly non-linear in $y_0$ and it
is convenient to begin with their $y_0$-linear approximation.
Expanding (\ref{LNGred}) in powers of $y_0$, we obtain
\be
\int \frac{dy_1dy_2}{(1-y_1^2-y_2^2)^{3/2}}
-\frac{1}{2}\int \left(
\frac{(\partial_iy_0)^2-
(y_i\partial_i y_0 - y_0)^2}{(1-y_1^2-y_2^2)^{3/2}}
+ \frac{3y_0^2}{(1-y_1^2-y_2^2)^{5/2}}\right)dy_1dy_2 + O(y_0^4)
\ee
The first (divergent) term is non-essential for equations of
motion. The $y_0$-quadratic term gives rise to $y_0$-linear
approximation to equations (\ref{NGH}) in the simple form:
\be
\Delta y_0 = 0
\label{NGlin}
\ee
where
\be
\Delta = \Delta_0 - {\cal D}^2 + {\cal D}
%,\nn \\
%\Delta_0 = 4\partial\bar\partial, \ \ \
%{\cal D} = z\partial + \bar z\bar\partial
\label{NGLaplace}
\ee
is expressed through the ordinary Laplace
\be
\Delta_0 = \frac{\partial^2}{\partial y_1^2}
+\frac{\partial^2}{\partial y_2^2} =
4\frac{\partial^2}{\partial z\partial\bar z} =
4\partial\bar\partial
\ee
and dilatation operators
\be
{\cal D} = y_1\frac{\partial}{\partial y_1}+
y_2\frac{\partial}{\partial y_2} =
z\frac{\partial}{\partial z} +
\bar z\frac{\partial}{\partial \bar z}
= z\partial + \bar z\bar\partial
\ee

If there were no dilatation operators in (\ref{NGLaplace}),
like it happens in the flat $R^3$ space
(i.e. if we linearize not only w.r.t. $y_0$ but also
w.r.t. $y_1$ and $y_2$),
then the solution of the ordinary Laplace equation
$\Delta_0 y_0^{flat} = 0$ would be just a combination
of holomorphic and antiholomorphic functions,
\be
y_0^{flat} = \sum_{k\geq 0}{\rm Re}\left(\alpha_{k0} z^k\right)
\ee
However, in $AdS_3$ case the situation is different:
$\alpha_{kj} \neq 0$ for all $j\neq 0$ in (\ref{y0serC}).
Substitution of (\ref{y0serC}) into (\ref{NGlin})
gives rise to linearized version of recurrence relations,
\be
\alpha_{k,j+1}^{lin} = \frac{(k+2j)(k+2j-1)}{4(j+1)(k+j+1)}\,
\alpha_{kj}^{lin},
\label{rerelin}
\ee
which can be easily resolved to give:
\be
\alpha_{kj}%^{lin}
= \frac{k!(k+2j-2)!}{4^jj!(k-2)!(k+j)!}\,
\alpha_{k0} + O(\alpha^3) =
\frac{k(k-1)}{4^jj!}\frac{(k+2j-2)!}{(k+j)!}\,
\alpha_{k0} + O(\alpha^3)
\ee
(One easily recognizes here eq.(4.2) of \cite{malda3}
with $k=n/2$.)
Therefore in linear approximation
\be
y_0^{lin} =
\sum_{k\geq 0}  {\rm Re}\left(\alpha_{k0} z^k\right)\cdot
{\phantom .}_2\!F_1\left(\frac{k}{2},\frac{k-1}{2};k+1;\,
z\bar z\right)
\ee
is a combination of hypergeometric functions
\be
\!\!{\phantom .}_2\!F_1(a,b;c;x) =
\sum_{j\geq 0} \frac{\Gamma(a+j)\Gamma(b+j)}{j!\Gamma(c+j)}\,x^j
\ee

\subsection{Back to non-linear NG equations}

The full non-linear NG equation, implied by (\ref{LNGred}),
can be written in a form, which looks like a deformation of
%somewhat similar to
(\ref{NGlin}):
\be
\left\{\Big(1+y_0^2
%%+ (y_1^2+y_2^2-1)
+(y^2-1)(\partial_i y_0)^2 -
2y_0({\cal D}y_0)\Big)\Delta_0 - {\cal D}^2 + {\cal D}
%%- (y_1^2+y_2^2-1)
%+ (1-y^2)\partial_iy_0\partial_jy_0 \partial^2_{ij}
%+ 2y_0y_i\partial_jy_0
+\Big((1-y^2)\partial_iy_0 + 2y_iy_0\Big)\partial_jy_0\,
\partial^2_{ij}\right\} y_0 = 0
\ee
where $y^2=y_1^2+y_2^2$ and $i,j=1,2$,
or, in complex notation,
\be
\left\{\Big(1+y_0^2+2(z\bar z-1)\partial y_0\bar\partial y_0
- y_0{\cal D}y_0\Big) \partial\bar\partial
- \frac{1}{4}({\cal D}^2-{\cal D}) + \right. \nn \\
\left.\phantom{\frac{1}{4}}
+\Big((1-z\bar z)(\bar\partial y_0)^2 + zy_0\bar\partial y_0\Big)
\partial^2
+\Big((1-z\bar z)(\partial y_0)^2 + \bar zy_0\partial y_0\Big)
\bar\partial^2\right\} y_0 = 0
\label{nonlin}
\ee
If all terms with $y_0$ in curved brackets are neglected,
we return back to (\ref{NGlin}).
Note that the equation is at most cubic in $y_0$,
what implies that it can be obtained also from some
$\phi^4$-type action, somewhat less non-linear than NG one.

If equation (\ref{nonlin}) is solved iteratively, it gives
rise to more sophisticated recurrence relations.
In order to obtain them we rewrite (\ref{nonlin}) as
$\Delta y_0 = 4h$, where $h$ is formed by all $y_0$-cubic
terms in (\ref{nonlin}).
Then instead of (\ref{rerelin}) we get
\be
\alpha_{k,j+1}^{(h)} = \frac{(k+2j)(k+2j-1)}{4(j+1)(k+j+1)}\,
\alpha_{kj}^{(h)} + \frac{1}{(j+1)(k+j+1)}\,h_{kj},
\label{rereh}
\ee
At the next stage $h_{kj}$ are substituted by cubic combinations
of $\alpha_{k'j'}$ with lower values of
$k'$ and $j\,'$ and this provides
cubic recurrence relations for $\alpha_{kj}$,
which we do not write down explicitly in this paper.

\newpage

\section{Angles in the case of approximately imposed
boundary conditions
\label{andi}}
\setcounter{equation}{0}

\subsection{Approximation can damage IR properties
of the regularized action}

Before we proceed in s.\ref{bori} to construction
of the boundary ring ${\cal R}_\Pi$ for a given polygon $\Pi$,
consider a reversed problem: how can
polygon $\Pi$ be defined by a pair
of algebraically independent elements from ${\cal R}_\Pi$, say
\be
\Pi = \left\{ \begin{array}{c}
P_2 = 0, \\
P_\Pi = 0
\end{array} \right.
\label{PthrR}
\ee
There are only two equations because we assume that
there are just three $y$-variables, i.e. $y_3=0$.
For one of these equations one can always take $P_2=0$
because we assume existence of inscribed circle and thus
of distinguished element $P_2 \in {\cal R}_\Pi$.
In some of our approximate considerations we actually substitute
the second equation $P_\Pi = 0$ by some truncated
series for $y_0$, $y_0 - F(y_1,y_2)=0$, which does not
belong to ${\cal R}_\Pi$.
Thus instead of $\Pi$ we obtain some approximation:
\be
\tilde\Pi = \left\{ \begin{array}{c}
P_2 = y_0^2+1-y_1^2-y_2^2 = 0, \\
y_0 = F(y_1,y_2)
\end{array}
\right.
\label{appPthrR}
\ee
and instead of $\bar\Pi$
%(projection of $\Pi$ on the $y_1,y_2)$ plane)
-- a curve on the $y_1,y_2)$ plane
\be
\widetilde{\bar\Pi} =
\Big\{ G_{\Pi}(y_1,y_2) = 0\Big\}
\ee
In the case of (\ref{appPthrR}) this
\be
G_\Pi(y_1,y_2) = F^2(y_1,y_2)+1-y_1^2-y_2^2,
\label{Gdef}
\ee
but even if the second equation in (\ref{appPthrR}) is
not explicitly resolved w.r.t. $y_0$, there will be
a polynomial $G_{\Pi}(y_1,y_2)$, defining
$\widetilde{\bar\Pi}$.

Of course, in approximate treatment $\widetilde{\bar\Pi}$
is no longer a polygon, actually, for two reasons:
it is not made from straight segments and it does not
contain {\it angles}, generically $G=0$ is a smooth curve.
The latter deviation from polygonality can be most
disturbing for applications, like string/gauge duality,
which involve consideration of areas of our minimal surfaces.
Since in our approach
\be
r^2 = P_2\ \stackrel{(\ref{Gdef})}{=}\ G_2(y_1,y_2),
\ee
the area in question is
\be
{\cal A} = \int L_{NG}d^2y = \int_{G>0} \frac{H d^2y}{G}
\ee
with some non-singular function $H(y_1,y_2)$ in denominator.
This integral diverges at the boundary of integration domain,
where $G=0$, but this is generically a logarithmic
divergence: if integral is regularized in any of the two
obvious ways,
\be
{\cal A}[\varepsilon] = \int_{G > \varepsilon}\frac{H d^2y}{G}
\ee
or
\be
{\cal A}(\epsilon) =  \int_{G>0} \frac{H d^2y}{G^{1-\epsilon}},
\ee
to be called $\varepsilon$- and $\epsilon$-regularizations
in what follows,
we generically get
\be
{\cal A}[\varepsilon] \sim \log\varepsilon \oint
\sqrt{h^{[\varepsilon]}}dl + A_{finite}^{[varepsion]}
\label{varepar}
\ee
or
\be
{\cal A}(\epsilon) \sim \frac{1}{\epsilon}\oint
\sqrt{h^{(\epsilon)}}dl + A_{finite}^{(varepsion)}
\label{epar}
\ee
However, if the resulting metrics $h$ are themselves singular,
divergence can become quadratic, and this is what actually
happens if the curve $\widetilde{\bar\Pi}:\ G=0$ is singular:
has {\it angles}.
Then additional terms,
\be
\sum_{angles} (\log\varepsilon)^2\cdot \kappa(angle)
\ \ \ {\rm and}  \ \ \
\sum_{angles} \frac{1}{\epsilon^2}\cdot \kappa(angle)
\ee
appear at the r.h.s. of (\ref{varepar}) and (\ref{epar})
respectively.
Since $\kappa(angle) \sim \sin(angle)$,
smoothening of the curve  has a drastic
effect on divergencies of regularized area, which are
interpreted as IR singularities in string/gauge duality studies.
This smoothening can be of course actually considered as an
alternative (or, rather, supplementary) regularization, but
using it can further obscure the problem, which is already
sufficiently complicated.
Instead one can require that the angles -- sources of
dominant (quadratic) IR divergencies -- are preserved by
our approximate schemes.
This imposes a new kind of restrictions on the free parameters
of formal series solutions and provide an alternative way
to fix some of them
(which gives values slightly different from other approaches).

\subsection{Angles and discriminants}

Singularities in algebraic geometry are analytically
described in terms of discriminants and resultants,
see \cite{NOLAL,nolal2} for a modernized presentation of
these methods, of which only a standard elementary part
will be used in this paper.

The curve $G(y_1,y_2)=0$ possesses angles whenever
repeated discriminant vanishes,
\be
{\rm discrim}_{y_2}\left(
{\rm discrim}_{y_1}\Big(G(y_1,y_2)\Big)
\right)
\label{didi}
= 0
\ee
Indeed, as a function of $y_1$ the polynomial
$G(y_1,y_2)$ can be decomposed into a product
\be
G(y_1,y_2) = \prod_{\nu} \Big(y_1-\lambda_\nu(y_2)\Big)
\label{evdeco}
\ee
Each eigenvalue $\lambda_\nu(y_2)$ describes a branch
of our curve.
Branches intersect whenever the two eigenvalues coincide,
i.e. when discriminant \cite{disc}
\be
D(G;y_2) = {\rm discrim}_{y_1}(G) \sim \prod_{\mu<\nu}
\Big(\lambda_\mu(y_2) - \lambda_\nu(y_2)\Big)^2
\label{discrim}
\ee
vanishes.
For given function $G$ this condition defines some
points on the $(y_1,y_2)$ plane, a variety of complex
codimension one.
However, there are two different situations:
two branches can {\it merge} and they can indeed {\it intersect}.
Merging is in the degree of discriminant's zero at the
intersection.
If two branches are indeed intersecting at some non-vanishing
angle at a point $y_2=y_{20}$, we expect that
\be
\lambda_\mu(y_2) - \lambda_\nu(y_2) =
(\lambda'_\mu-\lambda'_\nu)(y_2-y_{20})
\ee
where the difference of $\lambda$-derivatives at point $y_{20}$
is the tangent of the intersection angle.
However, this implies that
discriminant in (\ref{discrim}) behaves as $(y_2-y_{20})^2$,
i.e. has a {\it double} zero.
This is not usual, normally discriminant zeroes are of the first
order, then $\delta\lambda \sim \sqrt{y_2-y_{20}}$ and
the branches {\it merge} smoothly, tangents to the
curves $y_1 = \lambda_\mu(y_1)$ and $y_2 = \lambda_\nu(y_2)$
coincide (as it happens, for example, when the two real roots
of quadratic polynomial merge and then decouple into two complex
conjugate ones: the difference between the two roots has a square
root singularity what means that the tangents get both vertical
and thus coincide!).
Thus the condition that two branches intersect at non-vanishing
angle, i.e. that $\widetilde{\bar\Pi}$ has angles, is that
discriminant $D(G;y_2)$ possesses double zeroes, i.e. that
{\it its own} discriminant vanishes:
\be
{\rm discrim}_{y_2}\Big(D(G;y_2)\Big) = 0
\ee
This is exactly the equation (\ref{didi}) -- and it is a
restriction on the shape of the function $G(y_1,y_2)$.

\subsection{A way to proceed in $\varepsilon$-regularization}

Making use of decomposition (\ref{evdeco}), we can write
\be
\frac{1}{G} = \frac{1}{\prod_i (y_1 - \lambda_i(y_2))}
= \sum_i \frac{1}{y_1 - \lambda_i}
\prod_{j\neq i}\frac{1}{\lambda_j-\lambda_i}
\ee
Divergent part of integral over $y_1$ is thus
\be
\int \frac{dy_1}{G(y_1,y_2)} \sim \log\varepsilon
\sum_i  \prod_{j\neq i}\frac{1}{\lambda_j-\lambda_i}
\ee
and the remaining integral over $y_2$ diverges whenever some
$\lambda_j(y_2) = \lambda_k(y_2)$, i.e. at $y_2$
which are roots of the discriminant ${\rm Disrim}_{y_1}(G)$.
It is also clear that these are the {\it angles} of our
boundary, $G(y_1,y_2)=0$ which consists of lines
$y_1=\lambda_i(y_2)$, at intersection points they form
angles, and these angles produce quadratic divergencies.
Linear divergencies come from the sides (lines themselves)
and we are interested in separating the finite piece.

The basic example is $G = (1-y_1^2)(1-y_2^2)$, then
it is easy to observe the $(\log \varepsilon)^2$.

Similarly one can analyze $\epsilon$-regularization.

It is unclear how to extract the finite part. Probably this
could be done numerically, but for this the divergent parts
should first be subtracted "by hands".

\subsection{$Z_n$-symmetric examples}

We illustrate above consideration with the help of
a few examples.
For the sake of simplicity we pick up the
$Z_n$-symmetric configurations, analyzed in \cite{malda3}.

Plots for $y_1(y_2)$ are obtained by solving
\be
P_2=y_0^2+1-y_1^2-y_2^2 = 0
\ee
with
\be
y_0 = c_n K_{n/2} = 2^{1-n/2} c_n {\rm Im} (y_1+\i y_2)^{n/2}
\label{y0Kn}
\ee
The following is a small piece of calculations behind
s.4.3.4 of \cite{malda3}.

\subsubsection{$n=4$}

\begin{figure}\begin{center}
%{\includegraphics[width=100pt,height=100pt]
%{./pics/discn402.eps}}
{\includegraphics[width=100pt,height=100pt]
{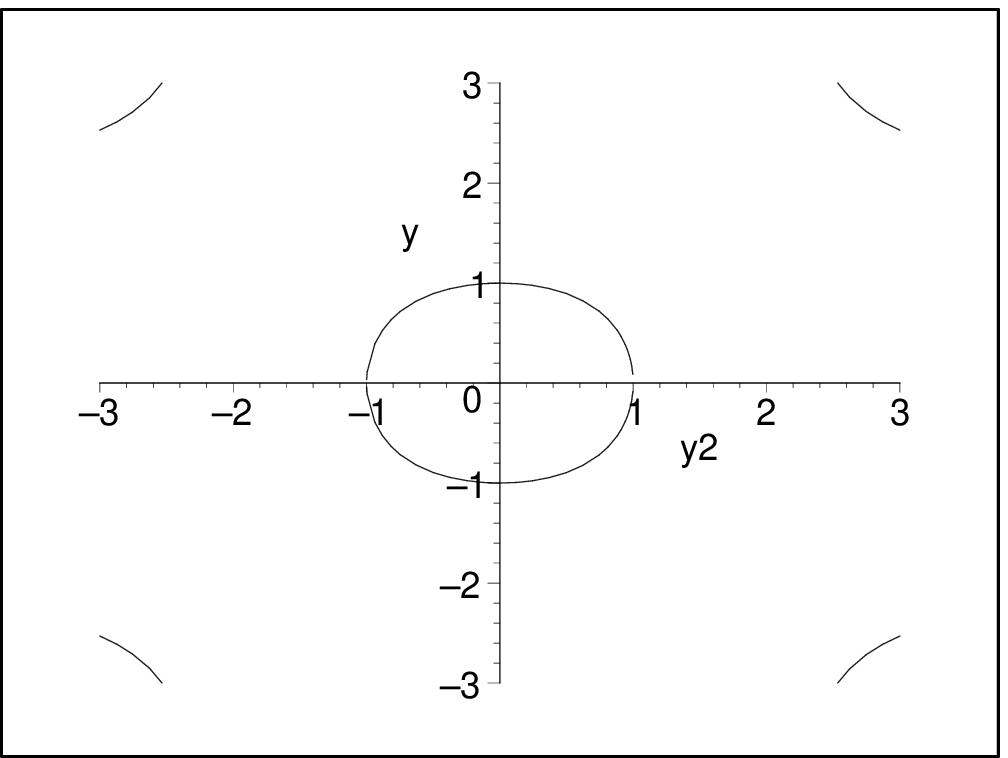}}
{\includegraphics[width=100pt,height=100pt]
{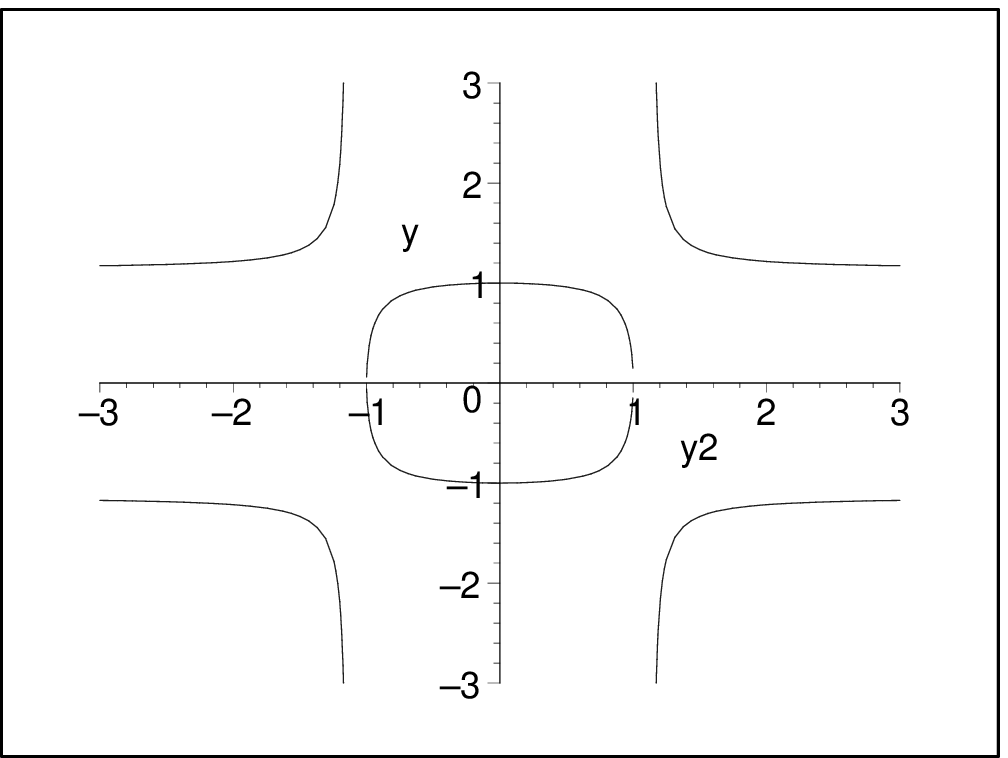}}
{\includegraphics[width=100pt,height=100pt]
{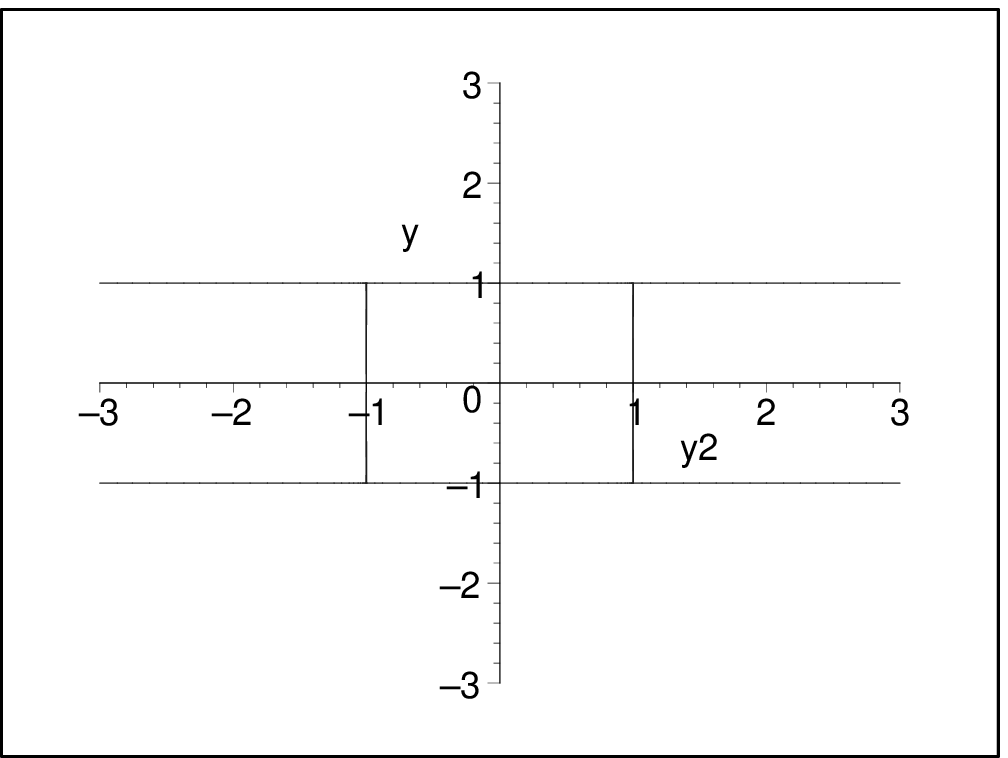}}
{\includegraphics[width=100pt,height=100pt]
{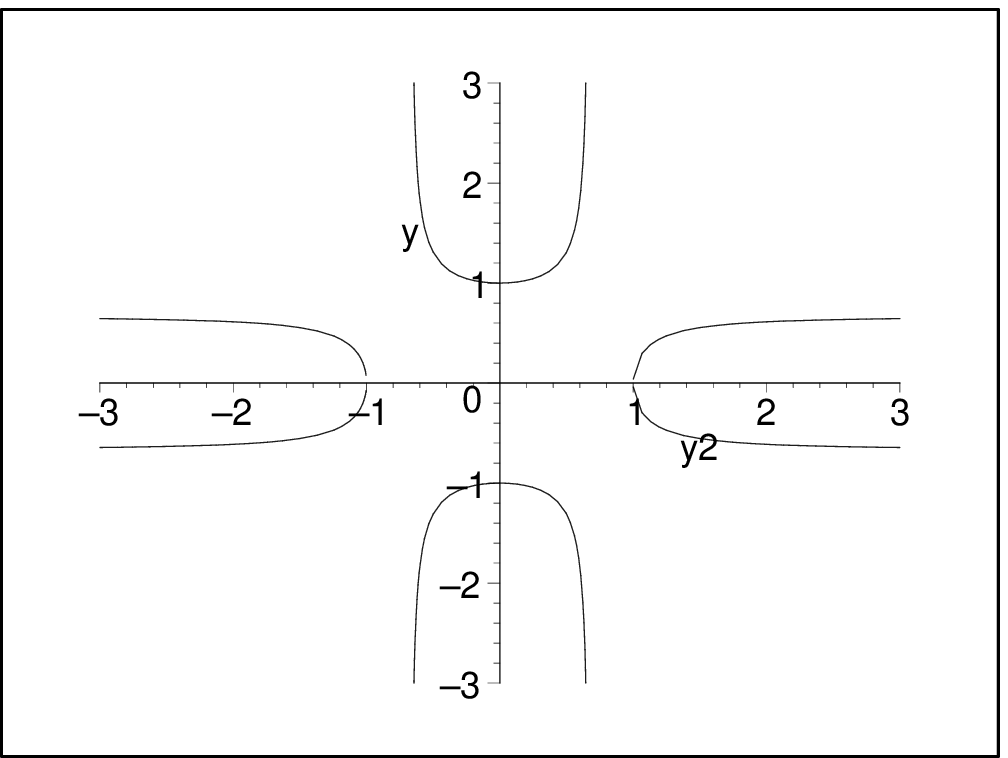}}
\caption{{\footnotesize
The plot of the function $y_1(y_2)$ in (\ref{n4y1vy2})
at different values of $c$=
0.5, 0.87, 1 and 1.5.
It is clearly seen that the unit square is formed at exactly
$c=1$, as predicted by (\ref{n4didi}.
}}
\label{squarediresult}
\end{center}\end{figure}

In this case the equation
\be
G(y_1,y_2)=(cy_1y_2)^2+1-y_1^2-y_2^2=0
\label{n4y1vy2eq}
\ee
is easily resolved:
\be
y_1 = \pm \sqrt{\frac{y_2^2-1}{c^2y_2^2-1}}
\label{n4y1vy2}
\ee
and plots of this function at different values of $c$
are shown in Fig.\ref{squarediresult}.
Distinguished point $c=1$ is clearly see.
In terms of discriminants we have:
\be
D(G;y_2) = {\rm discrim}_{y_1}(G) =
4(c^2y_2^2-1)(y_2^2-1)
\ee
(the two branches in (\ref{n4y1vy2}) merge when discriminant
vanishes,
either at zero or at infinity, when $y_2=\pm c^{-1}$
and $y_2=\pm 1$ respectively), and
\be
{\rm discrim}_{y_2}D(G;y_2) = 65536c^2(c^2-1)^4
\label{n4didi}
\ee
(double discriminant vanishes when branches intersect:
at $c=\pm 1$ they do so at four points, thus zero is of
the fourth power --
the vertices of our square,-- while at $c=0$ an intersection
at two points takes place at infinity).

\bigskip
\bigskip

\subsubsection{$n=6$}

This time the
plots for $y_1(y_2)$ obtained by solving
\be
P_2=y_0^2+1-y_1^2-y_2^2 = 0
\ee
with
\be
y_0 = c_3K_3 = cy_2(3y_1^2-y_2^2)
\label{y0K3}
\ee
so that $c=\frac{1}{4}c_3$.

\begin{figure}\begin{center}
{\includegraphics[width=150pt,height=150pt]
{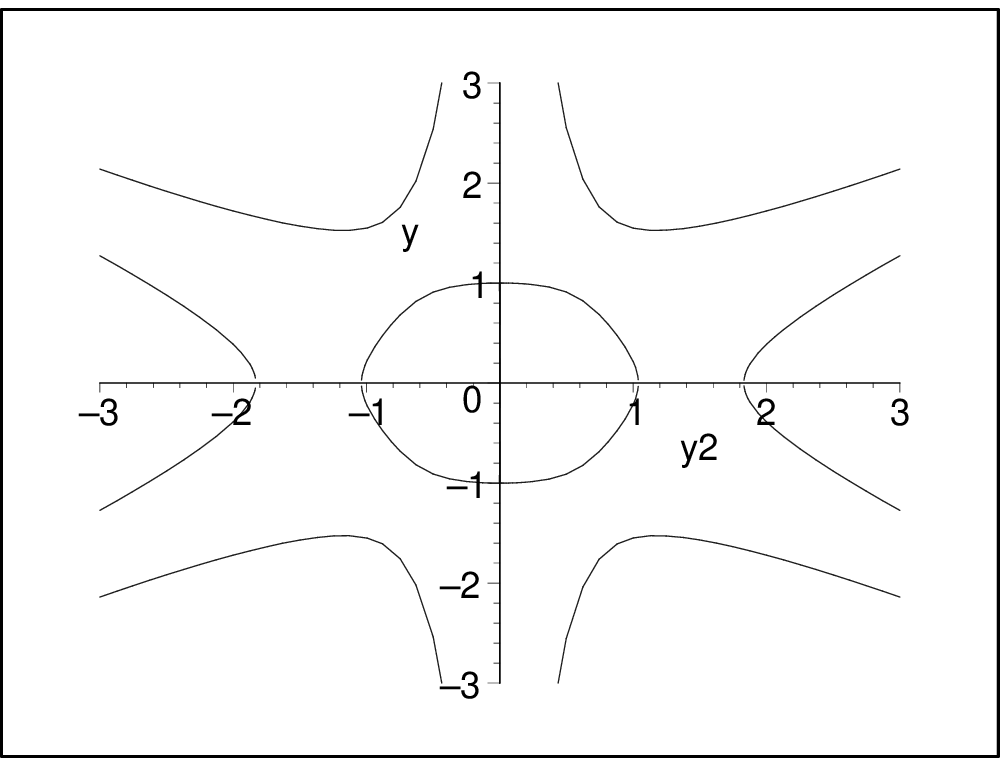}}
{\includegraphics[width=150pt,height=150pt]
{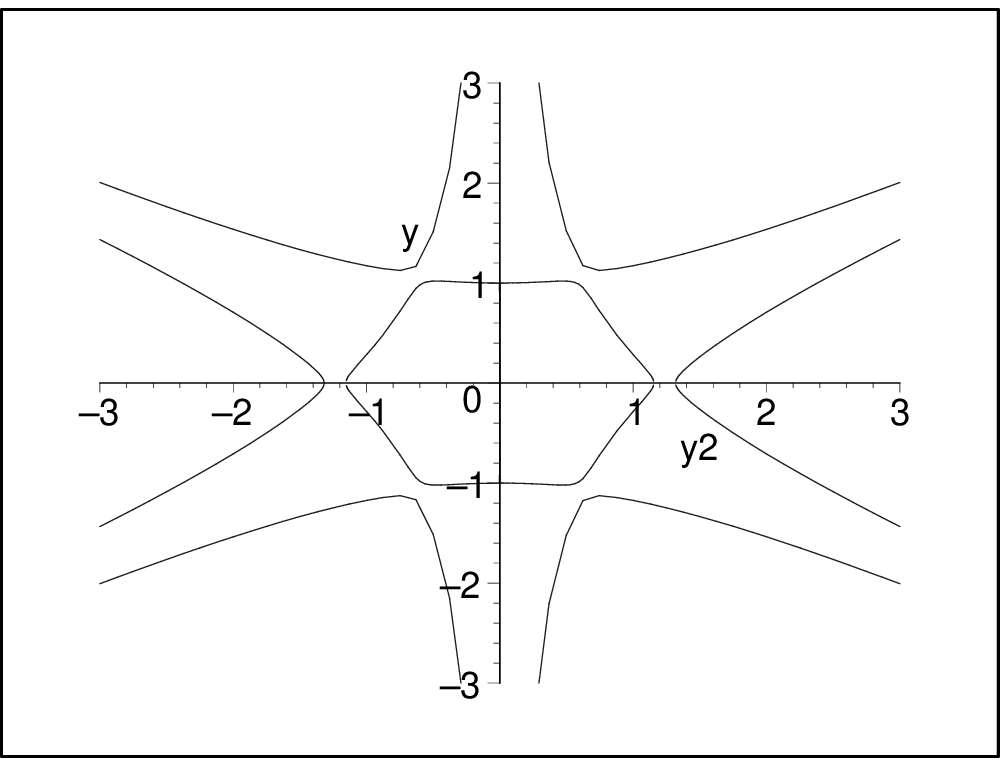}}
\caption{{\footnotesize
The plot of $y_1(y_2)$ at $c=1/4$ (left) and at
$c=\frac{2\sqrt{3}}{9}-\frac{1}{100}$ (right).
In the left picture the central domain is far from being
a polygon: at this value of $c$ it looks almost like a circle
(and will get even closer to this shape for smaller $|c|$).
The right picture shows what happens in a close vicinity
of the critical value of $c=\frac{2\sqrt{3}}{9}$.
The central domain still does not possess angles, see also
Fig.\ref{hexadiresultenlarged}, but is already close to that.
Note that parameter $c$ here is different from $c_{00}^{(6)}$
in \cite{malda3}:
$c = \frac{1}{4}c_{00}^{(6)}$.
}}
\label{hexadiresultleft}
\end{center}\end{figure}

\begin{figure}\begin{center}
{\includegraphics[width=150pt,height=150pt]
{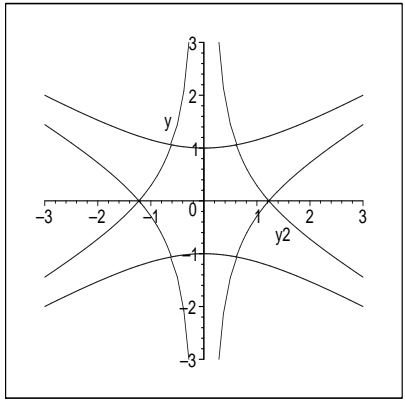}}
%\Fig{hexadiresult303}
%{150,150}
\caption{{\footnotesize
The plot of $y_2(y_1)$ at the critical value of
$c=\frac{2\sqrt{3}}{9}$.
Angles are well seen at the intersections of different
branches, the central domain looks similar to a hexagonal
polygon. Despite angles exist, the sides are not exactly
straight: (\ref{y0K3}) satisfies boundary conditions
(and also NG equations) only approximately, this value
of $c$ is distinguished by existence of angles.
}}
\label{hexadiresult303}
\end{center}\end{figure}

\begin{figure}\begin{center}
{\includegraphics[width=150pt,height=150pt]
{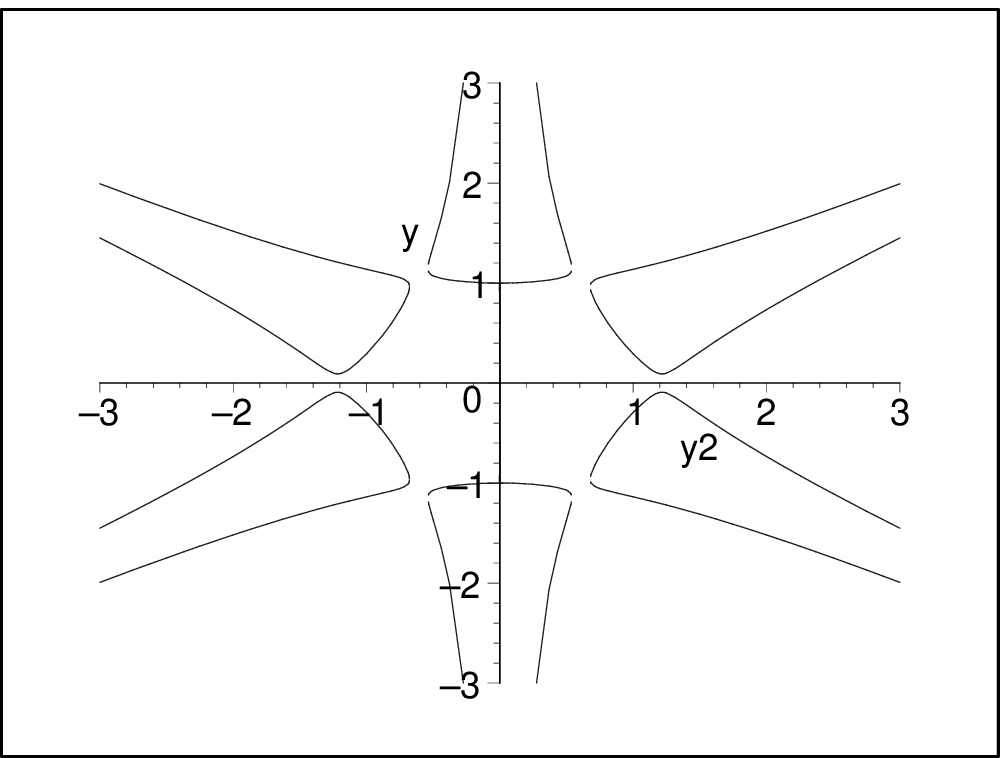}}
{\includegraphics[width=150pt,height=150pt]
{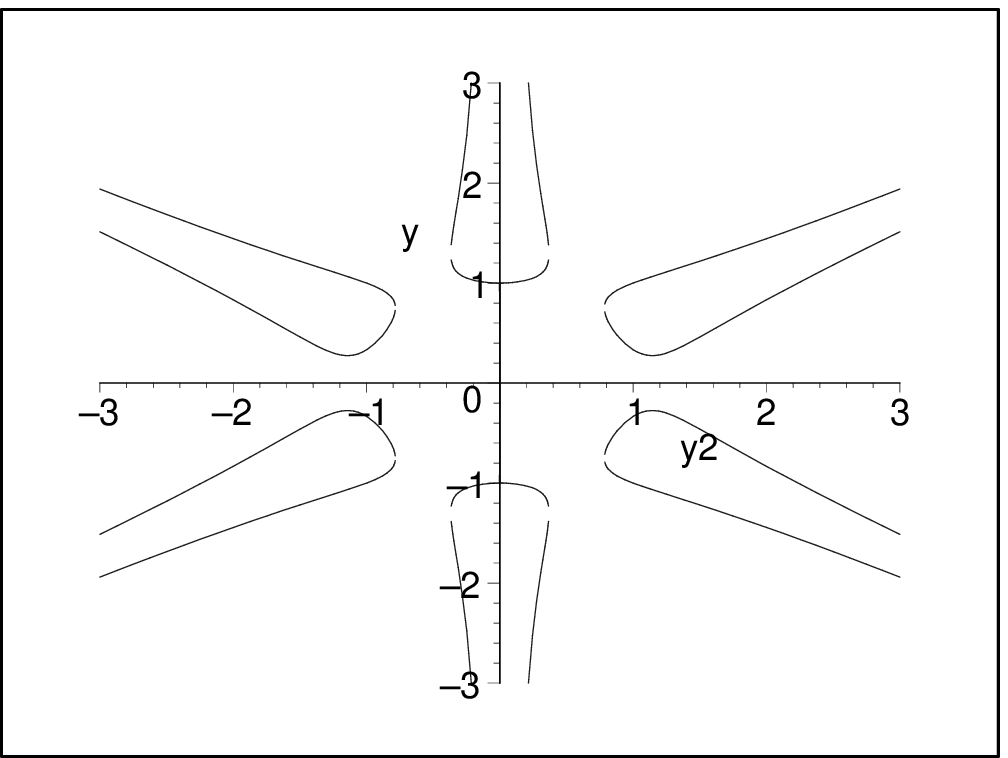}}
\caption{{\footnotesize
The plot of $y_1(y_2)$
at $c=\frac{2\sqrt{3}}{9}+\frac{1}{100}$ (left), in
close vicinity of the critical value of $\frac{2\sqrt{3}}{9}$,
and at $c=\frac{1}{2}$ (right), a little further away.
Different branches are now intersecting at complex values
$y$-variables, and the central domain is no longer closed.
}}
\label{hexadiresultright}
\end{center}\end{figure}

\begin{figure}\begin{center}
{\includegraphics[width=150pt,height=150pt]
{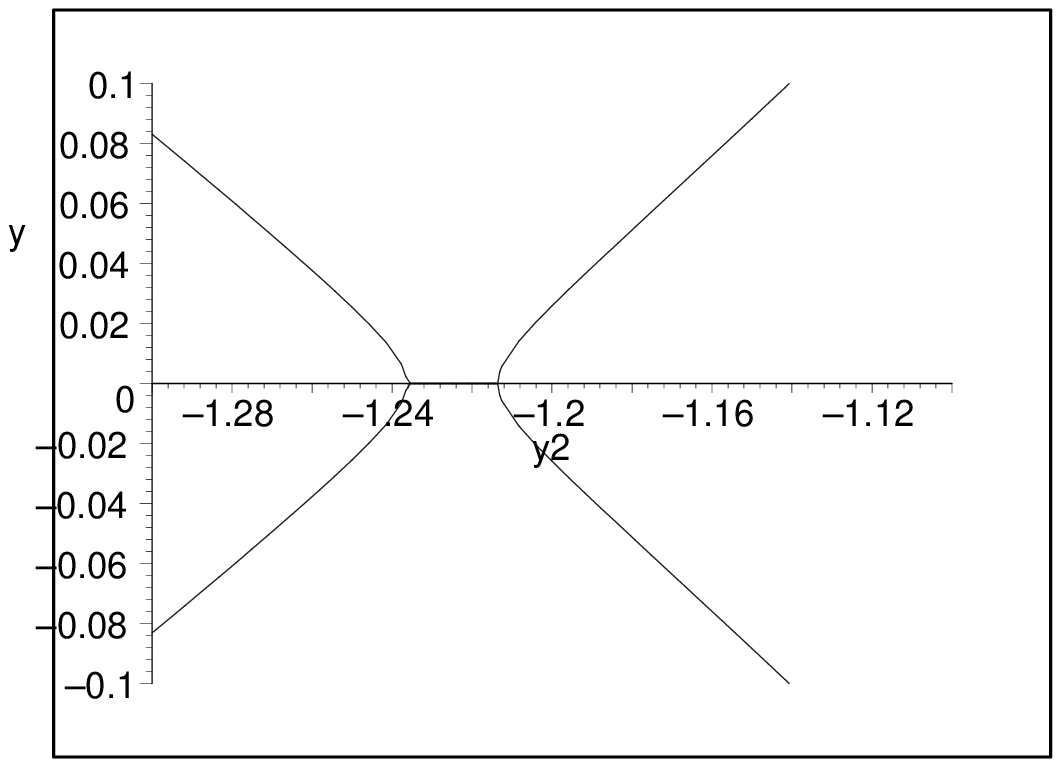}}
{\includegraphics[width=150pt,height=150pt]
{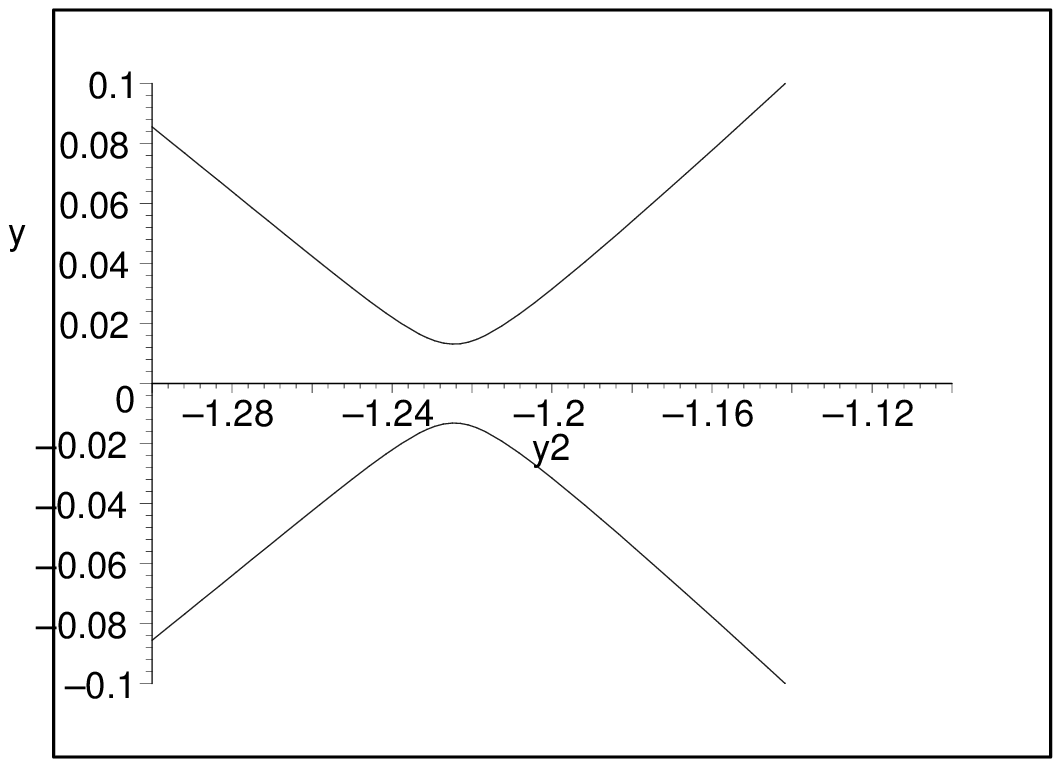}}
\caption{{\footnotesize
Enlarged pictures, showing the vicinity of the branches
merging point at
$c=\frac{2\sqrt{3}}{9}-\frac{2}{1000}$ (left picture)
and
$c=\frac{2\sqrt{3}}{9}+\frac{2}{1000}$ (right picture)
-- i.e. at the very close vicinity of the critical point
$c=\frac{2\sqrt{3}}{9}$. Clearly, no angles are present
at "microscopic" level. They appear exactly at the critical
point, where the two branches intersect.
}}
\label{hexadiresultenlarged}
\end{center}\end{figure}

Discriminant
\be
{\rm discrim}_{y_1}(G) =
144c^2y_2^2(c^2y_2^6-y_2^2+1)(48c^2y_2^4-36c^2y_2^2+1)^2
\ee
Since powers appear at the r.h.s.,
repeated discriminant w.r.t. $y_2$ is vanishing
and we need to look at the individual factors at the r.h.s.:
\be
{\rm discrim}_{y_2}(c^2y_2^6-y_2^2+1) = -64c^6(27c^2-4)^2,\nn\\
{\rm discrim}_{y_2}(48c^2y_2^4-36c^2y_2^2+1) =
1769472c^6(27c^2-4)^2,\nn\\
{\rm resultant}_{y_2}(c^2y_2^6-y_2^2+1,+48c^2y_2^4-36c^2y_2^2+1)=
c^8(216c^2+49)^4
\ee
The interesting critical values of $c$ are zeroes of $27c^2-4$,
i.e. $c=\pm \frac{2\sqrt{3}}{9} = \pm 0.38490\ldots$.
Figs.\ref{hexadiresultleft}-\ref{hexadiresultenlarged}
show exact meaning of these calculations and preceding
argumentation.

%discrn3.mws
%discrn3corr.mws

\bigskip
\bigskip

\bigskip
\bigskip

\bigskip
\bigskip

\bigskip
\bigskip

\subsubsection{$n=8$}

\begin{figure}\begin{center}
{\includegraphics[width=150pt,height=150pt]
{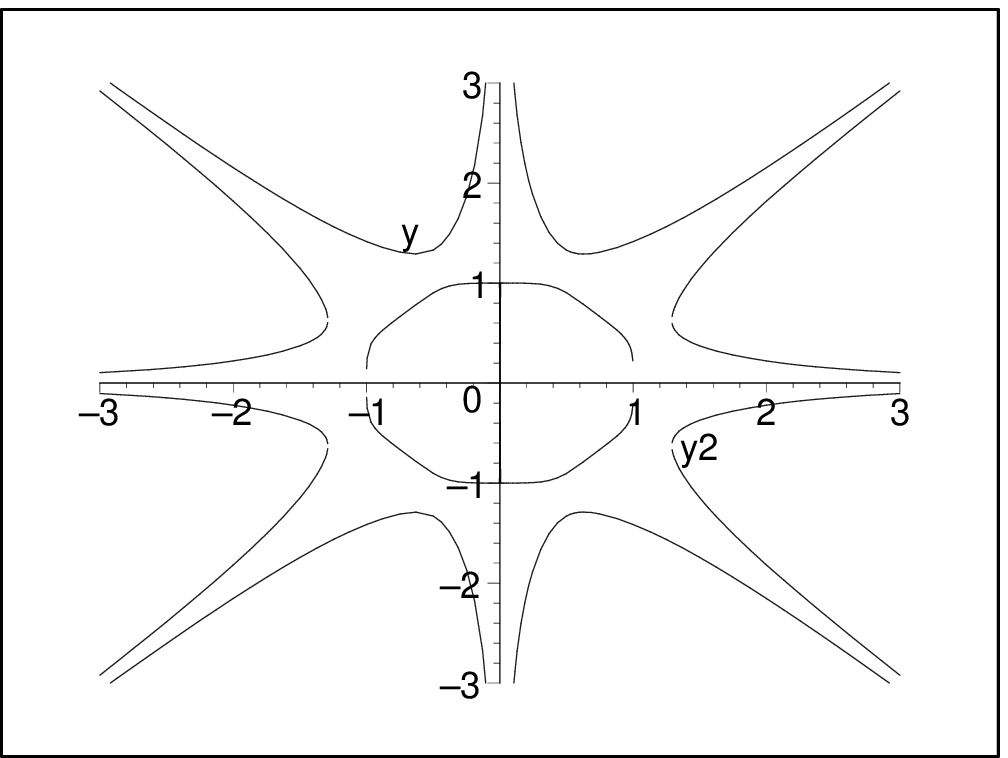}}
{\includegraphics[width=150pt,height=150pt]
{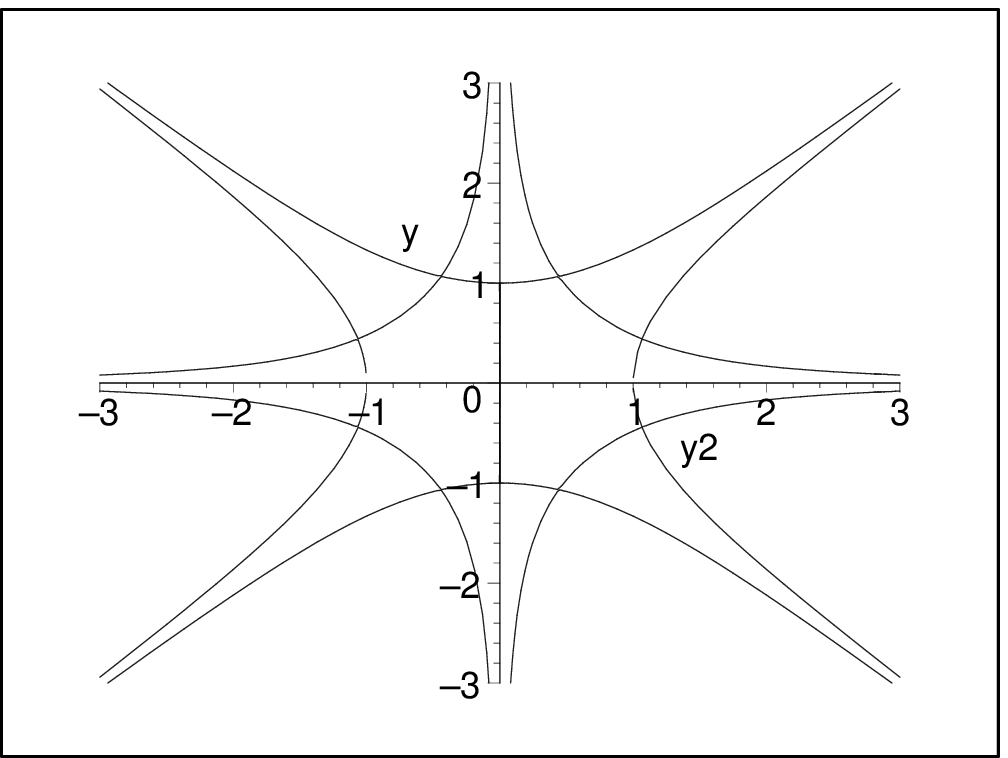}}
{\includegraphics[width=150pt,height=150pt]
{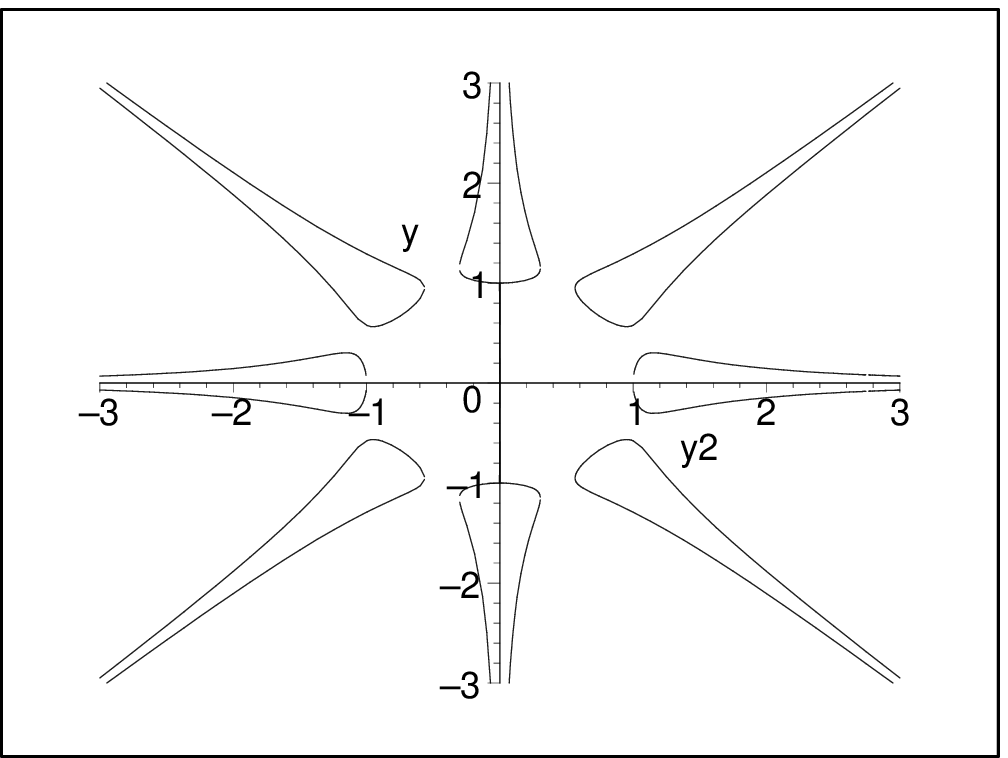}}
\caption{{\footnotesize
The analogues of
Figs.\ref{hexadiresultleft}-\ref{hexadiresultright}
for $n=8$ with
$\ y_0 = cK_4 = cy_1y_2(y_1^2-y_2^2)\ $ at $\ c=1$,
$\ c= \frac{3\sqrt{3}}{4}$ (the critical value) and
$\ c=\frac{3}{2}$.
At critical value the branches intersect at non-trivial
angles, but the sides of emerging octagon are not straight:
boundary conditions (and NG equations) are matched
only approximately. The sides look "more straight" in the
left picture -- for $c$ below the critical point, where
angles are less pronounced: this illustrates the thesis
that different criteria lead to slightly different
values of the matching parameter $c$.
This choice of parameter is different from $c_{00}^{(8)}$
in \cite{malda3}:
$c = \frac{1}{2}c_{00}^{(8)}$.
}}
\label{octodiresult}
\end{center}\end{figure}

This time
\be
y_0 = c_4K_4 = cy_1y_2(y_1^2-y_2^2),
\ee
so that $c = \frac{1}{2}c_4$,
the plots for $y_2(y_1)$ are shown in Fig.\ref{octodiresult}
and discriminants are:
\be
{\rm discrim}_{y_1}(G) =
64c^6y_2^6(y_2^2-1)g^2(y_2,c),\nn\\
g(y_2,c)=
4-27c^2y_2^2+90c^2y_2^4-71c^2y_2^6-4c^4y_2^{10}+8c^4y_2^{12},\nn\\
{\rm discrim}_{y_2} g = 137438953472
c^{44}(16c^2-27)^4(243c^2+4913)^6
\ee
so that the relevant zero is $c=\frac{3\sqrt{3}}{4}$.

\bigskip
\bigskip

\subsection{Exact solutions to (\ref{didi})}

\begin{figure}\begin{center}
{\includegraphics[width=150pt,height=150pt]
{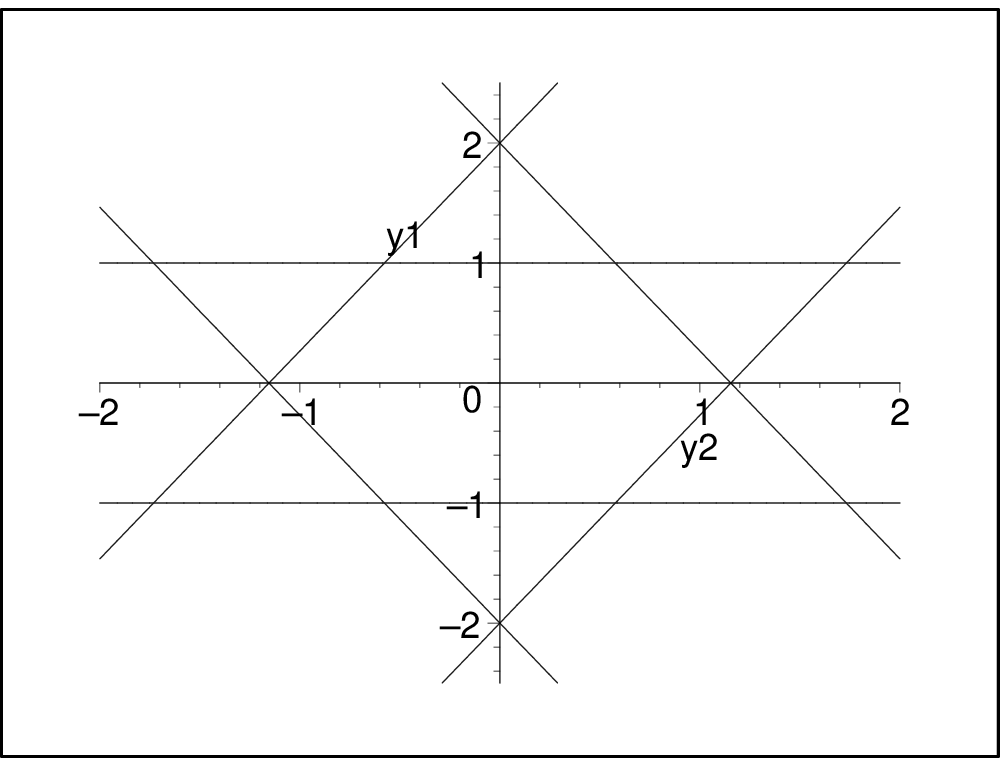}}
{\includegraphics[width=150pt,height=150pt]
{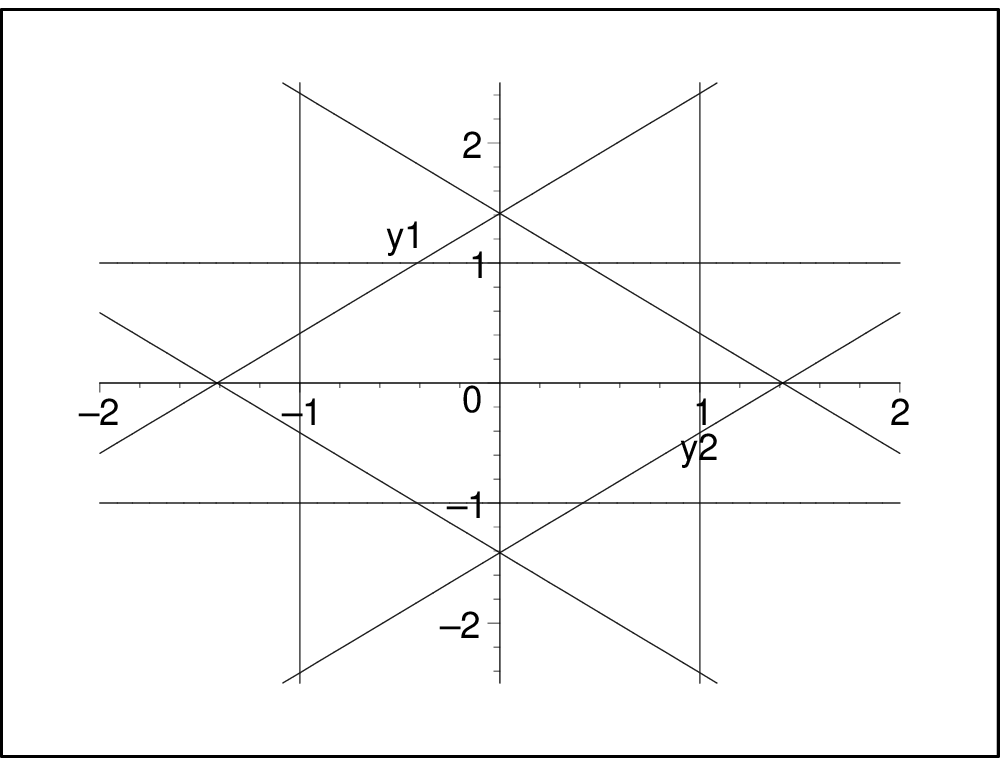}}
\caption{{\footnotesize
The analogues of
Figs.\ref{hexadiresultleft}-\ref{hexadiresultright}
for $n=6$ with
$\ y_0 = \frac{K_3}{(1-\frac{1}{4}y^2)}$
and of Fig.\ref{octodiresult}
for $n=8$ with $\ y_0 = \frac{K_4}{(1-\frac{1}{2}y^2)}$
which satisfy the boundary condition exactly.
Ideal hexagon and octagon
with sharp angles and straight sides
are clearly seen in the pictures.
}}
\label{discrexactresult}
\end{center}\end{figure}

The angle-sharpening problem can actually be reversed:
one can consider (\ref{didi}) as an equation for $G(y_1,y_2)$.
In \cite{malda3} we already showed exact solutions to
this problem:
\be
G_\Pi = K_\Pi^2 + (1-y^2)Q_\Pi^2 = P_\Pi =
\prod_{\stackrel{{\rm segments}}{{\rm of}\ \Pi}}
P_|,
\label{factorG}
\ee
are totally decomposed into a product of linear
functions, associated with individual segments,
see (\ref{Ppolsdef}).
The corresponding analogues of
Figs.\ref{hexadiresultleft}-\ref{octodiresult}
are just $6$ or $8$ straight lines which form
the regular hexagon and octagon at the intersection,
see Fig.\ref{discrexactresult}.
In formulas for (\ref{factorG}) this looks like:
%are totally factorized (decomposed), for example
\be
n=4:& (y_1y_2)^2+(1-y^2) = (1-y_1)(1+y_1)(1-y_2)(1+y_2),\nn\\
n=6: & \left(\frac{y_2(3y_1^2-y_2^2)}{4}\right)^2 +
(1-y^2)\left(1-\frac{y^2}{4}\right)^2 = \nn \\ =&
(1-y_1)(1-cy_1-sy_2)(1+cy_1-sy_2)(1+y_1)(1+cy_1+sy_2)(1-cy_1+sy_2),
\nn\\
&{\rm with}\ \ \ c=\frac{1}{2}, \ \ \ s=\frac{\sqrt{3}}{2}, \nn\\
%& {\rm and\ so\ on}
%-1/16(y1+y2\frac{\sqrt{3}}{2}-2)*(y1-y2*3^(1/2)-2)
%*(y1+y2*3^(1/2)+2)*(y1-y2*3^(1/2)+2)*(y1-1)*(y1+1)
&\ldots
\ee
These examples are provided by the knowledge of boundary rings,
their perturbation like (\ref{calPanzapert}) should give rise
to more solutions and (\ref{didi}) can serve as one more
property of ${\cal P}_\Pi$,
to be added to the list in s.\ref{goal}.

\bigskip
\bigskip

\newpage

\section{NG solution for generic skew quadrilateral
\label{quadrila}}
\setcounter{equation}{0}

Solutions to the $\sigma$-model and NG equations with such
boundary conditions were considered in
\cite{mmt1} and \cite{mmt2} respectively.
Though the single-parametric rhombus family,
originally introduced in \cite{Kru,am1}, is sufficient
for direct application to string-gauge duality studies,
generic solutions are definitely interesting from the
point of view of Plateau problem.
The difficulty is that in \cite{mmt2} NG solution is
not represented in the resolved form, as $y_0(y_1,y_2)$,
it is left in a parametric representation, inherited from
the $\sigma$-model solution of \cite{mmt1}.
The situation is similar to the rhombic solution, which
is transformed from the parametric representation of
\cite{Kru,am1,mmt1} to resolved expression only in
s.2.6 of \cite{malda3}.

\subsection{Solutions from \cite{mmt1,mmt2}}

For $n=4$ coordinate system can always be rotated so, that
the boundary conditions and thus a solution (the one
which does not correspond to spontaneously broken $Z_2$-symmetry
$y_3\rightarrow -y_3$) have $y_3=0$.
The skew quadrilateral $\Pi$ is formed by four null-vectors
only provided $\bar\Pi$ possesses an inscribed circle, thus
the conditions (\ref{ads3}) can always be imposed.
It is only important to remember that in this form it
requires the special choice of coordinate system:
$y_1=y_2=0$ at the center of the circle, and $y_0=0$
at its tangent points with the sides of the quadrilateral
(if $y_0$ vanishes at any of these points, it automatically
does so at the other three).
Thus NG solution is described by a single function $y_0(y_1,y_2)$.

In \cite{mmt1,mmt2} it is instead described in a very different
way: $r$ and ${\bf y} = (y_0;y_1,y_2)$ are expressed through the
variables $z=1/r$ and ${\bf v} = z{\bf y}$, which are actually
the embedding (most natural) coordinates for $AdS$ $\sigma$-model.
In these variables generic solution looks simple:
\be
z = z_1(e^{\vec k_1\vec u} + e^{-\vec k_1\vec u})
+ z_2(e^{\vec k_2\vec u} + e^{-\vec k_2\vec u}), \nn \\
{\bf v} = {\bf v}_1 e^{\vec k_1\vec u}
+ {\bf v}_3 e^{-\vec k_1\vec u}
+ {\bf v}_2 e^{\vec k_2\vec u} + {\bf v}_4 e^{-\vec k_2\vec u}
\label{mmtsols}
\ee
Remaining parameters are constrained by NG equations and
boundary conditions.
The latter imply that
\be
\frac{{\bf v}_{a+1}}{z_{a+1}} - \frac{{\bf v}_a}{z_a}
= {\bf p}_a,
\ \ \ \ a = 1,2,3,4
\label{n4bc}
\ee
where ${\bf p}_a$ are the four null-vectors, forming the sides
of our polygon $\Pi$ (i.e. external momenta of the four gluons).
The former imply that
\be
z_1 = z_3  = \frac{1}{\sqrt{2s}}
= \frac{1}{2\sqrt{{\bf p}_1{\bf p}_2}}, \ \ \
z_2=z_4 = \frac{1}{\sqrt{2t}}
= \frac{1}{2\sqrt{{\bf p}_2{\bf p}_3}}
= \frac{1}{2\sqrt{{\bf p}_1{\bf p}_4}}, \nn\\
s = ({\bf p}_1+{\bf p}_2)^2 = 2{\bf p}_1{\bf p}_2, \ \ \
t = ({\bf p}_2+{\bf p}_3)^2 = 2{\bf p}_2{\bf p}_3
\ee
Our usual variables are:
\be
r = \frac{1}{z}, \ \ \ {\bf y} = \frac{{\bf v}}{z}
\label{yvsvz}
\ee

\subsection{From ${\bf y}(\vec u)$ to $y_0(y_1,y_2)$}

Our goal is to express $y_0$ through $y_1$ and $y_2$, i.e.
to eliminate two variables $\vec u$ from the three-component
vector equation (\ref{mmtsols}) for ${\bf y} = z^{-1}{\bf v}$.
Our strategy is to reformulate the problem in terms of polynomials
and then solve it with the standard methods of
{\it non-linear algebra} \cite{nolal2}.
In result we obtain $y_0$ as a solution to {\it quadratic}
equation, which will be afterwards compared with the results from
boundary ring considerations.

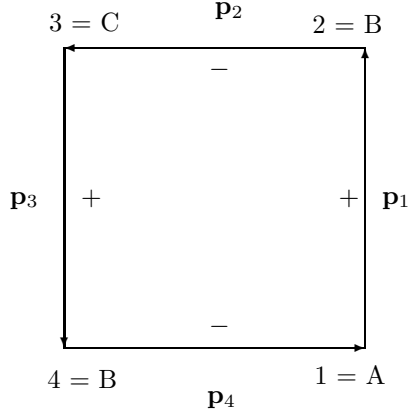
\begin{figure}
\begin{center}
%%
%%
%TeXCAD Picture [figsquare.tex]. Options:
%\grade{\on}
%\emlines{\off}
%\epic{\off}
%\beziermacro{\on}
%\reduce{\on}
%\snapping{\off}
%\pvinsert{% Your \input, \def, etc. here}
%\quality{8.000}
%\graddiff{0.005}
%\snapasp{1}
%\zoom{4.0000}
\unitlength 1mm % = 2.845pt
\linethickness{0.4pt}
\ifx\plotpoint\undefined\newsavebox{\plotpoint}\fi
% GNUPLOT compatibility
\begin{picture}(85,52.25)(0,0)
\put(27.75,47.75){\vector(0,-1){40}}
\put(67.75,47.75){\vector(-1,0){40}}
\put(67.75,7.75){\vector(0,1){40}}
\put(27.75,7.75){\vector(1,0){40}}
\put(60.75,49.75){2 = B}
\put(61,3){1 = A}
\put(25.75,50){3 = C}
\put(25.5,2.5){4 = B}
\put(70,27){${\bf p}_1$}
\put(20.5,27){${\bf p}_3$}
\put(47.75,52.25){${\bf p}_2$}
\put(46.75,0.5){${\bf p}_4$}
\thicklines
\put(64.25,26.75){+}
\put(47,44){$-$}
\put(47,9.75){$-$}
\put(30,26.75){+}
\end{picture}
\caption{{\footnotesize
Convention for labeling sides and vertices of
the quadrilateral, a square is used as an example.
Pluses and minuses stand for $y_0$ increasing (+)
or decreasing (-) along the vector.
}}
\label{figquadrisquare}
\end{center}
\end{figure}

Our equations become polynomial in terms of
$U \equiv e^{\vec k_1\vec u}$ and $W \equiv e^{\vec k_2\vec u}$:
\be
z_1({\bf y} - {\bf y}_A)U + z_2({\bf y} - {\bf y}_B)W
+ z_1({\bf y} - {\bf y}_C)U^{-1}
+ z_2({\bf y} - {\bf y}_D)W^{-1}={\bf 0}
\label{mmtsols2}
\ee
where the four vertices are now denoted by $A,B,C,D$,
see Fig.\ref{figquadrisquare}, and
${\bf y}_a = \frac{{\bf v}_a}{z_a}$,
with $a=A,B,C,D$, $z_A=z_C=z_1$, $z_B=z_D=z_2$,
are the values of ${\bf y}$ at these vertices.
Of course, resolvability of the system (\ref{mmtsols2})
in four variables $U,U^{-1},W,W^{-1}$
requires that the $4\times 4$ determinant vanishes -- and
this is guaranteed by the possibility to choose all $3$-components
of ${\bf y}$ and ${\bf y}_a$ vanishing, so that vectors in
(\ref{mmtsols2}) have only three components, $0,1,2$.
However, since of the four variables $U,U^{-1},W,W^{-1}$
only two are algebraically independent the vanishing of
$4\times 4$ determinant is  not the only resolvability condition.
The more restrictive discriminantal constraint can be derived
as follows.

Take any pair of the three equations in (\ref{mmtsols2}) and
eliminate $W^{-1}$ or $W$:
\be
z_1\Big({\bf K}_{AD}U + {\bf K}_{CD}U^{-1}\Big)
+ z_2{\bf K}_{BD}W = {\bf 0}, \nn \\
z_1\Big({\bf K}_{AB}U - {\bf K}_{BC}U^{-1}\Big)
- z_2{\bf K}_{BD}W^{-1} = {\bf 0}
\label{mmtsols3}
\ee
Here $K_{ab}^\lambda = \epsilon^{\lambda\mu\nu}K_{ab}^{\mu\nu}$
with
\be
%K_{ab}^\lambda
%= \epsilon^{\lambda\mu\nu}K_{ab}^{\mu\nu},\ \ \ \
K^{\mu\nu}_{ab}
= (y^\mu - y_a^\mu)(y^\nu - y_b^\nu) -
(y^\mu - y_b^\mu)(y^\nu - y_a^\nu) =
y^\mu(y_a^\nu-y_b^\nu) + y^\nu(y_b^\mu-y_a^\mu)
+ (y_a^\mu y_b^\nu - y_b^\mu y_a^\nu)
\ee
and $\lambda,\mu,\nu = 0,1,2$
is  {\it linear} in $y$-variables
and antisymmetric in $ab$.

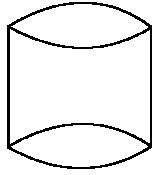
\begin{figure}\begin{center}
%%
%%
%TeXCAD Picture [Caydia.tex]. Options:
%\grade{\on}
%\emlines{\off}
%\epic{\off}
%\beziermacro{\on}
%\reduce{\on}
%\snapping{\off}
%\pvinsert{% Your \input, \def, etc. here}
%\quality{8.000}
%\graddiff{0.005}
%\snapasp{1}
%\zoom{8.0000}
\unitlength 1mm % = 2.845pt
\linethickness{0.4pt}
\ifx\plotpoint\undefined\newsavebox{\plotpoint}\fi
% GNUPLOT compatibility
\begin{picture}(29,32.063)(0,0)
\put(10,9.875){\line(0,1){15.875}}
\put(29,9.875){\line(0,1){16}}
\qbezier(10,25.75)(20.5,32.063)(29,25.625)
\qbezier(29,25.625)(20.188,20.125)(10.125,25.625)
\qbezier(10,9.75)(20.25,15.75)(29,9.75)
\qbezier(29,9.75)(20,4.063)(10,9.625)
\end{picture}
%%
%%
%{\includegraphics[width=150pt,height=150pt]
%{Caydia.pcx}}
\caption{{\footnotesize
Feynman diagram for the Cayley hyperdeterminant (\ref{D23}).
Tensor $T_{\alpha\beta\gamma}$ stands at the
valence-three vertices, while propagators are $\epsilon$-symbols.
See \cite{nolal2} for more explanations.}}
\label{Caydia}
\end{center}\end{figure}

Picking any component of the first
and any component of the second equation in (\ref{mmtsols3})
we can use $WW^{-1}=1$ to obtain {\it nine} equations:
\be
z_2^2 {\bf K}_{BD}\otimes {\bf K}_{BD} = z_1^2
\Big({\bf K}_{AD}U + {\bf K}_{CD}U^{-1}\Big)\otimes
\Big({\bf K}_{BC}U - {\bf K}_{AB}U^{-1}\Big)
\ee
or
\be
z_1^2{\bf K}_{AD}\otimes {\bf K}_{AB} U^4
+ \Big(z_1^2{\bf K}_{CD}\otimes {\bf K}_{AB}
-z_1^2{\bf K}_{AD}\otimes{\bf K}_{BC}
%-\frac{z_2^2}{z_1^2}
+z_2^2{\bf K}_{BD}\otimes {\bf K}_{BD}\Big) U^2
-z_1^2{\bf K}_{CD}\otimes {\bf K}_{BC} = {\bf 0}\times {\bf 0}
\label{mmtsols4}
\ee
Consistency of any pair of these equations is a non-trivial
condition on ${\bf K}$
(all $36$ pairs are giving rise to equivalent $y_0(y_1,y_2)$!).
According to \cite{nolal2},
\be
\sum_{\beta,\gamma=\pm}^2
T_{\alpha\beta\gamma}x_\beta x_\gamma = 0
\ee
is resolvable system of two equations (with $\alpha=1,2$)
for two variables $x_+,x_-$ iff its resultant $R_{2|2}$
-- which in this case coincides with the Cayley discriminant
or "hyperdeterminant"
\cite{Cay}, see Fig.\ref{Caydia},-- vanishes:
\be
D_{2|3}(T) = \varepsilon^{\alpha\alpha''}
\varepsilon^{\alpha'\alpha'''}
\epsilon^{\beta\beta'}\epsilon^{\gamma\gamma'}
\epsilon^{\beta''\beta'''}\epsilon^{\gamma''\gamma'''}
T_{\alpha\beta\gamma}
T_{\alpha'\beta'\gamma'}
T_{\alpha''\beta''\gamma''}
T_{\alpha'''\beta'''\gamma'''} =
\nn \\
= (T_{1++}T_{2--}-T_{1--}T_{2++})^2 +
4(T_{1+-}T_{2++}-T_{1++}T_{+-})(T_{1+-}T_{2--}-T_{1--}T_{2+-})
= 0
\label{D23}
\ee
Of course, this is nothing but the condition that two quadratic
equations have a common root and can be derived by elementary
means, say, from explicit knowledge of the formula for the roots.
In our case $x_+=U^2$, $x_-=1$, and tensor $T_{\alpha\beta\gamma}$
is made out of ${\bf K}\otimes{\bf K}$.
Discriminant $D_{2|3}$ is bilinear in both components of $T_{1..}$
and $T_{2..}$, while ${K^\lambda}$ is linear in the complementary
$y$-variables (i.e. in $y^\mu$ with $\mu\neq \lambda$.
Thus discriminantal condition can be made
quadratic in $y_0$ if we choose as a pair of equations from
(\ref{mmtsols4}) either
$K^0K^1$ and $K^0K^0$  or $K^0K^2$ and $K^0K^0$.
Indeed, $K^0$ is independent of $y_0$, while $K^1$ and $K^2$
are linear in $y_0$, thus the corresponding discriminants
will be quadratic.
Instead, both expressions are {\it a priori} asymmetric
in $y_1$ and $y_2$, one can also consider a linear combination
$K^0(\mu K^1+\nu K^2)$ to put this asymmetry under control.

\bigskip
\bigskip

{\bf Example:}
In the case of the {\bf square} we have,
see Fig.\ref{figquadrisquare}:
\be
{\bf p}_1 = (2;0,2),\ \ \
{\bf p}_2 = (-2;-2,0), \ \ \
{\bf p}_3 = (2;0,-2), \ \ \
{\bf p}_4 = (-2;2,0), \ \ \ \ \ \
z_1 = z_2 = \frac{1}{4}
\ee
and
\be
{\bf y}_A = (-1;1,-1), \ \ \  %(1;1,1), \ \ \
{\bf y}_B = (1;1,1), \ \ \    %(-1;-1,1), \ \ \
{\bf y}_C = (-1;-1,1), \ \ \  %(1;-1,-1), \ \ \
{\bf y}_D = (1;-1,-1), \ \ \  %(-1;1,-1), \ \ \
\label{squarevertices}
\ee
so that
\be
y_0 = \frac{-U+W-U^{-1}+W^{-1}}{U+W+U^{-1}+W^{-1}},\nn\\
y_1 = \frac{U+W-U^{-1}-W^{-1}}{U+W+U^{-1}+W^{-1}},\nn\\
y_2 = \frac{-U+W+U^{-1}-W^{-1}}{U+W+U^{-1}+W^{-1}}
\ee
These equations are simple enough to be solved
directly:
\be
U = \sqrt{\frac{(1+y_1)(1-y_2)}{(1-y_1)(1+y_2)}},\ \ \ \
W = \sqrt{\frac{(1+y_1)(1+y_2)}{(1-y_1)(1-y_2)}}
\ee
and in this case $y_0(y_1,y_2)$ is a solution to the linear
equation:
\be
y_0=y_1y_2
\label{squareq}
\ee
However, equation is essentially quadratic already in the
case of rhombus \cite{malda3}.

\subsection{Evaluating hyperdeterminant}

In general resolving  eqs.(\ref{mmtsols2}) is rather tedious,
moreover (\ref{mmtsols4}) provides $U$ and $W$ as solutions to
biquadratic equations, which are of limited practical use.
However, since we need $y_0(y_1,y_2)$, there is no need to
find $U$ and $W$: this function is defined by discriminantal
condition and what we actually need is evaluation of
hyperdeterminant.
This is a straightforward calculation with a nice answer:
\be
D_{2|3} \sim \big\{{\bf P}_+{\bf Q}_+{\bf Q}_-\big\}%_{++-}
\big\{{\bf P}_-{\bf Q}_+{\bf Q}_-\big\}%_{-+-}
- \big\{{\bf P}_+{\bf P}_-{\bf Q}_+\big\}%_{+-+}
\big\{{\bf P}_+{\bf P}_-{\bf Q}_-\big\}%_{+--}
%\sim \nn \\
%\sim \Big\{({\bf P}_+{\bf P}_-){\bf Q}_+^2{\bf Q}_-^2
%-({\bf P}_+{\bf P}_-)({\bf Q}_+{\bf Q}_-)^2
%+({\bf P}_+{\bf Q}_+)({\bf P}_-{\bf Q}_-)({\bf Q}_+{\bf Q}_-)
%+({\bf P}_+{\bf Q}_-)({\bf P}_-{\bf Q}_+)({\bf Q}_+{\bf Q}_-)
%-\nn \\
%-({\bf P}_+{\bf Q}_+)({\bf P}_-{\bf Q}_+){\bf Q}_-^2
%-({\bf P}_+{\bf Q}_-)({\bf P}_-{\bf Q}_-){\bf Q}_+^2\Big\}
%\ \ - \ \ \Big\{ {\bf P} \leftrightarrow {\bf Q}\Big\}
\label{discreq}
\ee
where
$\big\{{\bf P}{\bf Q}{\bf R}\big\} \equiv \epsilon^{\lambda\mu\nu}
P^\lambda Q^\mu R^\nu$
%and $({\bf P}{\bf Q}) = P^\mu Q^\mu = -P_0Q_0 + P_1Q_1+P_2Q_2$ are
is the mixed product of three $3$-component vectors.
%and the scalar products of $3$-component vectors.
Proportionality coefficient between the first and the second lines
in (\ref{discreq}) is $-1$ for Minkovski signature.
Vectors ${\bf P}_\pm$ and ${\bf Q}_\pm$ are still another version
of parametrization of (\ref{mmtsols2}):
\be
U {\bf P}_+\ + U^{-1}{\bf P}_-\ + W{\bf Q}_+\ + W^{-1}{\bf Q}_- = 0
\ee
i.e.
\be
{\bf P}_+ = z_1({\bf y} - {\bf y}_A), \ \ \ \ \
{\bf P}_- = z_1({\bf y} - {\bf y}_C), \ \ \ \ \
{\bf Q}_+ = z_2({\bf y} - {\bf y}_B), \ \ \ \ \
{\bf Q}_- = z_2({\bf y} - {\bf y}_D)
\label{PQvsyv}
\ee
Note that $D_{2|3}$ itself is of the $16$-th power in
components of ${\bf P}$ and ${\bf Q}$,
moreover it depends on particular choice of a pair of equations
out of nine in (\ref{mmtsols4}).
However, all these $36$ versions of $D_{2|3}$ contain one and
the same factor (\ref{discreq}), which is the quadratic
equation for $y_0$ that we are looking for.
Quadraticity is obvious in the first line of (\ref{discreq})
and is obscure in representation through scalar products,
which is still also useful in applications.

\subsection{Examples
\label{exacon}}

Eq.(\ref{discreq}) provides $y_0(y_1,y_2)$ for generic
quadrilateral as a function of positions of its four vertices in
${\bf y}$-space.
%First of all, it requires expressing all ingredients in
%terms of independent {\it free} parameters.
According to (\ref{n4bc}) these $4\times 3=12$ components of
${\bf y}_a = \frac{{\bf v}_a}{z_a}$ are not {\it free} parameters
(i.e. can not be chosen in arbitrary way):
they are expressed
through $3\times 2-1=5$ components of the three independent
null-vectors, constrained by the inscribed circle condition
$l_1+l_3=l_2+l_4$.
Two of these five free parameters depend on the choice of the
general orientation and scale, so that finally the whole
pattern of boundary conditions is labeled by $3$ parameters
and they can be chosen in different ways.

Mixed products with ${\bf P}$ and ${\bf Q}$ from (\ref{PQvsyv})
are actually all {\it linear} in ${\bf y}$:
\be
\begin{array}{lrcccl}
ABD:\ \ & \big\{{\bf P}_+{\bf Q}_+{\bf Q}_-\big\} &=& z_1z_2^2
\Big({\bf y}\cdot\big([{\bf y}_A\times {\bf y}_B] +
[{\bf y}_B\times {\bf y}_D] + [{\bf y}_D\times {\bf y}_A]\big) &-&
\{{\bf y}_A{\bf y}_B{\bf y}_D\}\Big)\\
CBD:& \big\{{\bf P}_-{\bf Q}_+{\bf Q}_-\big\} &=& z_1z_2^2
\Big({\bf y}\cdot\big([{\bf y}_C\times {\bf y}_B] +
[{\bf y}_B\times {\bf y}_D] + [{\bf y}_D\times {\bf y}_C]\big) &+&
\{{\bf y}_B{\bf y}_C{\bf y}_D\}\Big)\\
ACB:&  \big\{{\bf P}_+{\bf P}_-{\bf Q}_+\big\} &=& z_1^2z_2
\Big({\bf y}\cdot\big([{\bf y}_A\times {\bf y}_C] +
[{\bf y}_C\times {\bf y}_B] + [{\bf y}_B\times {\bf y}_A]\big) &+&
\{{\bf y}_A{\bf y}_B{\bf y}_C\}\Big)\\
ACD:&  \big\{{\bf P}_+{\bf P}_-{\bf Q}_-\big\} &=& z_1^2z_2
\Big({\bf y}\cdot\big([{\bf y}_A\times {\bf y}_C] +
[{\bf y}_D\times {\bf y}_A] + [{\bf y}_C\times {\bf y}_D]\big) &-&
\{{\bf y}_A{\bf y}_C{\bf y}_D\}\Big)
\end{array}
\label{mipro}
\ee
Each line in (\ref{mipro}) can also be written as
a sum of four $3\times 3$ determinants, for example,
$$
ABD: \ \ \
z_1z_2^2\left(\,\left|\left|\begin{array}{ccc}
y_0 & y_1 & y_2 \\
y_{A0} & y_{A1} & y_{A2} \\
y_{B0} & y_{B1} & y_{B2}
\end{array}\right|\right|
+ \left|\left|\begin{array}{ccc}
y_0 & y_1 & y_2 \\
y_{B0} & y_{B1} & y_{B2} \\
y_{D0} & y_{D1} & y_{D2}
\end{array}\right|\right|
+ \left|\left|\begin{array}{ccc}
y_0 & y_1 & y_2 \\
y_{D0} & y_{D1} & y_{D2} \\
y_{A0} & y_{A1} & y_{A2}
\end{array}\right|\right|
- \left|\left|\begin{array}{ccc}
y_{A0} & y_{A1} & y_{A2} \\
y_{B0} & y_{B1} & y_{B2} \\
y_{D0} & y_{D1} & y_{D2}
\end{array}\right|\right|\,\right)
$$
It remains to substitute particular values of ${\bf y}_a$ and
$z_a$ in order to obtain concrete equations in concrete examples.

\subsubsection{Square}

From Fig.\ref{figquadrisquare} and (\ref{squarevertices}),
\be
{\bf y}_A = (-1;1,-1), \ \ \  %(1;1,1), \ \ \
{\bf y}_B = (1;1,1), \ \ \    %(-1;-1,1), \ \ \
{\bf y}_C = (-1;-1,1), \ \ \  %(1;-1,-1), \ \ \
{\bf y}_D = (1;-1,-1) \ \ \  %(-1;1,-1), \ \ \
\label{squarevertices2}
\ee
Substituting these vectors for lines in determinants,
we obtain for the first line in (\ref{mipro}):
%a sum of $3\times 3$ determinants:
$$
ABD: \ \ \
z_1z_2^2\left(\,\left|\left|\begin{array}{ccc}
y_0 & y_1 & y_2 \\
-1 & 1 & -1 \\
1 & 1 & 1
\end{array}\right|\right|
+ \left|\left|\begin{array}{ccc}
y_0 & y_1 & y_2 \\
1 & 1 & 1 \\
1 & -1 & -1
\end{array}\right|\right|
+ \left|\left|\begin{array}{ccc}
y_0 & y_1 & y_2 \\
1 & -1 & -1 \\
-1 & 1 & -1
\end{array}\right|\right|
- \left|\left|\begin{array}{ccc}
-1 & 1 & -1 \\
1 & 1 & 1 \\
1 & -1 & -1
\end{array}\right|\right|\,\right) =
$$
\vspace{-0.3cm}
\be
%= y_0(2+ 0 +2)+ y_1(0 + 2 +2)+ y_2(-2-2+ 0) - 4
= 4z_1z_2^2(y_0+y_1-y_2-1)
\ee
Similarly for the other three lines we get:
$$
CBD: \ \ \
z_1z_2^2\left(\,\left|\left|\begin{array}{ccc}
y_0 & y_1 & y_2 \\
-1 & -1 & 1\\
1 & 1 & 1
\end{array}\right|\right|
+ \left|\left|\begin{array}{ccc}
y_0 & y_1 & y_2 \\
1 & 1 & 1 \\
1 & -1 & -1                               %A:  -1 & 1 & -1
\end{array}\right|\right|                 %B:  1 & 1 & 1
+ \left|\left|\begin{array}{ccc}          %C:  -1 & -1 & 1
y_0 & y_1 & y_2 \\                        %D:  1 & -1 & -1
1 & -1 & -1 \\
-1 & -1 & 1
\end{array}\right|\right|
+ \left|\left|\begin{array}{ccc}
1 & 1 & 1 \\
-1 & -1 & 1 \\
1 & -1 & -1
\end{array}\right|\right|\,\right) =
$$
\vspace{-0.3cm}
\be
%= y_0(2+ 0 +2)+ y_1(0 + 2 +2)+ y_2(-2-2+ 0) - 4
= 4z_1z_2^2(-y_0+y_1-y_2+1)
\ee
$$
ACB: \ \ \
z_1^2z_2\left(\,\left|\left|\begin{array}{ccc}
y_0 & y_1 & y_2 \\
-1 & 1 & -1 \\
-1 & -1 & 1
\end{array}\right|\right|
+ \left|\left|\begin{array}{ccc}
y_0 & y_1 & y_2 \\
-1 & -1 & 1 \\
1 & 1 & 1
\end{array}\right|\right|
+ \left|\left|\begin{array}{ccc}
y_0 & y_1 & y_2 \\
1 & 1 & 1 \\
-1 & 1 & -1
\end{array}\right|\right|
+ \left|\left|\begin{array}{ccc}
-1 & 1 & -1 \\
1 & 1 & 1 \\
-1 & -1 & 1
\end{array}\right|\right|\,\right) =
$$
\vspace{-0.3cm}
\be
%= y_0(2+ 0 +2)+ y_1(0 + 2 +2)+ y_2(-2-2+ 0) - 4
= 4z_1^2z_2(-y_0+y_1+y_2-1)
\ee
$$
ACD: \ \ \
z_1^2z_2\left(\,\left|\left|\begin{array}{ccc}
y_0 & y_1 & y_2 \\
-1 & 1 & -1 \\
-1 & -1 & 1
\end{array}\right|\right|
+ \left|\left|\begin{array}{ccc}
y_0 & y_1 & y_2 \\
1 & -1 & -1 \\
-1 & 1 & -1
\end{array}\right|\right|
+ \left|\left|\begin{array}{ccc}
y_0 & y_1 & y_2 \\
-1 & -1 & 1 \\
1 & -1 & -1
\end{array}\right|\right|
- \left|\left|\begin{array}{ccc}
-1 & 1 & -1 \\
-1 & -1 & 1 \\
1 & -1 & -1
\end{array}\right|\right|\,\right) =
$$
\vspace{-0.3cm}
\be
%= y_0(2+ 0 +2)+ y_1(0 + 2 +2)+ y_2(-2-2+ 0) - 4
= 4z_1^2z_2(y_0+y_1+y_2+1)
\ee
Since in this case $z_1=z_2=\frac{1}{4}$ we finally
obtain for (\ref{discreq}) the familiar result (\ref{squareq}):
\be
{\cal S}_\Box \sim D_{2|3} =
\frac{1}{16^2}\Big(
(y_0+y_1-y_2-1)(-y_0+y_1-y_2+1)-(-y_0+y_1+y_2-1)(y_0+y_1+y_2+1)
\Big) = \nn \\ = \frac{1}{16^2}
\Big((y_1-y_2)^2-(y_0-1)^2-(y_1+y_2)^2 + (y_0+1)^2\Big)
= \frac{1}{64}(y_0 -y_1y_2) = 0
\ee

\subsubsection{Rhombus
\label{conrhomb}}

\begin{figure}
\begin{center}
%%
%%
%TeXCAD Picture [figrhomb3.tex]. Options:
%\grade{\on}
%\emlines{\off}
%\epic{\off}
%\beziermacro{\on}
%\reduce{\on}
%\snapping{\off}
%\pvinsert{% Your \input, \def, etc. here}
%\quality{8.000}
%\graddiff{0.005}
%\snapasp{1}
%\zoom{13.4543}
\unitlength 1mm % = 2.845pt
\linethickness{0.4pt}
\ifx\plotpoint\undefined\newsavebox{\plotpoint}\fi
% GNUPLOT compatibility
\begin{picture}(66.75,51.25)(0,0)
%\circle(34.5,24.75){25.928}
\put(47.464,24.75){\line(0,1){.6678}}
\put(47.447,25.418){\line(0,1){.666}}
\put(47.395,26.084){\line(0,1){.6625}}
\put(47.309,26.746){\line(0,1){.6572}}
\multiput(47.19,27.403)(-.030703,.130021){5}{\line(0,1){.130021}}
\multiput(47.036,28.053)(-.031133,.106889){6}{\line(0,1){.106889}}
\multiput(46.849,28.695)(-.031369,.090123){7}{\line(0,1){.090123}}
\multiput(46.63,29.326)(-.031474,.077339){8}{\line(0,1){.077339}}
\multiput(46.378,29.944)(-.031481,.067214){9}{\line(0,1){.067214}}
\multiput(46.095,30.549)(-.031411,.058953){10}{\line(0,1){.058953}}
\multiput(45.78,31.139)(-.031278,.052051){11}{\line(0,1){.052051}}
\multiput(45.436,31.711)(-.031091,.046174){12}{\line(0,1){.046174}}
\multiput(45.063,32.265)(-.033428,.044511){12}{\line(0,1){.044511}}
\multiput(44.662,32.8)(-.0329321,.0394429){13}{\line(0,1){.0394429}}
\multiput(44.234,33.312)(-.0324258,.0350018){14}{\line(0,1){.0350018}}
\multiput(43.78,33.802)(-.0341857,.0332851){14}{\line(-1,0){.0341857}}
\multiput(43.301,34.268)(-.0358548,.03148){14}{\line(-1,0){.0358548}}
\multiput(42.8,34.709)(-.0403078,.0318677){13}{\line(-1,0){.0403078}}
\multiput(42.276,35.123)(-.045387,.032228){12}{\line(-1,0){.045387}}
\multiput(41.731,35.51)(-.051258,.032561){11}{\line(-1,0){.051258}}
\multiput(41.167,35.868)(-.058154,.032865){10}{\line(-1,0){.058154}}
\multiput(40.585,36.197)(-.066411,.03314){9}{\line(-1,0){.066411}}
\multiput(39.988,36.495)(-.076534,.033385){8}{\line(-1,0){.076534}}
\multiput(39.375,36.762)(-.089316,.033598){7}{\line(-1,0){.089316}}
\multiput(38.75,36.997)(-.090928,.028953){7}{\line(-1,0){.090928}}
\multiput(38.114,37.2)(-.107682,.028269){6}{\line(-1,0){.107682}}
\multiput(37.468,37.37)(-.130794,.027222){5}{\line(-1,0){.130794}}
\put(36.814,37.506){\line(-1,0){.6601}}
\put(36.154,37.608){\line(-1,0){.6645}}
\put(35.489,37.676){\line(-1,0){.6671}}
\put(34.822,37.71){\line(-1,0){.668}}
\put(34.154,37.709){\line(-1,0){.6671}}
\put(33.487,37.674){\line(-1,0){.6644}}
\put(32.823,37.605){\line(-1,0){.6599}}
\multiput(32.163,37.502)(-.130744,-.027465){5}{\line(-1,0){.130744}}
\multiput(31.509,37.364)(-.107629,-.028469){6}{\line(-1,0){.107629}}
\multiput(30.863,37.193)(-.090874,-.029122){7}{\line(-1,0){.090874}}
\multiput(30.227,36.99)(-.078097,-.029543){8}{\line(-1,0){.078097}}
\multiput(29.602,36.753)(-.076472,-.033527){8}{\line(-1,0){.076472}}
\multiput(28.99,36.485)(-.06635,-.033263){9}{\line(-1,0){.06635}}
\multiput(28.393,36.186)(-.058093,-.032973){10}{\line(-1,0){.058093}}
\multiput(27.812,35.856)(-.051198,-.032656){11}{\line(-1,0){.051198}}
\multiput(27.249,35.497)(-.045327,-.032312){12}{\line(-1,0){.045327}}
\multiput(26.705,35.109)(-.0402486,-.0319424){13}{\line(-1,0){.0402486}}
\multiput(26.182,34.694)(-.0357963,-.0315465){14}{\line(-1,0){.0357963}}
\multiput(25.681,34.252)(-.0341239,-.0333485){14}{\line(-1,0){.0341239}}
\multiput(25.203,33.785)(-.0323608,-.0350619){14}{\line(0,-1){.0350619}}
\multiput(24.75,33.294)(-.0328589,-.0395039){13}{\line(0,-1){.0395039}}
\multiput(24.323,32.781)(-.033346,-.044573){12}{\line(0,-1){.044573}}
\multiput(23.923,32.246)(-.031005,-.046231){12}{\line(0,-1){.046231}}
\multiput(23.551,31.691)(-.031181,-.052109){11}{\line(0,-1){.052109}}
\multiput(23.208,31.118)(-.031301,-.059011){10}{\line(0,-1){.059011}}
\multiput(22.895,30.528)(-.031356,-.067272){9}{\line(0,-1){.067272}}
\multiput(22.613,29.922)(-.03133,-.077398){8}{\line(0,-1){.077398}}
\multiput(22.362,29.303)(-.031202,-.090181){7}{\line(0,-1){.090181}}
\multiput(22.143,28.672)(-.030935,-.106947){6}{\line(0,-1){.106947}}
\multiput(21.958,28.03)(-.030462,-.130078){5}{\line(0,-1){.130078}}
\put(21.806,27.38){\line(0,-1){.6574}}
\put(21.687,26.722){\line(0,-1){.6626}}
\put(21.602,26.06){\line(0,-1){.6661}}
\put(21.552,25.394){\line(0,-1){2.0014}}
\put(21.607,23.392){\line(0,-1){.6623}}
\multiput(21.694,22.73)(.03026,-.16423){4}{\line(0,-1){.16423}}
\multiput(21.815,22.073)(.030945,-.129964){5}{\line(0,-1){.129964}}
\multiput(21.97,21.423)(.031332,-.106831){6}{\line(0,-1){.106831}}
\multiput(22.158,20.782)(.031537,-.090065){7}{\line(0,-1){.090065}}
\multiput(22.379,20.152)(.031617,-.077281){8}{\line(0,-1){.077281}}
\multiput(22.632,19.534)(.031605,-.067155){9}{\line(0,-1){.067155}}
\multiput(22.916,18.929)(.03152,-.058894){10}{\line(0,-1){.058894}}
\multiput(23.231,18.34)(.031374,-.051993){11}{\line(0,-1){.051993}}
\multiput(23.577,17.768)(.031177,-.046116){12}{\line(0,-1){.046116}}
\multiput(23.951,17.215)(.033511,-.044449){12}{\line(0,-1){.044449}}
\multiput(24.353,16.682)(.0330053,-.0393817){13}{\line(0,-1){.0393817}}
\multiput(24.782,16.17)(.0324907,-.0349416){14}{\line(0,-1){.0349416}}
\multiput(25.237,15.68)(.0342474,-.0332216){14}{\line(1,0){.0342474}}
\multiput(25.716,15.215)(.0359131,-.0314135){14}{\line(1,0){.0359131}}
\multiput(26.219,14.776)(.0403669,-.0317928){13}{\line(1,0){.0403669}}
\multiput(26.744,14.362)(.045447,-.032144){12}{\line(1,0){.045447}}
\multiput(27.289,13.976)(.051319,-.032466){11}{\line(1,0){.051319}}
\multiput(27.854,13.619)(.058215,-.032757){10}{\line(1,0){.058215}}
\multiput(28.436,13.292)(.066472,-.033017){9}{\line(1,0){.066472}}
\multiput(29.034,12.995)(.076596,-.033243){8}{\line(1,0){.076596}}
\multiput(29.647,12.729)(.089378,-.033432){7}{\line(1,0){.089378}}
\multiput(30.272,12.495)(.106146,-.033582){6}{\line(1,0){.106146}}
\multiput(30.909,12.293)(.129281,-.033683){5}{\line(1,0){.129281}}
\multiput(31.556,12.125)(.16356,-.03372){4}{\line(1,0){.16356}}
\put(32.21,11.99){\line(1,0){.6603}}
\put(32.87,11.889){\line(1,0){.6646}}
\put(33.535,11.822){\line(1,0){.6672}}
\put(34.202,11.789){\line(1,0){.668}}
\put(34.87,11.791){\line(1,0){.667}}
\put(35.537,11.828){\line(1,0){.6642}}
\put(36.201,11.898){\line(1,0){.6597}}
\multiput(36.861,12.003)(.130692,.027707){5}{\line(1,0){.130692}}
\multiput(37.514,12.141)(.107576,.028669){6}{\line(1,0){.107576}}
\multiput(38.16,12.313)(.09082,.02929){7}{\line(1,0){.09082}}
\multiput(38.796,12.518)(.078042,.029688){8}{\line(1,0){.078042}}
\multiput(39.42,12.756)(.076409,.033669){8}{\line(1,0){.076409}}
\multiput(40.031,13.025)(.066288,.033386){9}{\line(1,0){.066288}}
\multiput(40.628,13.326)(.058032,.033081){10}{\line(1,0){.058032}}
\multiput(41.208,13.657)(.051137,.032751){11}{\line(1,0){.051137}}
\multiput(41.771,14.017)(.045267,.032396){12}{\line(1,0){.045267}}
\multiput(42.314,14.406)(.0401893,.032017){13}{\line(1,0){.0401893}}
\multiput(42.836,14.822)(.0357377,.0316129){14}{\line(1,0){.0357377}}
\multiput(43.337,15.264)(.0340619,.0334117){14}{\line(1,0){.0340619}}
\multiput(43.814,15.732)(.0322957,.0351219){14}{\line(0,1){.0351219}}
\multiput(44.266,16.224)(.0327856,.0395648){13}{\line(0,1){.0395648}}
\multiput(44.692,16.738)(.033263,.044634){12}{\line(0,1){.044634}}
\multiput(45.091,17.274)(.03373,.050497){11}{\line(0,1){.050497}}
\multiput(45.462,17.829)(.031085,.052167){11}{\line(0,1){.052167}}
\multiput(45.804,18.403)(.031192,.059069){10}{\line(0,1){.059069}}
\multiput(46.116,18.994)(.031231,.06733){9}{\line(0,1){.06733}}
\multiput(46.397,19.6)(.031187,.077456){8}{\line(0,1){.077456}}
\multiput(46.647,20.219)(.031035,.090239){7}{\line(0,1){.090239}}
\multiput(46.864,20.851)(.030736,.107004){6}{\line(0,1){.107004}}
\multiput(47.048,21.493)(.030221,.130134){5}{\line(0,1){.130134}}
\put(47.199,22.144){\line(0,1){.6576}}
\put(47.317,22.801){\line(0,1){.6628}}
\put(47.4,23.464){\line(0,1){1.2859}}
%\end
\put(34.5,24.5){\line(1,0){32.25}}
%\emline(34.75,24.75)(34.5,25)
\multiput(34.75,24.75)(-.03125,.03125){8}{\line(0,1){.03125}}
%\end
\put(34.75,24.5){\line(-1,0){31}}
\put(34.5,24.75){\line(0,1){26.5}}
\put(34.5,24){\line(0,-1){17}}
\put(63.25,26.75){$y_1$}
\put(29.75,49.25){$y_2$}
%\emline(52.028,6.987)(13.899,45.116)
\multiput(52.028,6.987)(-.03371260223,.03371260223){1131}
{\line(0,1){.03371260223}}
%\end
%\emline(47.94,37.906)(20.365,10.331)
\multiput(47.94,37.906)(-.03370995107,-.03370995107){818}
{\line(0,-1){.03370995107}}
%\end
%\emline(52.102,6.912)(45.339,35.379)
\multiput(52.102,6.912)(-.033649828,.141625101){201}
{\line(0,1){.141625101}}
%\end
%\emline(45.339,35.379)(15.608,43.406)
\multiput(45.339,35.379)(-.124916714,.033727513){238}
{\line(-1,0){.124916714}}
%\end
%\emline(15.608,43.406)(15.757,43.257)
\multiput(15.608,43.406)(.02973,-.02973){5}{\line(1,0){.02973}}
%\end
%\emline(15.757,43.257)(23.858,13.825)
\multiput(15.757,43.257)(.033616072,-.122128117){241}
{\line(0,-1){.122128117}}
%\end
%\emline(23.858,13.825)(52.102,6.987)
\multiput(23.858,13.825)(.139131374,-.033684438){203}
{\line(1,0){.139131374}}
%\end
\put(34.561,24.453){\line(-1,0){.074}}
\put(34.487,24.453){\line(0,1){.149}}
%\emline(34.413,24.602)(34.487,24.676)
\put(34.413,24.602){\line(1,0){.0743}}
%\end
%\emline(34.413,24.527)(52.697,29.284)
\multiput(34.413,24.527)(.12967418,.033736372){141}
{\line(1,0){.12967418}}
%\end
\qbezier(39.021,25.717)(39.244,25.011)(39.17,24.453)
\qbezier(37.906,24.453)(37.683,25.754)(36.865,26.906)
\put(44.737,25.196){{\footnotesize $\phi$}}
\put(39.244,27.203){{\footnotesize $\theta_B$}}
\qbezier(38.575,24.453)(38.501,22.781)(37.534,21.554)
\put(40.284,20.852){{\footnotesize $\tilde\theta_A$}}
\put(53.366,8.176){${\bf y}_A = \left(-\frac{1+b}{1-b};
\frac{B}{1-b},-\frac{B}{1-b}\right)$}
\put(48.089,34.933){${\bf y}_B = \left(\frac{1-b}{1+b};
\frac{B}{1+b},\frac{B}{1+b}\right)$}
\put(-25.615,40.041){${\bf y}_C = \left(-\frac{1+b}{1-b};
-\frac{B}{1-b},\frac{B}{1-b}\right)$}
\put(-20.021,13.825){${\bf y}_D = \left(\frac{1-b}{1+b};
-\frac{B}{1+b},-\frac{B}{1+b}\right)$}
%\put(34.487,24.453){\line(-0.2,-1){5.149}}
%\put(26,7){{\footnotesize $y_0=0$}}
\end{picture}
\caption{{\footnotesize
Rhombus in the standard parametrization, suggested
in \cite{am1}.
The values of $y_0$ are also shown, $y_0=0$ at four
tangent points.
The angle $\phi$ defines the direction of a normal to
the rhombus side.
Directions to the vertices are $\theta_A=-\frac{\pi}{4}$
(so that $\tilde\theta_A = 2\pi - \theta_A = \frac{\pi}{4}$),
$\theta_B=\frac{\pi}{4}$, $\theta_C=\frac{3\pi}{4}$,
$\theta_D=\frac{5\pi}{4}$.
Parameter $B = \sqrt{1+b^2}$.
External momenta ${\bf p}_a={\bf y}_{a+1} - {\bf y}_a$
are vectors along the sides, i.e. are given by differences
between the values that ${\bf y}$ takes at vertices.
Parameters $z_a$ are made from scalar products of these
vectors and therefore are derived from the data in the
picture.
}}
\label{figrhomb}
\end{center}
\end{figure}
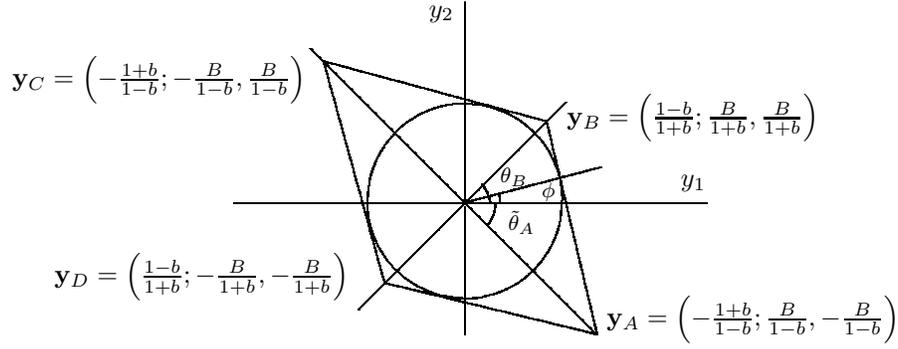

According to the table in s.2.6.3 of \cite{malda3},
see also Fig.\ref{figrhomb},
\be
{\bf y}_A = (-b_-,-B_-,B_-), \ \ \
%(-\frac{1+b}{1-b};\frac{B}{1-b},-\frac{B}{1-b}),
{\bf y}_B = (B_+;B_+,b_+), \ \ \    %(-1;-1,1), \ \ \
{\bf y}_C = (-b_-;B_-,-B_-), \ \ \  %(1;-1,-1), \ \ \
{\bf y}_D = (b_+;-B_+,-B_+), \ \ \  %(-1;1,-1), \ \ \
\label{rhombusvertices}
\ee
where
\be
b_- = \frac{1+b}{1-b}, \ \ \
b_+ = \frac{1-b}{1+b}, \ \ \
B_-=\frac{B}{1-b}, \ \ \
B_+ = \frac{B}{1+b}, \ \ \ B=\sqrt{1+b^2}
\label{bBpars}
\ee
The four lines in (\ref{mipro}) are now
$$
%ABD: \ \ \
z_1z_2^2\left(\,\left|\left|\begin{array}{ccc}
y_0 & y_1 & y_2 \\
-b_- & B_- & -B_- \\
b_+ & B_+ & B_+
\end{array}\right|\right|
+ \left|\left|\begin{array}{ccc}
y_0 & y_1 & y_2 \\
b_+ & B_+ & B_+ \\
b_+ & -B_+ & -B_+
\end{array}\right|\right|
+ \left|\left|\begin{array}{ccc}
y_0 & y_1 & y_2 \\
b_+ & -B_+ & -B_+ \\
-b_- & B_- & -B_-
\end{array}\right|\right|
- \left|\left|\begin{array}{ccc}
-b_- & B_- & -B_- \\
b_+ & B_+ & B_+ \\
b_+ & -B_+ & -B_+
\end{array}\right|\right|\,\right) =
$$
\vspace{-0.3cm}
\be
%\begin{array}{lc}
%ABD: &
= 4z_1z_2^2B_+
\left(B_-(y_0-b_+) +\frac{1}{2}(b_+ + b_-)(y_1-y_2)\right)
%\end{array}
\ee
$$
%CBD: \ \ \
z_1z_2^2\left(\,\left|\left|\begin{array}{ccc}
y_0 & y_1 & y_2 \\
-b_- & -B_- & B_-\\
b_+ & B_+ & B_+
\end{array}\right|\right|
+ \left|\left|\begin{array}{ccc}
y_0 & y_1 & y_2 \\
b_+ & B_+ & B_+ \\
b_+ & -B_+ & -B_+                               %A:  -1 & 1 & -1
\end{array}\right|\right|                 %B:  1 & 1 & 1
+ \left|\left|\begin{array}{ccc}          %C:  -1 & -1 & 1
y_0 & y_1 & y_2 \\                        %D:  1 & -1 & -1
b_+ & -B_+ & -B_+ \\
-b_- & -B_- & B_-
\end{array}\right|\right|
+ \left|\left|\begin{array}{ccc}
b_+ & B_+ & B_+ \\
-b_- & -B_- & B_- \\
b_+ & -B_+ & -B_+
\end{array}\right|\right|\,\right) =
$$
\vspace{-0.3cm}
\be
%\begin{array}{lc}
%CBD: &
= 4z_1z_2^2B_+
\left(-B_-(y_0-b_+) +\frac{1}{2}(b_+ + b_-)(y_1-y_2)\right)
%\end{array}
\ee
$$
%ACB: \ \ \
z_1^2z_2\left(\,\left|\left|\begin{array}{ccc}
y_0 & y_1 & y_2 \\
-b_- & B_- & -B_- \\
-b_- & -B_- & B_-
\end{array}\right|\right|
+ \left|\left|\begin{array}{ccc}
y_0 & y_1 & y_2 \\
-b_- & -B_- & B_- \\
b_+ & B_+ & B_+
\end{array}\right|\right|
+ \left|\left|\begin{array}{ccc}
y_0 & y_1 & y_2 \\
b_+ & B_+ & B_+ \\
-b_- & B_- & -B_-
\end{array}\right|\right|
+ \left|\left|\begin{array}{ccc}
-b_- & B_- & -B_- \\
b_+ & B_+ & B_+ \\
-b_- & -B_- & B_-
\end{array}\right|\right|\,\right) =
$$
\vspace{-0.3cm}
\be
%\begin{array}{lc}
%ACB: &
= 4z_1^2z_2B_-
\left(-B_+(y_0+b_-) +\frac{1}{2}(b_+ + b_-)(y_1+y_2)\right)
%\end{array}
\ee
$$
%ACD: \ \ \
z_1^2z_2\left(\,\left|\left|\begin{array}{ccc}
y_0 & y_1 & y_2 \\
-b_- & B_- & -B_- \\
-b_- & -B_- & B_-
\end{array}\right|\right|
+ \left|\left|\begin{array}{ccc}
y_0 & y_1 & y_2 \\
b_+ & -B_+ & -B_+ \\
-b_- & B_- & -B_-
\end{array}\right|\right|
+ \left|\left|\begin{array}{ccc}
y_0 & y_1 & y_2 \\
-b_- & -B_- & B_- \\
b_+ & -B_+ & -B_+
\end{array}\right|\right|
- \left|\left|\begin{array}{ccc}
-b_- & B_- & -B_- \\
-b_- & -B_- & B_- \\
b_+ & -B_+ & -B_+
\end{array}\right|\right|\,\right) =
$$
\vspace{-0.3cm}
\be
%\begin{array}{lc}
%ACD: &
= 4z_1^2z_2B_-
\left(B_+(y_0+b_-) +\frac{1}{2}(b_+ + b_-)(y_1+y_2)\right)
%\end{array}
\ee
Therefore we obtain for (\ref{discreq}):
$$
(4z_1z_2)^2\left\{ (z_2B_+)^2
\left(\frac{1}{4}(b_++b_-)^2(y_1-y_2)^2 -B_-^2(y_0-b_+)^2\right)
- (z_1B_-)^2
\left(\frac{1}{4}(b_++b_-)^2(y_1+y_2)^2 -B_+^2(y_0+b_-)^2\right)
\right\}=
$$
$$
= (4z_1z_2)^2\left\{(B_+B_-)^2\Big((z_1^2-z_2^2)y_0^2
+ 2(z_1^2b_-+z_2^2b_+)y_0 +(z_1b_-)^2-(z_2b_+)^2\Big) + \right.
$$
$$
\left.
+ \frac{1}{4}(b_++b_-)^2\left[
\left((z_2B_+)^2-(z_1B_-)^2\right)(y_1^2+y_2^2)
- 2\left((z_2B_+)^2+(z_1B_-)^2\right)y_1y_2
\right]\right\} \ \stackrel{(\ref{bBpars})}{=}\
$$
$$
= \left(\frac{2z_1z_2B^2}{1-b^2}\right)^2\left\{
(z_1^2-z_2^2)y_0^2+2y_0\frac{z_1^2(1+b)^2+z_2^2(1-b)^2}{1-b^2}
+ \left(\frac{z_1(1+b)}{1-b}\right)^2
-\left(\frac{z_2(1-b)}{1+b}\right)^2
+ \right. $$ $$ \left.
+ B^2\left[\left(\frac{z_2}{1+b}\right)^2
-\left(\frac{z_1}{1-b}\right)^2\right](y_1^2+y_2^2) -
2B^2\left[\left(\frac{z_2}{1+b}\right)^2
+\left(\frac{z_1}{1-b}\right)^2\right]y_1y_2
\right\}=
$$
\be
= \left(\frac{8z_1z_2zB^2}{1-b^2}\right)^2
\Big((1-b^2)y_0 + b(1-y_0^2) - (1+b^2)y_1y_2\Big)
\ee
provided
\be
z_1 = (1-b)z, \ \ \ z_2 = (1+b)z,
\ee
what is indeed the case for rhombus,
with
\be
z = \frac{1}{4\sqrt{1+b^2}},
\ee
see \cite{am1,mmt1}.

Thus we see that exact solution to NG equations
with rhombus in the role of the boundary $\Pi$ is
\be
{\cal S}_\diamond = y_1y_2 - \frac{1}{2}(1-y_0^2)\sin(2\phi)
- y_0\cos(2\phi) = 0
\label{rhomeq}
\ee
where
\be
\sin(2\phi) = \frac{2b}{1+b^2}, \ \ \
\cos(2\phi) = \frac{1-b^2}{1+b^2}
\ee
This is in accordance with eq.(2.54) of \cite{malda3}.

\subsubsection{Kite
\label{conkite}}

\begin{figure}
\begin{center}
%%
%%
%TeXCAD Picture [figkite.pic]. Options:
%\grade{\on}
%\emlines{\off}
%\epic{\off}
%\beziermacro{\on}
%\reduce{\on}
%\snapping{\off}
%\pvinsert{% Your \input, \def, etc. here}
%\quality{8.000}
%\graddiff{0.005}
%\snapasp{1}
%\zoom{11.3137}
\unitlength 1mm % = 2.845pt
\linethickness{0.4pt}
\ifx\plotpoint\undefined\newsavebox{\plotpoint}\fi
% GNUPLOT compatibility
\begin{picture}(66.75,51.25)(0,0)
%\circle(34.5,24.75){25.928}
\put(47.464,24.75){\line(0,1){.6678}}
\put(47.447,25.418){\line(0,1){.666}}
\put(47.395,26.084){\line(0,1){.6625}}
\put(47.309,26.746){\line(0,1){.6572}}
\multiput(47.19,27.403)(-.030703,.130021){5}{\line(0,1){.130021}}
\multiput(47.036,28.053)(-.031133,.106889){6}{\line(0,1){.106889}}
\multiput(46.849,28.695)(-.031369,.090123){7}{\line(0,1){.090123}}
\multiput(46.63,29.326)(-.031474,.077339){8}{\line(0,1){.077339}}
\multiput(46.378,29.944)(-.031481,.067214){9}{\line(0,1){.067214}}
\multiput(46.095,30.549)(-.031411,.058953){10}{\line(0,1){.058953}}
\multiput(45.78,31.139)(-.031278,.052051){11}{\line(0,1){.052051}}
\multiput(45.436,31.711)(-.031091,.046174){12}{\line(0,1){.046174}}
\multiput(45.063,32.265)(-.033428,.044511){12}{\line(0,1){.044511}}
\multiput(44.662,32.8)(-.0329321,.0394429){13}{\line(0,1){.0394429}}
\multiput(44.234,33.312)(-.0324258,.0350018){14}{\line(0,1){.0350018}}
\multiput(43.78,33.802)(-.0341857,.0332851){14}{\line(-1,0){.0341857}}
\multiput(43.301,34.268)(-.0358548,.03148){14}{\line(-1,0){.0358548}}
\multiput(42.8,34.709)(-.0403078,.0318677){13}{\line(-1,0){.0403078}}
\multiput(42.276,35.123)(-.045387,.032228){12}{\line(-1,0){.045387}}
\multiput(41.731,35.51)(-.051258,.032561){11}{\line(-1,0){.051258}}
\multiput(41.167,35.868)(-.058154,.032865){10}{\line(-1,0){.058154}}
\multiput(40.585,36.197)(-.066411,.03314){9}{\line(-1,0){.066411}}
\multiput(39.988,36.495)(-.076534,.033385){8}{\line(-1,0){.076534}}
\multiput(39.375,36.762)(-.089316,.033598){7}{\line(-1,0){.089316}}
\multiput(38.75,36.997)(-.090928,.028953){7}{\line(-1,0){.090928}}
\multiput(38.114,37.2)(-.107682,.028269){6}{\line(-1,0){.107682}}
\multiput(37.468,37.37)(-.130794,.027222){5}{\line(-1,0){.130794}}
\put(36.814,37.506){\line(-1,0){.6601}}
\put(36.154,37.608){\line(-1,0){.6645}}
\put(35.489,37.676){\line(-1,0){.6671}}
\put(34.822,37.71){\line(-1,0){.668}}
\put(34.154,37.709){\line(-1,0){.6671}}
\put(33.487,37.674){\line(-1,0){.6644}}
\put(32.823,37.605){\line(-1,0){.6599}}
\multiput(32.163,37.502)(-.130744,-.027465){5}{\line(-1,0){.130744}}
\multiput(31.509,37.364)(-.107629,-.028469){6}{\line(-1,0){.107629}}
\multiput(30.863,37.193)(-.090874,-.029122){7}{\line(-1,0){.090874}}
\multiput(30.227,36.99)(-.078097,-.029543){8}{\line(-1,0){.078097}}
\multiput(29.602,36.753)(-.076472,-.033527){8}{\line(-1,0){.076472}}
\multiput(28.99,36.485)(-.06635,-.033263){9}{\line(-1,0){.06635}}
\multiput(28.393,36.186)(-.058093,-.032973){10}{\line(-1,0){.058093}}
\multiput(27.812,35.856)(-.051198,-.032656){11}{\line(-1,0){.051198}}
\multiput(27.249,35.497)(-.045327,-.032312){12}{\line(-1,0){.045327}}
\multiput(26.705,35.109)(-.0402486,-.0319424){13}{\line(-1,0){.0402486}}
\multiput(26.182,34.694)(-.0357963,-.0315465){14}{\line(-1,0){.0357963}}
\multiput(25.681,34.252)(-.0341239,-.0333485){14}{\line(-1,0){.0341239}}
\multiput(25.203,33.785)(-.0323608,-.0350619){14}{\line(0,-1){.0350619}}
\multiput(24.75,33.294)(-.0328589,-.0395039){13}{\line(0,-1){.0395039}}
\multiput(24.323,32.781)(-.033346,-.044573){12}{\line(0,-1){.044573}}
\multiput(23.923,32.246)(-.031005,-.046231){12}{\line(0,-1){.046231}}
\multiput(23.551,31.691)(-.031181,-.052109){11}{\line(0,-1){.052109}}
\multiput(23.208,31.118)(-.031301,-.059011){10}{\line(0,-1){.059011}}
\multiput(22.895,30.528)(-.031356,-.067272){9}{\line(0,-1){.067272}}
\multiput(22.613,29.922)(-.03133,-.077398){8}{\line(0,-1){.077398}}
\multiput(22.362,29.303)(-.031202,-.090181){7}{\line(0,-1){.090181}}
\multiput(22.143,28.672)(-.030935,-.106947){6}{\line(0,-1){.106947}}
\multiput(21.958,28.03)(-.030462,-.130078){5}{\line(0,-1){.130078}}
\put(21.806,27.38){\line(0,-1){.6574}}
\put(21.687,26.722){\line(0,-1){.6626}}
\put(21.602,26.06){\line(0,-1){.6661}}
\put(21.552,25.394){\line(0,-1){2.0014}}
\put(21.607,23.392){\line(0,-1){.6623}}
\multiput(21.694,22.73)(.03026,-.16423){4}{\line(0,-1){.16423}}
\multiput(21.815,22.073)(.030945,-.129964){5}{\line(0,-1){.129964}}
\multiput(21.97,21.423)(.031332,-.106831){6}{\line(0,-1){.106831}}
\multiput(22.158,20.782)(.031537,-.090065){7}{\line(0,-1){.090065}}
\multiput(22.379,20.152)(.031617,-.077281){8}{\line(0,-1){.077281}}
\multiput(22.632,19.534)(.031605,-.067155){9}{\line(0,-1){.067155}}
\multiput(22.916,18.929)(.03152,-.058894){10}{\line(0,-1){.058894}}
\multiput(23.231,18.34)(.031374,-.051993){11}{\line(0,-1){.051993}}
\multiput(23.577,17.768)(.031177,-.046116){12}{\line(0,-1){.046116}}
\multiput(23.951,17.215)(.033511,-.044449){12}{\line(0,-1){.044449}}
\multiput(24.353,16.682)(.0330053,-.0393817){13}{\line(0,-1){.0393817}}
\multiput(24.782,16.17)(.0324907,-.0349416){14}{\line(0,-1){.0349416}}
\multiput(25.237,15.68)(.0342474,-.0332216){14}{\line(1,0){.0342474}}
\multiput(25.716,15.215)(.0359131,-.0314135){14}{\line(1,0){.0359131}}
\multiput(26.219,14.776)(.0403669,-.0317928){13}{\line(1,0){.0403669}}
\multiput(26.744,14.362)(.045447,-.032144){12}{\line(1,0){.045447}}
\multiput(27.289,13.976)(.051319,-.032466){11}{\line(1,0){.051319}}
\multiput(27.854,13.619)(.058215,-.032757){10}{\line(1,0){.058215}}
\multiput(28.436,13.292)(.066472,-.033017){9}{\line(1,0){.066472}}
\multiput(29.034,12.995)(.076596,-.033243){8}{\line(1,0){.076596}}
\multiput(29.647,12.729)(.089378,-.033432){7}{\line(1,0){.089378}}
\multiput(30.272,12.495)(.106146,-.033582){6}{\line(1,0){.106146}}
\multiput(30.909,12.293)(.129281,-.033683){5}{\line(1,0){.129281}}
\multiput(31.556,12.125)(.16356,-.03372){4}{\line(1,0){.16356}}
\put(32.21,11.99){\line(1,0){.6603}}
\put(32.87,11.889){\line(1,0){.6646}}
\put(33.535,11.822){\line(1,0){.6672}}
\put(34.202,11.789){\line(1,0){.668}}
\put(34.87,11.791){\line(1,0){.667}}
\put(35.537,11.828){\line(1,0){.6642}}
\put(36.201,11.898){\line(1,0){.6597}}
\multiput(36.861,12.003)(.130692,.027707){5}{\line(1,0){.130692}}
\multiput(37.514,12.141)(.107576,.028669){6}{\line(1,0){.107576}}
\multiput(38.16,12.313)(.09082,.02929){7}{\line(1,0){.09082}}
\multiput(38.796,12.518)(.078042,.029688){8}{\line(1,0){.078042}}
\multiput(39.42,12.756)(.076409,.033669){8}{\line(1,0){.076409}}
\multiput(40.031,13.025)(.066288,.033386){9}{\line(1,0){.066288}}
\multiput(40.628,13.326)(.058032,.033081){10}{\line(1,0){.058032}}
\multiput(41.208,13.657)(.051137,.032751){11}{\line(1,0){.051137}}
\multiput(41.771,14.017)(.045267,.032396){12}{\line(1,0){.045267}}
\multiput(42.314,14.406)(.0401893,.032017){13}{\line(1,0){.0401893}}
\multiput(42.836,14.822)(.0357377,.0316129){14}{\line(1,0){.0357377}}
\multiput(43.337,15.264)(.0340619,.0334117){14}{\line(1,0){.0340619}}
\multiput(43.814,15.732)(.0322957,.0351219){14}{\line(0,1){.0351219}}
\multiput(44.266,16.224)(.0327856,.0395648){13}{\line(0,1){.0395648}}
\multiput(44.692,16.738)(.033263,.044634){12}{\line(0,1){.044634}}
\multiput(45.091,17.274)(.03373,.050497){11}{\line(0,1){.050497}}
\multiput(45.462,17.829)(.031085,.052167){11}{\line(0,1){.052167}}
\multiput(45.804,18.403)(.031192,.059069){10}{\line(0,1){.059069}}
\multiput(46.116,18.994)(.031231,.06733){9}{\line(0,1){.06733}}
\multiput(46.397,19.6)(.031187,.077456){8}{\line(0,1){.077456}}
\multiput(46.647,20.219)(.031035,.090239){7}{\line(0,1){.090239}}
\multiput(46.864,20.851)(.030736,.107004){6}{\line(0,1){.107004}}
\multiput(47.048,21.493)(.030221,.130134){5}{\line(0,1){.130134}}
\put(47.199,22.144){\line(0,1){.6576}}
\put(47.317,22.801){\line(0,1){.6628}}
\put(47.4,23.464){\line(0,1){1.2859}}
%\end
\put(34.5,24.5){\line(1,0){32.25}}
%\emline(34.75,24.75)(34.5,25)
\multiput(34.75,24.75)(-.03125,.03125){8}{\line(0,1){.03125}}
%\end
\put(34.75,24.5){\line(-1,0){31}}
\put(34.5,24.75){\line(0,1){26.5}}
\put(34.5,24){\line(0,-1){17}}
\put(63.25,27.75){$y_1$}
\put(29.75,49.25){$y_2$}
%\emline(34.471,7.955)(17.236,21.92)
\multiput(34.471,7.955)(-.0416321927,.033732751){414}
{\line(-1,0){.0416321927}}
%\end
%\emline(17.236,21.92)(34.471,50.381)
\multiput(17.236,21.92)(.0337294086,.055696767){511}
{\line(0,1){.055696767}}
%\end
\put(34.471,50.381){\line(3,-5){16.971}}
%\emline(51.442,22.097)(34.56,7.955)
\multiput(51.442,22.097)(-.0401956533,-.0336717515){420}
{\line(-1,0){.0401956533}}
%\end
\qbezier(31.908,45.962)(34.427,44.769)(37.123,45.874)
\qbezier(32.262,9.634)(34.427,10.341)(36.416,9.634)
\put(31.731,41.719){$\alpha$}
\put(35.593,41.719){$\alpha$}
\put(29.698,7.513){$\beta$}
\put(37.156,7.69){$\beta$}
%\emline(34.56,24.484)(48.083,32.792)
\multiput(34.56,24.484)(.0547506769,.0336376708){247}
{\line(1,0){.0547506769}}
%\end
\qbezier(37.653,26.34)(38.449,25.677)(38.537,24.484)
\put(42.338,26.428){$\phi_1$}
%\emline(34.648,24.307)(59.22,20.86)
\multiput(34.648,24.307)(.238562725,-.033467433){103}
{\line(1,0){.238562725}}
%\end
\qbezier(42.25,24.484)(42.647,23.511)(42.161,23.246)
\put(39.952,19.594){{\footnotesize $\tilde\theta_A$}}
\qbezier(37.035,24.484)(36.858,26.428)(34.56,26.958)
\put(36.946,28.373){{\footnotesize $\theta_B$}}
%\vector[middle](46.934,36.681)(46.05,32.439)
\put(46.492,34.56){\vector(-1,-4){.07}}
\multiput(46.934,36.681)(-.0327364,-.1571348){27}
{\line(0,-1){.1571348}}
%\end
\put(47.818,39.068){{\footnotesize $y_0=0$}}
\put(52.326,17.501){${\bf y}_A =
\left(-\frac{1-\cos(\alpha+\beta)}{\sin(\alpha+\beta)};\
\frac{\sin\alpha+\sin\beta}{\sin(\alpha+\beta)},\
\frac{\cos\beta-\cos\alpha}{\sin(\alpha+\beta)}\right)$}
\put(39.421,49.674){${\bf y}_B =
\left(\frac{\cos\alpha}{\sin\alpha};\ 0,\
\frac{1}{\sin\alpha}\right) = \left(P; 0, p\right) $}
\put(-50.276,14.887){${\bf y}_C =
\left(-\frac{1-\cos(\alpha+\beta)}{\sin(\alpha+\beta)};\
-\frac{\sin\alpha+\sin\beta}{\sin(\alpha+\beta)},\
\frac{\cos\beta-\cos\alpha}{\sin(\alpha+\beta)}\right)$}
\put(29.875,2.74){${\bf y}_D =
\left(\frac{\cos\beta}{\sin\beta};\ 0,\
-\frac{1}{\sin\beta}\right) = \left(Q; 0, q\right)$}
\put(68.326,11.501){$= \left(-K; k_+, k_-\right)$}
\put(-28.326,8.887){$= \left(-K; -k_+, k_-\right)$}
%\put(39.421,49.674){$= \left(P; 0, p\right)$}
%\put(69.875,2.74){$= \left(Q; 0, q\right)$}
\end{picture}
%%
%%
%%\input{./pics/figkite.tex}
%{\includegraphics[width=200pt,height=200pt]
%{./pics/rho.jpg}}
%\input{./pics/FigZnpol.tex}
\caption{{\footnotesize
Kite-like polygon $\bar\Pi$ with only one $Z_2$-symmetry,
$y_1\rightarrow -y_1$.
Kites form a two-dimensional family, parameterized by
$\alpha$ and $\beta$.
Angles at four vertices are
$\pi - \alpha-\beta$ at $A$ and $C$,  $2\alpha$ at $B$
and $2\beta$ at $D$ and directions to vertices
are $\theta_A = 2\pi-\tilde\theta_A =\frac{\alpha-\beta}{2}$,
$\theta_B=\frac{\pi}{2}$, $\theta_C = \pi-\frac{\alpha-\beta}{2}$
and $\theta_D = \frac{3\pi}{2}$.
The four normal directions are:
$\phi_1 = \alpha$, $\phi_2 = \pi-\phi_1 = \pi-\alpha$,
$\phi_3 = \pi+\beta$ and $\phi_4 = 2\pi-\beta$.
Rhombus is a particular sub-family with $\beta=\alpha$.
Note that this picture is rotated by an angle $\frac{\pi}{4}$
as compared to Fig.\ref{figrhomb}.
}}
\label{figkite}
\end{center}
\end{figure}
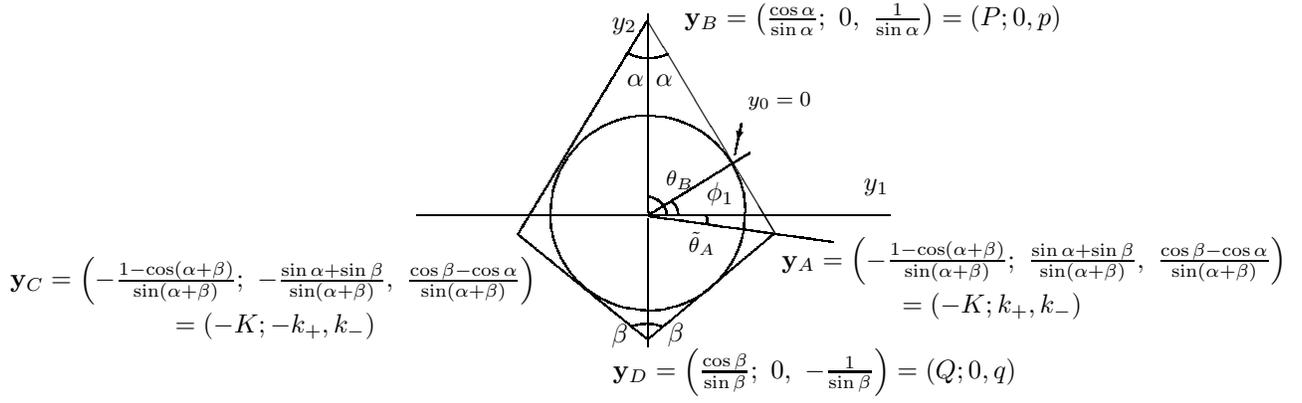

Kites form a two-dimensional family of polygons ${\bar\Pi}$,
which possess only one $Z_2$-symmetry, $y_1\leftrightarrow -y_1$.
We parameterize them by two angle variables $\alpha$ and $\beta$,
which are {\it halves} of the angles at two non-equivalent
vertices, see Fig.\ref{figkite}.
Rhombi with symmetry, enhanced to $Z_2\times Z_2$,
$y_2\leftrightarrow -y_2$ in addition to
$y_1\leftrightarrow -y_1$ are a one-parametric sub-family
of kites with $\alpha=\beta$.
Note that for comparison with the results of s.\ref{conrhomb}
one should also make a rotation of the $(y_1,y_2)$ plane
by $\frac{\pi}{4}$.
After this rotation the square solution (\ref{squareq})
turns into
\be
{\cal S}'_\Box = 2y_0 +y_1^2 -y_2^2=0
\label{squareqrot}
\ee
and rhombic solution (\ref{rhomeq}) -- into
\be
{\cal S}'_\diamond = 2y_0\cos(2\phi)
+ (1-y_0^2)\sin(2\phi) +y_1^2 -y_2^2=0
\label{rhomeqrot}
\ee
or
\be
{\cal S}'_\diamond \sim
2(1-b^2)y_0 + 2b(1-y_0^2) + (1+b^2)(y_1^2-y_2^2)=0
\label{rhomeqrotb}
\ee
with $\phi = \frac{\pi}{4}-\alpha$ and
\be
b = \frac{|\cos\alpha-\sin\alpha|}{\cos\alpha+\sin\alpha}
\ee

It is a simple geometrical exercise to express the values
of ${\bf y}$ at the kite vertices through $\alpha$ and $\beta$.
It is only important to remember that we put the radius
of inscribed circle equal to one.
It follows that the ordinates of the vertices $B$ and $D$
are $y_{2B}=\cot\alpha$ and $y_{2D}=\cot\beta$,
while the corresponding values of $y_0$ are
$y_{0B} = \frac{1}{\sin\alpha}$ and $y_{0D}=\frac{1}{\sin\beta}$,
because $y_0$ vanishes at the tangent points with the unit circle.
Further, the two side lengths $l_1=l_{AB}$ and
$l_4 = l_{DA}$ of the kite are related through
\be
l_1\cos\alpha + l_4\cos\beta
= \frac{1}{\sin\alpha} + \frac{1}{\sin\beta}, \nn \\
l_1\sin\alpha = l_4\sin\beta = y_{1A}
\ee
The most convenient variables for actual calculations are
$t = \tan\frac{\alpha}{2}$ and $t' = \tan\frac{\beta}{2}$,
i.e. trigonometric functions of the {\it quarters} of
the kite's angles with values bound between $0$ and $1$:
$0<t,t'<1$.
Unfortunately, they are much less convenient for consideration
of particular degenerations, in particular for
%rhombus $t'=t = ??$ and for
the square $t'=t=
\tan\frac{\pi}{8} = \sqrt{\frac{\sqrt{2}-1}{\sqrt{2}+1}}$.
In terms of these variables
\be
\sin\alpha = \frac{2t}{1+t^2}, \ \ \
\cos\alpha = \frac{1-t^2}{1+t^2}, \ \ \
\sin\beta = \frac{2t'}{1+{t'}^2}, \ \ \
\cos\beta = \frac{1-{t'}^2}{1+{t'}^2}
\label{alphat}
\ee
and
\be
\begin{array}{ccccc}
{\bf y}_A =
\left(-\frac{1-\cos(\alpha+\beta)}{\sin(\alpha+\beta)};\
\frac{\sin\alpha+\sin\beta}{\sin(\alpha+\beta)},\
\frac{\cos\beta-\cos\alpha}{\sin(\alpha+\beta)}\right)
&=& \big(-K;\ k_+,\ k_-\big) &=& %\frac{1}{1-tt'}
\left( -\frac{t+t'}{1-tt'};\ \frac{1+tt'}{1-tt'},\
\frac{t-t'}{1-tt'}\right), \\ &&\\
%\big(-K;\ k_+,\ k_-\big), \\ && \\
{\bf y}_B = \left(\frac{\cos\alpha}{\sin\alpha};\ 0,\
\frac{1}{\sin\alpha}\right) &=&
%\frac{1}{2t}
\big(P;\ 0,\ p\big) &=&
\left(\frac{1-t^2}{2t};\ 0,\ \frac{1+t^2}{2t}\,\right),\\ && \\
{\bf y}_C =
\left(-\frac{1-\cos(\alpha+\beta)}{\sin(\alpha+\beta)};\
-\frac{\sin\alpha+\sin\beta}{\sin(\alpha+\beta)},\
\frac{\cos\beta-\cos\alpha}{\sin(\alpha+\beta)}\right)
&=& \big(-K;\ -k_+,\ k_-\big) &=& %\frac{1}{1-tt'}
\left( -\frac{t+t'}{1-tt'};\ -\frac{1+tt'}{1-tt'},\
\frac{t-t'}{1-tt'}\right), \\ &&\\
%&=& \big(-K;\ -k_+,\ k_-\big),\\ && \\
{\bf y}_D = \left(\frac{\cos\beta}{\sin\beta};\ 0,\
-\frac{1}{\sin\beta}\right) &=&
\big(Q;\ 0,\ q\big) &=&
%\frac{1}{2t'}
\left(\frac{1-{t'}^2}{2t'};\ 0,\ -\frac{1+{t'}^2}{2t'}\right)
\end{array}
\label{kitevertices3}
\ee
It follows that
\be
z_1=z_A=z_C = \frac{1}{2}
\Big({\bf y}_{C}-{\bf y}_B)({\bf y}_B-{\bf y}_A)\Big)^{-1/2}
= \frac{1}{2\sqrt{2}}\frac{1-tt'}{1+tt'} ,\nn \\
z_2=z_B=z_D = \frac{1}{2}
\Big({\bf y}_{C}-{\bf y}_B)({\bf y}_D-{\bf y}_C)\Big)^{-1/2}
= \frac{\sqrt{2tt'}}{2(1+tt')}
\ee
In terms of condensed notation, introduced in
(\ref{kitevertices3}), the four lines in (\ref{mipro}) are now:
$$
ABD: \ \ \
z_1z_2^2\left(\,\left|\left|\begin{array}{ccc}
y_0 & y_1 & y_2 \\
-K & k_+ & k_- \\
P & 0 & p
\end{array}\right|\right|
+ \left|\left|\begin{array}{ccc}
y_0 & y_1 & y_2 \\
P & 0 & p \\
Q & 0 & q
\end{array}\right|\right|
+ \left|\left|\begin{array}{ccc}
y_0 & y_1 & y_2 \\
Q & 0 & q \\
-K & k_+ & k_-
\end{array}\right|\right|
- \left|\left|\begin{array}{ccc}
-K & k_+ & k_- \\
P & 0 & p \\
Q & 0 & q
\end{array}\right|\right|\,\right) =
$$
\vspace{-0.5cm}
\be
%= y_0(2+ 0 +2)+ y_1(0 + 2 +2)+ y_2(-2-2+ 0) - 4
= z_1z_2^2
\left\{k_+\Big((p-q)y_0 - (P-Q)y_2 + (Pq-Qp)\Big) +
\Big(K(p-q)+(P-Q)k_--(Pq-Qp)\Big)y_1\right\} \nn
%(y_0+y_1-y_2-1)
\ee
$$
CBD: \ \ \
z_1z_2^2\left(\,\left|\left|\begin{array}{ccc}
y_0 & y_1 & y_2 \\
-K & -k_+ & k_-\\
P & 0 & p
\end{array}\right|\right|
+ \left|\left|\begin{array}{ccc}
y_0 & y_1 & y_2 \\
P & 0 & p \\
Q & 0 & q
\end{array}\right|\right|
+ \left|\left|\begin{array}{ccc}
y_0 & y_1 & y_2 \\
Q & 0 & q \\
-K & -k_+ & k_-
\end{array}\right|\right|
+ \left|\left|\begin{array}{ccc}
P & 0 & p \\
-K & -k_+ & k_- \\
Q & 0 & q
\end{array}\right|\right|\,\right) =
$$
\vspace{-0.5cm}
\be
= z_1z_2^2
\left\{-k_+\Big((p-q)y_0 - (P-Q)y_2 + (Pq-Qp)\Big) +
\Big(K(p-q)+(P-Q)k_--(Pq-Qp)\Big)y_1\right\} \nn
%(-y_0+y_1-y_2+1)
\ee
$$
ACB: \ \ \
z_1^2z_2\left(\,\left|\left|\begin{array}{ccc}
y_0 & y_1 & y_2 \\
-K & k_+ & k_- \\
-K & -k_+ & k_-
\end{array}\right|\right|
+ \left|\left|\begin{array}{ccc}
y_0 & y_1 & y_2 \\
-K & -k_+ & k_- \\
P & 0 & p
\end{array}\right|\right|
+ \left|\left|\begin{array}{ccc}
y_0 & y_1 & y_2 \\
P & 0 & p \\
-K & k_+ & k_-
\end{array}\right|\right|
+ \left|\left|\begin{array}{ccc}
-K & k_+ & k_- \\
P & 0 & p \\
-K & -k_+ & k_-
\end{array}\right|\right|\,\right) =
$$
\vspace{-0.5cm}
\be
= 2k_+ z_1^2z_2
\Big((k_--p)y_0+(K+P)y_2-(Kp+Pk_-)\Big)\nn
%(-y_0+y_1+y_2-1)
\ee
$$
ACD: \ \ \
z_1^2z_2\left(\,\left|\left|\begin{array}{ccc}
y_0 & y_1 & y_2 \\
-K & k_+ & k_- \\
-K & -k_+ & k_-
\end{array}\right|\right|
+ \left|\left|\begin{array}{ccc}
y_0 & y_1 & y_2 \\
Q & 0 & q \\
-K & k_+ & k_-
\end{array}\right|\right|
+ \left|\left|\begin{array}{ccc}
y_0 & y_1 & y_2 \\
-K & -k_+ & k_- \\
Q & 0 & q
\end{array}\right|\right|
- \left|\left|\begin{array}{ccc}
-K & k_+ & k_- \\
-K & -k_+ & k_- \\
Q & 0 & q
\end{array}\right|\right|\,\right) =
$$
\vspace{-0.5cm}
\be
%= y_0(2+ 0 +2)+ y_1(0 + 2 +2)+ y_2(-2-2+ 0) - 4
= 2k_+z_1^2z_2
\Big((k_--q)y_0+(K+Q)y_2-(Kq+Qk_-)\Big)\nn
%(y_0+y_1+y_2+1)
\ee
Thus (\ref{discreq}) becomes:
\be
(z_1z_2)^2\left\{
z_2^2\Big(K(p-q)+(P-Q)k_--(Pq-Qp)\Big)^2y_1^2
-(k_+z_2)^2\Big((p-q)y_0 - (P-Q)y_2 + (Pq-Qp)\Big)^2
- \right. \nn \\ \left.
-(2k_+z_1)^2\Big((k_--p)y_0+(K+P)y_2-(Kp+Pk_-)\Big)
\Big((k_--q)y_0+(K+Q)y_2-(Kq+Qk_-)\Big)
\right\}
\nn
\ee
and finally
\be
{\cal S}_{kite} = D_{2|3} =
\frac{1}{128(1+tt')^2}\Big\{
\left((1-tt')^2-2(t^2+{t'}^2)\right)y_0^2
-4(t-t')(t+t')y_0y_2 +\nn\\
%+4(t+t')y_0\big(1-tt'-(t-t')y_2\big) + \nn \\
+ (1+tt')^2(y_1^2-y_2^2) - 2(t-t')^2y_2^2
+4(1-tt')\Big((t+t')y_0+(t-t')y_2\Big)
- \big(1-6tt' + (tt')^2\big)
\Big\}
\label{kitesol}
\ee
This expression can be also rewritten as
$$
64{\cal S}_{kite} = \frac{1}{2}(y_1^2+y_2^2-y_0^2-1)
+ \frac{(1-t^2)(1-{t'}^2)}{(1+tt')^2}y_0^2
%-\left(1+\frac{(t-t')^2}{(1+tt')^2}\right)
-\frac{(1+t^2)(1+{t'}^2)}{(1+tt')^2}y_2^2
+ \frac{4tt'}{(1+tt')^2} - $$ $$
-\frac{2(t^2-{t'}^2)}{(1+tt')^2}y_0y_2
+ \frac{2(1-tt')}{(1+tt')^2}
\Big((t+t')y_0+(t-t')y_2\Big)
$$
%$$= \nn \\
%= -\frac{1}{2}P_2 + \frac{(1+t^2)(1+{t'}^2)}{(1+tt')^2}
%\Big(y_0^2\cos\alpha\cos\beta - y_2^2 +
%y_0y_2(\cos\alpha-\cos\beta)
%+y_0\sin(\alpha+\beta) + y_2\sin(\alpha-\beta)
%+\sin\alpha\sin\beta\Big)
%$$
or, making use of (\ref{alphat}) to convert back to original
angular variables:
\be
\frac{1}{2}\Big(128{\cal S}_{kite} + P_2\Big)\cos(\alpha-\beta)
=\nn\\
= y_0^2\cos\alpha\cos\beta - y_2^2 +
y_0y_2(\cos\alpha-\cos\beta)
+y_0\sin(\alpha+\beta) + y_2(\sin\alpha-\sin\beta)
+\sin\alpha\sin\beta = \nn\\
= y_0^2\cos\alpha\cos\beta + y_0y_2(\cos\alpha-\cos\beta)
+y_0\sin(\alpha+\beta) +(\sin\alpha-y_2)(\sin\beta+y_2)
\label{kitesol2}
\ee

\subsubsection{A version of parametrization for generic
quadrilateral case \label{geskqu}}

\begin{figure}
\begin{center}
%%
%%
%TeXCAD Picture [quadril.pic]. Options:
%\grade{\on}
%\emlines{\off}
%\epic{\off}
%\beziermacro{\on}
%\reduce{\on}
%\snapping{\off}
%\pvinsert{% Your \input, \def, etc. here}
%\quality{8.000}
%\graddiff{0.005}
%\snapasp{1}
%\zoom{11.3138}
\unitlength 1mm % = 2.845pt
\linethickness{0.4pt}
\ifx\plotpoint\undefined\newsavebox{\plotpoint}\fi % GNUPLOT compatibility
\begin{picture}(76,83.25)(0,0)
%\circle(40.5,36.5){38.552}
\put(59.776,36.5){\line(0,1){.8998}}
\put(59.755,37.4){\line(0,1){.8978}}
\put(59.692,38.298){\line(0,1){.8939}}
\multiput(59.587,39.192)(-.02929,.177611){5}{\line(0,1){.177611}}
\multiput(59.441,40.08)(-.031291,.146709){6}{\line(0,1){.146709}}
\multiput(59.253,40.96)(-.032662,.124361){7}{\line(0,1){.124361}}
\multiput(59.024,41.83)(-.033627,.107363){8}{\line(0,1){.107363}}
\multiput(58.755,42.689)(-.030882,.084541){10}{\line(0,1){.084541}}
\multiput(58.446,43.535)(-.031631,.075461){11}{\line(0,1){.075461}}
\multiput(58.099,44.365)(-.032193,.067744){12}{\line(0,1){.067744}}
\multiput(57.712,45.178)(-.0326031,.0610776){13}{\line(0,1){.0610776}}
\multiput(57.288,45.972)(-.0328888,.0552399){14}{\line(0,1){.0552399}}
\multiput(56.828,46.745)(-.0330695,.0500682){15}{\line(0,1){.0500682}}
\multiput(56.332,47.496)(-.0331599,.0454405){16}{\line(0,1){.0454405}}
\multiput(55.801,48.223)(-.0331717,.0412641){17}{\line(0,1){.0412641}}
\multiput(55.237,48.925)(-.0331139,.0374667){18}{\line(0,1){.0374667}}
\multiput(54.641,49.599)(-.0329938,.0339917){19}{\line(0,1){.0339917}}
\multiput(54.015,50.245)(-.0345445,.0324145){19}{\line(-1,0){.0345445}}
\multiput(53.358,50.861)(-.0380211,.0324758){18}{\line(-1,0){.0380211}}
\multiput(52.674,51.445)(-.0418189,.0324695){17}{\line(-1,0){.0418189}}
\multiput(51.963,51.997)(-.0459945,.0323871){16}{\line(-1,0){.0459945}}
\multiput(51.227,52.516)(-.05062,.0322184){15}{\line(-1,0){.05062}}
\multiput(50.468,52.999)(-.055788,.0319504){14}{\line(-1,0){.055788}}
\multiput(49.687,53.446)(-.06162,.0315661){13}{\line(-1,0){.06162}}
\multiput(48.886,53.856)(-.068279,.031043){12}{\line(-1,0){.068279}}
\multiput(48.066,54.229)(-.083584,.033386){10}{\line(-1,0){.083584}}
\multiput(47.23,54.563)(-.094501,.03272){9}{\line(-1,0){.094501}}
\multiput(46.38,54.857)(-.107916,.031808){8}{\line(-1,0){.107916}}
\multiput(45.517,55.112)(-.124896,.030555){7}{\line(-1,0){.124896}}
\multiput(44.642,55.326)(-.147217,.028806){6}{\line(-1,0){.147217}}
\multiput(43.759,55.499)(-.2226,.03285){4}{\line(-1,0){.2226}}
\put(42.869,55.63){\line(-1,0){.8956}}
\put(41.973,55.72){\line(-1,0){.8988}}
\put(41.074,55.767){\line(-1,0){.9}}
\put(40.174,55.773){\line(-1,0){.8993}}
\put(39.275,55.737){\line(-1,0){.8967}}
\put(38.378,55.659){\line(-1,0){.892}}
\multiput(37.486,55.539)(-.177091,-.032288){5}{\line(-1,0){.177091}}
\multiput(36.601,55.377)(-.125279,-.028943){7}{\line(-1,0){.125279}}
\multiput(35.724,55.175)(-.108317,-.030414){8}{\line(-1,0){.108317}}
\multiput(34.857,54.932)(-.094915,-.0315){9}{\line(-1,0){.094915}}
\multiput(34.003,54.648)(-.084007,-.032306){10}{\line(-1,0){.084007}}
\multiput(33.163,54.325)(-.074916,-.032902){11}{\line(-1,0){.074916}}
\multiput(32.339,53.963)(-.06719,-.033333){12}{\line(-1,0){.06719}}
\multiput(31.533,53.563)(-.0605178,-.0336309){13}{\line(-1,0){.0605178}}
\multiput(30.746,53.126)(-.051031,-.0315633){15}{\line(-1,0){.051031}}
\multiput(29.98,52.652)(-.0464081,-.0317916){16}{\line(-1,0){.0464081}}
\multiput(29.238,52.144)(-.0422339,-.0319278){17}{\line(-1,0){.0422339}}
\multiput(28.52,51.601)(-.0384365,-.0319831){18}{\line(-1,0){.0384365}}
\multiput(27.828,51.025)(-.0349594,-.0319666){19}{\line(-1,0){.0349594}}
\multiput(27.164,50.418)(-.0334291,-.0335636){19}{\line(0,-1){.0335636}}
\multiput(26.529,49.78)(-.033594,-.0370368){18}{\line(0,-1){.0370368}}
\multiput(25.924,49.114)(-.0337008,-.0408331){17}{\line(0,-1){.0408331}}
\multiput(25.351,48.419)(-.0317579,-.0423618){17}{\line(0,-1){.0423618}}
\multiput(24.811,47.699)(-.033712,-.0496378){15}{\line(0,-1){.0496378}}
\multiput(24.305,46.955)(-.033598,-.0548115){14}{\line(0,-1){.0548115}}
\multiput(23.835,46.187)(-.0333876,-.0606524){13}{\line(0,-1){.0606524}}
\multiput(23.401,45.399)(-.033063,-.067324){12}{\line(0,-1){.067324}}
\multiput(23.004,44.591)(-.032601,-.075047){11}{\line(0,-1){.075047}}
\multiput(22.646,43.765)(-.031969,-.084136){10}{\line(0,-1){.084136}}
\multiput(22.326,42.924)(-.031118,-.095041){9}{\line(0,-1){.095041}}
\multiput(22.046,42.069)(-.029979,-.108439){8}{\line(0,-1){.108439}}
\multiput(21.806,41.201)(-.033179,-.146293){6}{\line(0,-1){.146293}}
\multiput(21.607,40.323)(-.031577,-.177219){5}{\line(0,-1){.177219}}
\put(21.449,39.437){\line(0,-1){.8925}}
\put(21.333,38.545){\line(0,-1){.897}}
\put(21.258,37.648){\line(0,-1){1.7995}}
\put(21.235,35.848){\line(0,-1){.8986}}
\put(21.286,34.95){\line(0,-1){.8952}}
\multiput(21.38,34.055)(.026999,-.177974){5}{\line(0,-1){.177974}}
\multiput(21.515,33.165)(.029397,-.1471){6}{\line(0,-1){.1471}}
\multiput(21.691,32.282)(.031056,-.124772){7}{\line(0,-1){.124772}}
\multiput(21.909,31.409)(.032241,-.107788){8}{\line(0,-1){.107788}}
\multiput(22.166,30.546)(.0331,-.094369){9}{\line(0,-1){.094369}}
\multiput(22.464,29.697)(.033722,-.083449){10}{\line(0,-1){.083449}}
\multiput(22.802,28.863)(.031317,-.068153){12}{\line(0,-1){.068153}}
\multiput(23.177,28.045)(.0318133,-.0614928){13}{\line(0,-1){.0614928}}
\multiput(23.591,27.245)(.0321741,-.0556592){14}{\line(0,-1){.0556592}}
\multiput(24.041,26.466)(.0324214,-.0504902){15}{\line(0,-1){.0504902}}
\multiput(24.528,25.709)(.0325715,-.0458641){16}{\line(0,-1){.0458641}}
\multiput(25.049,24.975)(.0326372,-.0416882){17}{\line(0,-1){.0416882}}
\multiput(25.604,24.266)(.0326283,-.0378904){18}{\line(0,-1){.0378904}}
\multiput(26.191,23.584)(.0325529,-.0344141){19}{\line(0,-1){.0344141}}
\multiput(26.809,22.93)(.0341239,-.032857){19}{\line(1,0){.0341239}}
\multiput(27.458,22.306)(.0375994,-.0329632){18}{\line(1,0){.0375994}}
\multiput(28.135,21.713)(.041397,-.0330057){17}{\line(1,0){.041397}}
\multiput(28.838,21.152)(.0455733,-.0329772){16}{\line(1,0){.0455733}}
\multiput(29.568,20.624)(.0502006,-.0328681){15}{\line(1,0){.0502006}}
\multiput(30.321,20.131)(.0553716,-.0326667){14}{\line(1,0){.0553716}}
\multiput(31.096,19.674)(.0612081,-.0323576){13}{\line(1,0){.0612081}}
\multiput(31.891,19.253)(.067873,-.031921){12}{\line(1,0){.067873}}
\multiput(32.706,18.87)(.075588,-.031328){11}{\line(1,0){.075588}}
\multiput(33.537,18.525)(.084665,-.030542){10}{\line(1,0){.084665}}
\multiput(34.384,18.22)(.107498,-.033196){8}{\line(1,0){.107498}}
\multiput(35.244,17.954)(.124491,-.032162){7}{\line(1,0){.124491}}
\multiput(36.115,17.729)(.146833,-.030701){6}{\line(1,0){.146833}}
\multiput(36.996,17.545)(.177727,-.028577){5}{\line(1,0){.177727}}
\put(37.885,17.402){\line(1,0){.8943}}
\put(38.779,17.301){\line(1,0){.8981}}
\put(39.678,17.242){\line(1,0){1.7996}}
\put(41.477,17.249){\line(1,0){.8976}}
\put(42.375,17.315){\line(1,0){.8935}}
\multiput(43.268,17.424)(.177492,.030003){5}{\line(1,0){.177492}}
\multiput(44.156,17.574)(.146582,.03188){6}{\line(1,0){.146582}}
\multiput(45.035,17.765)(.124229,.033161){7}{\line(1,0){.124229}}
\multiput(45.905,17.997)(.095313,.030274){9}{\line(1,0){.095313}}
\multiput(46.763,18.27)(.084417,.031221){10}{\line(1,0){.084417}}
\multiput(47.607,18.582)(.075334,.031934){11}{\line(1,0){.075334}}
\multiput(48.435,18.933)(.067614,.032465){12}{\line(1,0){.067614}}
\multiput(49.247,19.323)(.0609462,.0328482){13}{\line(1,0){.0609462}}
\multiput(50.039,19.75)(.0551074,.0331104){14}{\line(1,0){.0551074}}
\multiput(50.811,20.213)(.0499349,.0332703){15}{\line(1,0){.0499349}}
\multiput(51.56,20.712)(.045307,.0333422){16}{\line(1,0){.045307}}
\multiput(52.285,21.246)(.0411305,.0333372){17}{\line(1,0){.0411305}}
\multiput(52.984,21.813)(.0373334,.0332641){18}{\line(1,0){.0373334}}
\multiput(53.656,22.411)(.0338589,.03313){19}{\line(1,0){.0338589}}
\multiput(54.299,23.041)(.0322755,.0346744){19}{\line(0,1){.0346744}}
\multiput(54.912,23.7)(.0323229,.0381512){18}{\line(0,1){.0381512}}
\multiput(55.494,24.386)(.0323013,.041949){17}{\line(0,1){.041949}}
\multiput(56.043,25.099)(.0322021,.0461242){16}{\line(0,1){.0461242}}
\multiput(56.558,25.837)(.0320149,.050749){15}{\line(0,1){.050749}}
\multiput(57.039,26.599)(.0317261,.0559158){14}{\line(0,1){.0559158}}
\multiput(57.483,27.382)(.0313183,.0617463){13}{\line(0,1){.0617463}}
\multiput(57.89,28.184)(.033566,.074621){11}{\line(0,1){.074621}}
\multiput(58.259,29.005)(.033051,.083717){10}{\line(0,1){.083717}}
\multiput(58.59,29.842)(.032341,.094632){9}{\line(0,1){.094632}}
\multiput(58.881,30.694)(.031374,.108043){8}{\line(0,1){.108043}}
\multiput(59.132,31.558)(.030053,.125017){7}{\line(0,1){.125017}}
\multiput(59.342,32.433)(.028215,.147331){6}{\line(0,1){.147331}}
\multiput(59.511,33.317)(.03196,.22273){4}{\line(0,1){.22273}}
\put(59.639,34.208){\line(0,1){.8959}}
\put(59.725,35.104){\line(0,1){1.3958}}
%\end
\put(60.25,15.5){\line(0,1){57.75}}
\put(60.25,73.25){\line(0,1){0}}
%\emline(60.25,73.25)(12.75,41.75)
\multiput(60.25,73.25)(-.05085653105,-.03372591006){934}{\line(-1,0){.05085653105}}
%\end
%\emline(12.75,41.75)(28.75,19.5)
\multiput(12.75,41.75)(.0336842105,-.0468421053){475}{\line(0,-1){.0468421053}}
%\end
%\emline(28.75,19.5)(59.5,11.25)
\multiput(28.75,19.5)(.1255102041,-.0336734694){245}{\line(1,0){.1255102041}}
%\end
%\emline(59.5,11.25)(60.25,11)
\multiput(59.5,11.25)(.09375,-.03125){8}{\line(1,0){.09375}}
%\end
\put(60.25,15.25){\line(0,-1){4}}
%\emline(40.5,36.5)(60,72.75)
\multiput(40.5,36.5)(.0337370242,.062716263){578}{\line(0,1){.062716263}}
%\end
\put(1.25,36.5){\line(1,0){74.75}}
%\emline(40.75,36.25)(60,11.5)
\multiput(40.75,36.25)(.0337127846,-.0433450088){571}{\line(0,-1){.0433450088}}
%\end
%\emline(40.5,36.25)(28.75,19.5)
\multiput(40.5,36.25)(-.0336676218,-.0479942693){349}{\line(0,-1){.0479942693}}
%\end
%\emline(40.5,36.5)(13.25,41.75)
\multiput(40.5,36.5)(-.174679487,.033653846){156}{\line(-1,0){.174679487}}
%\end
\put(72.5,40){$y_1$}
%\emline(40.5,36.75)(40.25,83.25)
\multiput(40.5,36.75)(-.03125,5.8125){8}{\line(0,1){5.8125}}
%\end
\put(40.75,36){\line(0,-1){32}}
\put(34.5,78.75){$y_2$}
\put(63.5,67.75){$\alpha_B=\frac{\pi}{2}-\varphi_B$}
\put(52,73.75){$\alpha_B$}
\put(61,75){$B$}
\put(63.75,14){$\alpha_A=\frac{\pi}{2}-\varphi_A$}
\put(61.75,9){$A$}
\put(55.77,6.8){$\alpha_A$}
\qbezier(55.75,70.5)(57.5,67.125)(60.25,68.25)
\qbezier(56.25,12)(57.625,14.25)(60.5,14.5)
\put(58,80){$\theta_B=\varphi_B$}
\put(45,40.25){$\varphi_B$}
\put(47,31.5){$\varphi_A$}
\put(55,.5){$\theta_A = 2\pi-\varphi_A$}
\qbezier(44.595,31.217)(46.602,33.446)(46.528,36.271)
\qbezier(42.96,41.028)(44.818,38.872)(45.19,36.419)
\put(38.676,41.73){$\varphi_B$}
\put(32.676,39.73){$\varphi_C$}
\put(30.676,33.73){$\varphi_C$}
\put(28.676,25.73){$\varphi_D$}
\put(32.15,22.73){$\varphi_D$}
\put(41.676,26.73){$\varphi_A$}
%\emline(40.433,36.271)(26.163,58.271)
\multiput(40.433,36.271)(-.0337352246,.0520094563){423}{\line(0,1){.0520094563}}
%\end
%\emline(40.433,36.271)(20.96,21.852)
\multiput(40.433,36.271)(-.0454976636,-.0336892523){428}{\line(-1,0){.0454976636}}
%\end
%\emline(40.433,36.122)(34.338,14.419)
\multiput(40.433,36.122)(-.033674033,-.119906077){181}{\line(0,-1){.119906077}}
%\end
\put(19.473,60.65){$\phi_2=\pi-2\alpha_B$}
\put(-14.176,41.163){$\theta_C=2\varphi_B+\varphi_C$}
\put(9,37.5){$C$}
\put(-4.811,26.568){$\phi_3= 2\varphi_B+2\varphi_C$}
\put(-8.11,22.568){$= 2\pi - 2\varphi_A-2\varphi_D$}
\put(62.555,32.109){$\phi_1=0$}
\put(28.068,10.473){$\phi_4=2\pi-2\varphi_A$}
\put(67,5.176){${\bf y}_A =\Big(-\tan\varphi_A;
\ 1,\ -\tan\varphi_A\Big)$ }
\put(67,74){${\bf y}_B =\Big(\tan\varphi_B;
\ 1,\ \tan\varphi_B\Big)$ }
\put(-43,52.838){${\bf y}_C =
\Big(-\tan\varphi_C;\ 1-l_{BC}\sin(2\varphi_B),$}
\put(-20,48.838){$\tan\phi_B + l_{BC}\cos(2\varphi_B)\Big)$ }
\put(-47,4.798){${\bf y}_D = \Big(\tan\varphi_D;
\ 1-l_{AD}\sin(2\varphi_A),
\ -\tan\varphi_A - l_{AD}\cos(2\varphi_A)\Big)$ }
\put(-6.027,13.849){$\theta_D=2\pi-2\varphi_A-\varphi_D$}
\put(25,17){$D$}
\qbezier(38.626,39.067)(40.305,40.172)(41.984,38.979)
\qbezier(37.918,40.128)(35.886,39.023)(35.62,37.388)
\qbezier(34.648,37.565)(34.692,34.736)(36.327,33.322)
\qbezier(35.267,32.35)(35.709,31.245)(36.858,31.024)
\qbezier(36.858,31.024)(37.521,29.964)(38.714,30.317)
\qbezier(39.156,31.643)(41.542,31.466)(43.575,32.704)
\qbezier(15.114,43.31)(15.07,41.366)(14.142,39.775)
\qbezier(27.842,20.683)(29.256,20.108)(30.317,19.003)
\end{picture}
%%
%%
%\input{./pics/quadri2.tex}
%{\includegraphics[width=200pt,height=200pt]
%{./pics/rho.jpg}}
%\input{./pics/FigZnpol.tex}
\caption{{\footnotesize
Generic skew quadrilateral $\bar\Pi$ is parameterized by
four angles: $\varphi_{A,B,C,D}$, subjected to
constraint $2\varphi_A+2\varphi_B+2\varphi_C+2\varphi_D=2\pi$.
For circle of unit radius the side lengths are
$l_1=l_{AB} = \tan\varphi_A + \tan\varphi_B$ etc.
Rotation freedom is fixed by requiring that the first
segment $AB$ is vertical: $\phi_1 = 0$.
Then the other three normal directions are:
$\phi_2 = \phi_{BC} = 2\varphi_B$
$\phi_3=\phi_{CD} = 2\varphi_B + 2\varphi_C =
2\pi -2\varphi_A -2\varphi_D$,
$\phi_4 = \phi_{DA} = 2\pi - 2\varphi_D$
and direction towards the vertices are:
$\theta_A = 2\pi-\varphi_A$
$\theta_B = \varphi_B$, $\theta_C = 2\varphi_B+\varphi_C$,
$\theta_D = 2\pi - 2\varphi_A - \varphi_D$.
The angles of quadrilateral are
$2\alpha_A = \pi-2\varphi_A$,
$2\alpha_B = \pi-2\varphi_B$,
$2\alpha_C = \pi-2\varphi_C$ and
$2\alpha_D = \pi-2\varphi_D$
($\alpha_C$ and $\alpha_D$ are not shown).
}}
\label{figquadri}
\end{center}
\end{figure}
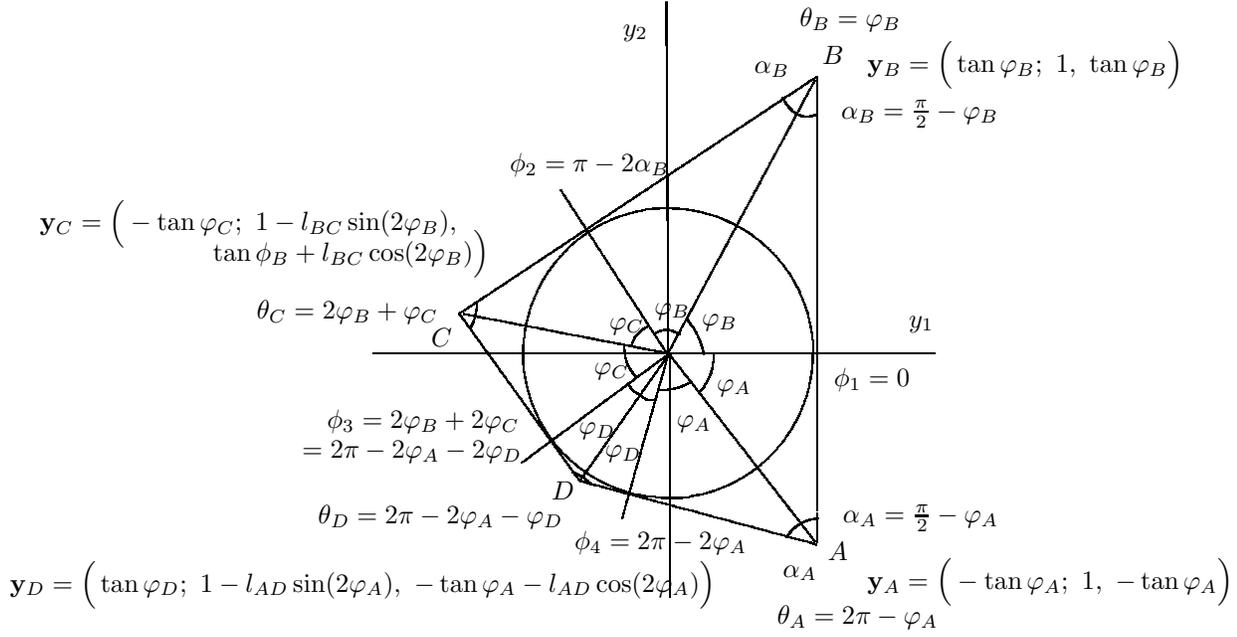

In the case of generic quadrilateral (with inscribed circle)
we have, see Fig.\ref{figquadri}:
\be
l_{AB} = \tan \varphi_A + \tan \varphi_B, \ \ \
{\bf y}_B-{\bf y}_A =
\Big(\sigma_{AB}l_{AB};\ -l_{AB}\sin\phi_{AB},\
l_{AB}\cos\phi_{AB}\Big)
\ee
and similarly for all other sides.
Therefore, assuming that the first side $AB$ is parallel
to ordinate axis, we can parameterize all vertices by
four angles $\varphi_a$ constrained by a single relation:
\be
\varphi_A+\varphi_B+\varphi_C+\varphi_D=\pi
\label{tancon}
\ee
Then
\be
{\bf y}_A = \Big(-\tan\varphi_A;\ 1,\ -\tan\varphi_A\Big),\nn\\
{\bf y}_B = \Big(\tan\varphi_B;\ 1,\ \tan\varphi_B\Big),\nn\\
{\bf y}_C = \Big(-\tan\varphi_C;\
1-l_{BC}\sin(2\varphi_B),\
\tan\phi_B + l_{BC}\cos(2\varphi_B)\Big),\nn\\
{\bf y}_D = \Big(\tan\varphi_D;\
1-l_{AD}\sin(2\varphi_A),\
-\tan\varphi_A - l_{AD}\cos(2\varphi_A)\Big)
\ee
One should further substitute
\be
\cos(2\varphi) = \frac{1-\tan^2\varphi}{1+\tan^2\varphi},
\ \ \ \sin(2\varphi) = \frac{2\tan\varphi}{1+\tan^2\varphi},
\ee
then the constraint (\ref{tancon}) is a simple relation
\be
t_A+t_B+t_C+t_D = t_At_Bt_C + t_At_Bt_D + t_At_Ct_D + t_Bt_Ct_D,
\label{tancon2}
\ee
linear in all $t$-variables.
If, say, $t_D$ is expressed through the three other variables, then
\be
1+t_D^2 = \frac{(1+t_A^2)(1+t_B^2)(1+t_C^2)}
{t_At_B + t_Bt_C + t_Ct_A-1)^2}
\ee
and
\be
z_1 = \sqrt{\frac{1+t_B^2}{8(t_A+t_B)(t_B+t_C)}},\ \ \ \
z_2 = \sqrt{\frac{t_At_B + t_Bt_C + t_Ct_A-1}{8(t_A+t_B)(t_B+t_C)}}
\ee
Evaluation of discriminant (\ref{discreq}) is straightforward and
results in:
$$
{\cal S}_{quadri} = D_{2|3} \sim
y_0^2\Big(-t_At_B-t_Bt_C + t_At_C + (2t_At_C-1)t_B^2\Big)
+\Big(2-t_At_B-t_Bt_C-t_Ct_A+t_B^2\Big)
+ $$ $$
+ y_1^2\Big(2+t_At_B-3t_Bt_C+t_At_C - t_B^2\Big)
+ y_2^2\Big(-3t_At_B+t_Bt_C-t_At_C + (2t_At_C+1)\Big)
+ $$ $$
+2y_1y_2\Big(-t_A+2t_B+t_C + (t_A-t_C)t_B^2+2t_At_Bt_C\Big)
+2y_0y_1\Big(t_A+t_C+2(-t_A+t_C)t_B^2 - 2t_At_Bt_C\Big)
+ $$
\be
+ 4y_0y_2t_At_B(1-t_Bt_C)
-2y_0(t_A+t_C)(1+t_B^2)
-4y_1(1-t_Bt_C)
+2y_2\Big(t_A-2t_B-t_C + (t_A+t_C)t_B^2\Big)
\label{quadrisol}
\ee
Omitted overall coefficient (unneeded for our purposes) is
\be
\frac{(z_1z_2)^2(t_A+t_B)(t_B+t_C)}{2(1+t_B^2)
(t_At_B+t_Bt_C+t_Ct_A-1)^2}
\ee
This is a rather long expression and it is asymmetric in
its variables, because use {\it independent} variables,
with $t_D$ excluded.
Actually this formula possesses cyclic symmetry under
$(ABCD)\rightarrow (BCDA) \rightarrow\ldots $and is also
invariant under permutations of {\it opposite} vertices
$B\leftrightarrow D$ and $A\leftrightarrow C$.
Only the last of these three symmetries is explicit in
(\ref{quadrisol}).

Particular case of square corresponds to $t_A=t_B=t_C
=\tan\frac{\pi}{4} = 1$, then (\ref{quadrisol}) becomes
\be
{\cal S}_{quadri} \ \stackrel{t_A=\ldots =1}{\longrightarrow}\
-8(y_0-y_1y_2) \sim {\cal S}_\Box,
\ee
as expected.

Comparison with the rhombus case is a little more involved.
For rhombus $t_A=t_B^{-1}=t_C=t_D^{-1}$:
pairs of opposite angles are equal, the sum of adjacent
angles is $\pi$ (this is true for any parallelogram, but
inscribed circle condition leaves only rhombi for our
consideration).
Expressed through $t_B=t$, eq.(\ref{quadrisol}) becomes:
\be
{\cal S}_{quadri} \ \stackrel{t_A=t_C=t_B^{-1}}{\longrightarrow}\
-4\left(t+\frac{1}{t}\right)\{y_0-y_1y_2+\frac{1}{4}
\left(t-\frac{1}{t})(y_0^2+y_1^2-y_2^2-1)\right\}
\sim {\cal S}_\diamond
\label{quadrirhom}
\ee
In order to compare this expression with other versions
of ${\cal S}_\diamond$ originated by \cite{am1}, we should
rotate it in the $(y_1,y_2)$ plane to switch from the choice of
vertical side $AB$, implied in \ref{quadrisol}, to
$\theta_B = \frac{\pi}{4}$, implied in (\ref{rhomeq}).
This means that we should rotate by angle $\phi_1=\phi_{AB}$,
which is related to $t=t_B=\tan(\varphi_B)$ with
$\varphi_B = \frac{\pi}{4}-\phi_{AB}$ by
\be
\tan(2\phi) = \cot(2\varphi_B) = \frac{\cos^2\varphi_B-
\sin^2\varphi_B}{2\sin\varphi_B\cos\varphi_B} =
-\frac{1}{2}\left(t-\frac{1}{t}\right)
\ee
Substituting $(y_1,y_2)\rightarrow
(y_1\cos\phi+y_2\sin\phi,\ -y_1\sin\phi+y_2\cos\phi)$ into
(\ref{quadrirhom}) we convert the r.h.s. into
\be
y_0 - y_1y_2\cos(2\phi) + \frac{1}{2}(y_1^2-y_2^2)\sin(2\phi)
-\frac{1}{2}\tan(2\phi)\Big(y_0^2+2y_1y_2\sin(2\phi)+
(y_1^2-y_2^2)\cos(2\phi)-1\Big) =\nn \\ =
\frac{1}{\cos(2\phi)}\left(y_0 \cos(2\phi) - y_1y_2
+\frac{1}{2}(1-y_0^2)\sin(2\phi)\right)
\ \stackrel{\ref{rhomeq})}{=}\
-\frac{1}{\cos(2\phi)}{\cal S}_\diamond
\ee

One more way to represent ${\cal S}_{quadri}$ is to
express it through canonical elements $P_2$ and
${\cal L}_{quadri}$, which is linear in $y_0$
with coefficient one:
\be
{\cal S}_{quadri} \sim {\cal L}_{quadri} + \mu_{quadri}P_2
\ee
From (\ref{quadrisol})
\be
\mu_{quadri} = -\frac{t_At_C-(t_A+t_C)t_B+(2t_At_C-1)t_B^2}
{2(t_A+t_C)(1+t_B^2)}
\ee
It turns into $\mu_\Box = 0$ for the square (when all four $t_a=1$)
and into $\mu_\diamond = \frac{1}{2}\left(t-\frac{1}{t}\right)
= -\frac{1}{2}\tan(2\phi_{AB})$ for rhombus.

\subsection{Intermediate conclusion
\label{intercon}}

The main result of this section is that {\it exact solution}
to our Plateau problem for generic skew quadrilateral $\Pi$
is reduced to {\it quadratic} equation in $y$-variables:
\be
{\cal S}_\Pi(y_0;y_1,y_2) = 0
\label{exasol}
\ee
Moreover, it is quadratic in $y_0$.
Only in the case of the square, $\bar\Pi=\Box$, i.e. for
$Z_4$-symmetric configuration, it further reduces to a linear
(\ref{squareq}).
This means that {\it such} elements,
more sophisticated than linear, but still simple,
should be of primary interest for us in the study of
the boundary ring at least at $n=4$.
This new experience implies certain modification of
research direction, suggested in sections \ref{goal} and
\ref{plan} on the base of $Z_n$-symmetric considerations,
shifting attention from $y_0$-linearity of the desired
boundary ring elements.

In the next section \ref{bori} we continue discussion of the
boundary ring structure, originated in \cite{malda3}.
Not-surprisingly, ${\cal S}_\Pi$ is not immediately distinguished
as an element of ${\cal R}_\Pi$ -- it belongs to the intersection
of the ring with the space of NG solutions and can not be found
by considerations of the ring only, -- but it can be
easily found within the {\it simple} classes of elements in
${\cal R}_\Pi$.
A systematic approach can be to classify the elements of
${\cal R}_\Pi$ of a given degree in $y$-variables,
and then use them as anzatze for solutions to Plateau problem.
Such anzatze will contain a few free parameters ("moduli"),
because degree does not fix the element of ${\cal R}_\Pi$
unambiguously.
They can be either perturbed, substituted into NG equations
and analyzed by the methods of s.{comprec} or, instead,
as suggested in \cite{mmt1}, used to evaluate
the regularized action, which can be
afterwards minimized w.r.t. the remaining "moduli".
This provides two approximate methods, which can occasionally
produce exact answers (and then coincide).
It would be particularly interesting to analyze in detail the
families ${\cal F}_{n/2}$ of degree $n/2$ in ${\cal R}_\Pi$.
Not only exact solutions ${\cal S}_\Pi$ at $n=2$ and $n=4$
belong to ${\cal F}_{n/2}$,
such families looks distinguished in the theory of the rings
themselves: $n/2$ looks like the lowest degree necessary to
distinguish between the ring itself and its sub-rings,
associated with unifications of $\Pi$ with additional lines.
%General consideration of these problems is postponed to
%other publications.

\newpage

\section{Boundary ring for polygons
\label{bori}}
\setcounter{equation}{0}

This section is devoted to simple arithmetics of the polygon
boundary ring and is a first step towards their systematic
consideration on the lines, implied by s.\ref{intercon}.
Essential simplification of ${\cal R}_\Pi$ is provided by
conditions (\ref{ads3}) and we continue to impose them.
Then $P_2 = y_0^2+1-y_1^2-y_2^2 = y_0^2+1-z\bar z$ is
always an element of ${\cal R}_\Pi$, but we need more.
The situation is not quite simple because generically there
are no "generators" in the rings of polynomials of many
variables,\footnote{This is the same simple algebro-geometric
statement, which is the origin of the old puzzle
in the foundations of first-quantized string theory,
see \cite{LM}.}
instead a sophisticated structure arises of complementary
maximal ideals and "dual" descriptions.
We do not go in details of abstract algebra
in this paper\footnote{It deserves emphasizing once again,
that we are interested in not-generic "singular" situation,
what is best illustrated by s.\ref{andi} above,
and all the associated peculiarities are essential.}
and concentrate on the down-to-earth consideration of
low-degree elements in ${\cal R}_\Pi$, to provide concrete
information for further developments.
Our "universal" $P_2$ is of degree two, but the other
"obvious" polynomials $P_\Pi$,
considered in \cite{malda3} and listed in (\ref{Ppolsdef}),
are of the "high" degree $n$.
At the same time, ${K}_{n/2}$  in (\ref{Kn2def})
and ${\cal S}$ in s.\ref{exacon} are of degree $n/2$ and
still belong to ${\cal R}_\Pi$.

In order to put the situation under control we fully use the
specifics of our boundary ring: the fact that $\Pi$ consists
of intersecting straight segments
(actually, entire lines, if we are interested in polynomial
boundary rings)
and thus can be constructed from elementary rings for
individual straight lines.
This allows to introduce complex-valued elements
${\cal C}_\Pi \in {\cal R}_\Pi$,
which, like $P_\Pi$, are multiplicative characters, i.e.
are products of the elementary ${\cal C}_|$
for individual segments.
Of course, they are also of degree $n$ in $y$-variables.
Then we demonstrate how the relevant real-valued elements of
lower degree can be extracted in a generalizable fashion.

\subsection{A single null segment
\label{1segsec}}

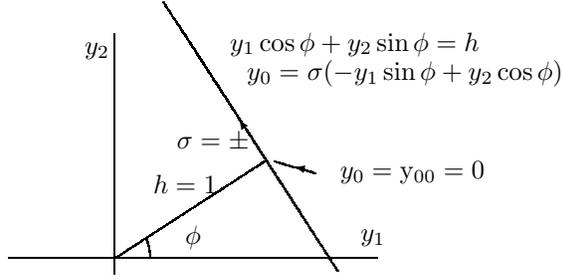
\begin{figure}
\begin{center}
%%
%%
%TeXCAD Picture [1segment.pic]. Options:
%\grade{\on}
%\emlines{\off}
%\epic{\off}
%\beziermacro{\on}
%\reduce{\on}
%\snapping{\off}
%\pvinsert{% Your \input, \def, etc. here}
%\quality{8.000}
%\graddiff{0.005}
%\snapasp{1}
%\zoom{11.3137}
\unitlength 1mm % = 2.845pt
\linethickness{0.4pt}
\ifx\plotpoint\undefined\newsavebox{\plotpoint}\fi
% GNUPLOT compatibility
\begin{picture}(50.392,38.271)(0,0)
\put(1.267,4.209){\line(1,0){49.125}}
\put(15.379,2.076){\line(0,1){31.953}}
%\emline(21.876,38.271)(45.078,2.475)
\multiput(21.876,38.271)(.03372383721,-.05202906977){688}
{\line(0,-1){.05202906977}}
%\end
%\emline(15.513,4.199)(35.532,17.192)
\multiput(15.513,4.199)(.0518626943,.0336606218){386}
{\line(1,0){.0518626943}}
%\end
\put(11.403,31.377){$y_2$}
\put(48.127,6.452){$y_1$}
\put(20.539,12.816){${\footnotesize h=1}$}
\put(30.759,31.775){$y_1\cos\phi + y_2\sin\phi = h$}
\put(33.013,27.93){$y_0 = \sigma(-y_1\sin\phi + y_2\cos\phi)$}
\qbezier(19.623,6.717)(20.152,5.325)(20.152,4.199)
\put(24.793,6.054){$\phi$}
%\vector[middle](40.173,10.032)(23.865,35.223)
\put(32.019,22.628){\vector(-2,3){.07}}
\multiput(40.173,10.032)(-.0336942149,.0520475207){484}
{\line(0,1){.0520475207}}
%\end
\put(23.709,18.71){${\footnotesize \sigma=\pm}$}
%\vector[middle](42.162,15.468)(36.592,17.058)
\put(39.377,16.263){\vector(-4,1){.07}}
\multiput(42.162,15.468)(-.11603125,.033125){48}
{\line(-1,0){.11603125}}
%\end
\put(45.343,14.805){$y_0 = {\rm y}_{00} = 0$}
\end{picture}
%%
%%
%%\input{./pics/1segment2.tex}
%{\includegraphics[width=200pt,height=200pt]
%{./pics/rho.jpg}}
%\input{./pics/FigZnpol.tex}
\caption{{\footnotesize
A single segment, a part of a straight line.
Shown is its projection  $\bar\Pi=|$ on the $(y_1,y_2)$ plane.
The line is light-like and thus is fully defined by
three parameters: angle $\phi$, distance $h$ and discrete
choice $\sigma = \pm$ of the $y_0$ direction w.r.t. direction
in the $(y_1,y_2)$ plane, denoted by arrow on the line.
Such straight line in the $3d$ space $(y_0;y_1,y_2)$ satisfies
two real-valued linear equations, which can be unified into
a single complex-valued ${\cal C}_| = 0$, eq.(\ref{P1}).
Note that $\phi$ is defined to be the direction of a {\it normal},
not of the line itself.
}}
\label{1segment}
\end{center}
\end{figure}

According to (\ref{segmcomp}),
\be
z = y_1+iy_2 = e^{i\phi}\Big(h+i\sigma (y_0-{\rm y}_{00})\Big),
\ee
where, see Fig.\ref{1segment},
$\phi$ is an angle between a normal to the
segment and the $y_1$-axis,
$h$ is the length of the normal from the origin
to its intersection point with the straight line which
contains our segment,
${\rm y}_{00}$ is the value of $y_0$ at this intersection point,
while $\sigma=\pm 1$, depending on the direction of $y_0$.
This relation defines an element of
the boundary ring,
\be
{\cal C}_|={\cal C}_|(\phi,\sigma|h,{\rm y}_{00})=
y_0-{\rm y}_{00} - i\sigma\left(h- e^{-i\phi}z\right)
\label{P1}
\ee
which vanishes along the segment. %\footnote{
In fact it vanishes on entire straight line,
which contains the segment.
Of course, this property is inherited by
boundary rings in all more complicated situations:
polynomials vanishing on the sides of a polygon
will do so on entire straight lines, containing these
segments.
This is a general feature of any approach based on polynomials,
though it is not necessarily preserved in transition from
algebraic geometry to functional analysis.
It deserves mentioning that solutions to
Plateau problem in flat Euclidean space are believed to
respect this property, see, for example, \cite{Oss}.

Actually the real and imaginary parts of (\ref{P1})
are the two independent generators of
${\cal R}_{\rm segment}$:
\be
{\rm Re}({\cal C}_|) = y_0-{\rm y}_{00}
- \sigma {\rm Im}(e^{-i\phi}z) =
y_0-{\rm y}_{00} + \sigma(sy_1- cy_2), \nn \\
\sigma {\rm Im}({\cal C}_|) = -h + {\rm Re}(e^{-i\phi}z) =
cy_1+sy_2-h
\label{1gene}
\ee
They are both {\it linear} in $y$-variables.
Our universal element $P_2$ is
a quadratic combination of these two generators:
\be
P_2 = (y_0-{\rm y}_{00})^2+h^2-y_1^2-y_2^2 = -|{\cal C}_||^2
+ 2\Big((y_0-{\rm y}_{00}){\rm Re}({\cal C}_|)
-h\,{\rm Im}({\cal C}_|)\Big)
\ee
For example, the boundary rings of coordinate axes
%$y_2$ and $y_1$
are produced by the complex generators
\be
\begin{array}{ccc}
y_2\ {\rm axis}: & \ \ &
{\cal C}_|(0,\sigma |0,0) = y_0+i\sigma z
= (y_0-\sigma y_2)+iy_1, \\
y_1\ {\rm axis}: & \ \ &
{\cal C}_|(\frac{\pi}{2},\sigma |0,0) = y_0 + \sigma z
= (y_0+\sigma y_1) + i\sigma y_2
\end{array}
\label{coaxgen}
\ee
Indeed, the normal to the $y_2$ axis is directed along the
$y_1$, i.e. $\phi = 0$, while normal to $y_1$ is directed
along $y_2$ so that $\phi'=\frac{\pi}{2}$.
Further, ${\cal C}_|(0,\sigma|0,0) = 0$ implies that $y_1=0$ and
$y_0 = \sigma y_2$, while
${\cal C}_|(\frac{\pi}{2},\sigma'|0,0) = 0$ --
that $y_2 = 0$ and $y_0 = -\sigma' y_1$.

For generic $\phi$ the real and imaginary parts of ${\cal C}_|$
are:
\be
{\rm Re} ({\cal C}_|) = 1-cy_1-sy_2
\ \stackrel{(\ref{Ppolsdef})}{=}\ P_|(y_1,y_2), \nn \\
{\rm Im} ({\cal C}_|) = \sigma y_0 + sy_1 - cy_2 \equiv
\sigma{\cal L}_|^\sigma
\label{Lsegm}
\ee
where $c = \cos\phi$, $s=\sin\phi$ and ${\cal L}$ is a linear
element from ${\cal R}_|$, satisfying the condition
(\ref{linearity1}).

It is clear from these examples that only the real and imaginary
part together, not any one of them separately, provides an
adequate description of the ring.
Perhaps more surprisingly, if we take any of these two elements
and supplement it by $P_2$, we do {\it not} obtain
a proper description of the boundary ring.
Indeed, a pair $\{P_|,P_2\}$ does not contain any information
about $\sigma$ and can not distinguish between the two
{\it different} boundary rings ${\cal R}_|^{\sigma=+}$ and
${\cal R}_|^{\sigma=-}$, associated with two different polygons
$\Pi$ which have the same projection $\bar\Pi$ on the $(y_1,y_2)$
plane.
As to the pair $\{{\cal L}_|^\sigma, P_2\}$, it specifies
$\sigma$ appropriately, however it does not distinguish between
two different $\bar\Pi$(!): two parallel, but different lines
with two different angle variables $\phi$ and $\phi+\pi$.
We return to discussion of this phenomenon in s.\ref{Lparpm}
below.

\subsection{From a single segment to generic polygon
\label{frosito}}

Given eq.(\ref{P1}), one can immediately construct a
complex element of the boundary ring for any collection
of intersecting straight lines:
\be
{\cal C}_{[+\ldots +]} =
\prod_{a=1}^n {\cal C}_|(\phi_a,\sigma_a|h_a,{\rm y}_{0a})
\label{propol}
\ee
Actually this formula is not unique, one can change
some entries in the product by complex conjugates:
actually there are $2^{n-1}$
non-equivalent possibilities,
\be
{\cal C}_{\underbrace{[\pm...\pm]}_n}
\{\phi_a,\sigma_a|h_a,{\rm y}_{0a}\},
\label{subscripttt}
\ee
where $\pm$ label the choice of ${\cal C}_1$
or $\overline{{\cal C}_|}$
at the given position in the product (\ref{propol}).
Any of them can be used for description of the boundary ring.
In what follows we concentrate on ${\cal C}_{[+\ldots +]}$,
which analytically depends on $z$,
and make additional simplifying assumptions.

When all $h_a$ are equal, $h_a=h$,
(this happens whenever projected polygon $\bar\Pi$
possesses an inscribed circle),
then also all ${\rm y}_{0a}$ are the same and can be shifted to
${\rm y}_{00}=0$, so that
\be
P_2 = y_0^2+h^2-y_1^2-y_2^2
\label{P2def}
\ee
is always an element of the boundary ring and
polynomials ${\cal C}$ can be divided by $P_2$,
like it was done in s.3.3 of \cite{malda3}, so that the
residue can be required to satisfy some constraint
of our choice.
For example, it can always be made linear in $y_0$
and satisfy the linearity condition (\ref{linearity1}).
As an example of a different choice,
$z$-analyticity implies that $P_2$ is not involved.
Most important, sometime the division procedure can be
used to decrease the degree of the bound ring element:
${\cal C}$ defined in (\ref{propol}) has degree $n$
in $y$-variables.

Since all $h_a$ are equal, we rescale $y$-variables to put $h=1$.
Thus in what follows in this section $h=1$, ${\rm y}_{0a}=0$
and subscript in (\ref{subscripttt}) is always $[+\ldots +]$.
Therefore all these labels will be omitted.
Instead,  to further simplify the formulas,
$\sigma$ will be often attached as
superscript to the corresponding $\phi$-variable.
Finally, in most cases (but not everywhere) we assume
that $y_0$ switches direction at the vertex, i.e.
$\sigma_{a+1} = -\sigma_a$ and $\sigma_a = (-)^{a-1}$
-- this, however, will always be mentioned {\it explicitly}.

\subsection{A chain of two null segments:
an angle (cusp or cross) and two parallel lines \label{ancusp}}

Consider first the two neighboring segments, with
different angles $\phi=\phi_1$ and $\phi'=\phi_2$, which
meet at a vertex and form an angle $2\alpha = \pi -(\phi'-\phi)$
(often called "cusp" in the literature on string/gauge
duality; since polynomials from the boundary ring will
vanish on entire two straight lines it could even better
be named "cross" in this context).

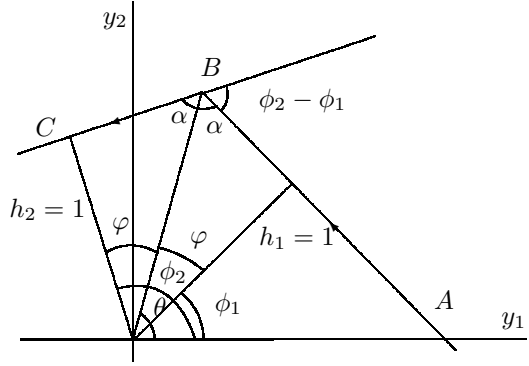
\begin{figure}
\begin{center}
%%
%%
%TeXCAD Picture [2angle2.tex]. Options:
%\grade{\on}
%\emlines{\off}
%\epic{\off}
%\beziermacro{\on}
%\reduce{\on}
%\snapping{\off}
%\pvinsert{% Your \input, \def, etc. here}
%\quality{8.000}
%\graddiff{0.005}
%\snapasp{1}
%\zoom{11.3137}
\unitlength 1mm % = 2.845pt
\linethickness{0.4pt}
\ifx\plotpoint\undefined\newsavebox{\plotpoint}\fi
% GNUPLOT compatibility
\begin{picture}(66.335,47.862)(0,0)
%\emline(-1.016,2.917)(66.335,3.049)
\multiput(-1.016,2.917)(16.83775,.033){4}{\line(1,0){16.83775}}
%\end
\put(13.965,-0){\line(0,1){47.862}}
%\vector[middle](56.922,1.458)(23.246,35.797)
\put(40.084,18.628){\vector(-1,1){.07}}
\multiput(56.922,1.458)(-.03370970971,.03437337337){999}
{\line(0,1){.03437337337}}
%\end
%\vector[middle](23.246,35.797)(-1.414,27.577)
\put(10.916,31.687){\vector(-3,-1){.07}}
\multiput(23.246,35.797)(-.101065574,-.033688525){244}
{\line(-1,0){.101065574}}
%\end
%\emline(13.965,2.917)(5.613,29.831)
\multiput(13.965,2.917)(-.0336774194,.1085241935){248}
{\line(0,1){.1085241935}}
%\end
%\emline(14.098,2.917)(35.046,23.6)
\multiput(14.098,2.917)(.0341172638,.0336856678){614}
{\line(1,0){.0341172638}}
%\end
%\emline(13.965,2.917)(23.114,35.665)
\multiput(13.965,2.917)(.0336360294,.1203970588){272}
{\line(0,1){.1203970588}}
%\end
\qbezier(20.462,9.016)(23.246,6.762)(23.379,2.917)
\put(16.705,6.132){$\theta$}
\put(23.705,30.406){$\alpha$}
\put(19.003,31.776){$\alpha$}
\put(24.837,6.806){$\phi_1$}
\put(30.731,33.941){$\phi_2-\phi_1$}
%\emline(23.335,35.709)(46.227,43.31)
\multiput(23.335,35.709)(.101292035,.033632743){226}
{\line(1,0){.101292035}}
%\end
\qbezier(25.633,33.322)(26.826,34.516)(26.428,36.593)
\qbezier(20.594,34.825)(22.23,32.969)(25.102,33.941)
%\put(17.296,40.333){$2\alpha=\pi+\phi_1-\phi_2$}
\qbezier(14.938,6.452)(16.794,5.038)(16.882,2.917)
\qbezier(11.932,9.723)(19.313,11.711)(22.274,2.917)
\put(17.501,10.96){$\phi_2$}
\put(63,5){$y_1$}
\put(10,45){$y_2$}
\put(1,30){$C$}
\put(54,7){$A$}
\put(23,38){$B$}
\put(-2.419,19.291){$h_2=1$}
\put(30.759,15.468){$h_1=1$}
\qbezier(10.341,14.496)(13.612,16.44)(17.236,14.496)
\qbezier(17.501,15.114)(20.374,14.496)(23.423,12.109)
\put(11.198,17.589){$\varphi$}
\put(21.567,15.822){$\varphi$}
\end{picture}
%%
%%
%%\input{./pics/2angle2.tex}
%{\includegraphics[width=200pt,height=200pt]
%{./pics/rho.jpg}}
%\input{./pics/FigZnpol.tex}
\caption{{\footnotesize
A pair of segments $AB$ and $BC$, which form an angle
$ABC$ of the size $2\alpha=\pi+\phi_1-\phi_2$.
Both sides of the angle are at the same distance
$h_1=h_2=1$ from the origin.
Shown also are the angle $\theta = \frac{\phi_1+\phi_2}{2}$,
which defines the direction
to the vertex $B$ of the angle and $\varphi = \theta-\phi_1 =
\phi_2-\theta = \frac{\pi}{2}-\alpha = \frac{\phi_2-\phi_1}{2}$.
}}
\label{2angle}
\end{center}
\end{figure}

\subsubsection{The case of $\ \sigma_2=-\sigma_1$}

With all above-mentioned restrictions we have
\be
{\cal C}_\angle =
{\cal C}\{\phi_2^-,\phi_1^+\} =
{\cal C}_|(\phi_2^-){\cal C}_|(\phi_1^+) =
\Big(y_0 + {\i}(1-e^{-{\i}\phi_1}z)\Big)
\Big(y_0 - {\i}(1-e^{-{\i}\phi_2}z)\Big) = \nn \\ =
1+y_0^2 - 2ze^{-\i\theta}(y_0\sin\varphi + \cos\varphi)
+ \left(ze^{-\i\theta}\right)^2 =
P_2 + z\bar z - 2ze^{-\i\theta}(y_0\sin\varphi + \cos\varphi)
+ \left(ze^{-\i\theta}\right)^2
\ee
where $\phi_1=\theta-\varphi$ and $\phi_2=\theta+\varphi$.

For example, at $\theta = \frac{\pi}{4}$ imaginary and real
part of ${\cal C}_\angle$ are
\be
{\rm Im} \left({\cal C}_\angle(\theta=\frac{\pi}{4})\right)
= (y_1-y_2)
\Big(\sqrt{2}(y_0\sin\varphi + \cos\varphi) - (y_1+y_2)\Big)
\ee
and
\be
{\rm Re} \left({\cal C}_\angle(\theta=\frac{\pi}{4})\right) =
1+y_0^2 - \sqrt{2}(y_0\sin\varphi + \cos\varphi)(y_1+y_2)
\ee
respectively.
These two elements of ${\cal R}_\angle$
are related by addition/subtraction of $P_2$,
one of them is quadratic while another linear in $y_0$,
however, the coefficient in front of $y_0$ is proportional
to $y_1-y_2$ and condition (\ref{linearity1}) is not satisfied.
However, this $(y_1-y_2)$ is a common factor in front of
entire expression, moreover it does not belong to ${\cal R}_\angle$
and can be simply thrown away -- thus giving rise to an
$y$-linear element of ${\cal R}_\angle$.

Since $\theta=\frac{\pi}{4}$ is not a restriction
($\theta$ can be changed by overall rotation),
this linear element always exists in ${\cal R}_\angle^{-+}$.
Because it is a procedure  that we repeatedly use below,
we formulate it once again.
Subtracting $P_2$, one can convert ${\cal C}_\angle$ into
an $y_0$-linear element of the boundary ring:
\be
%\tilde P\{\phi_2^-,\phi_1^+\} =
{\cal C}\{\phi_2^-,\phi_1^+\}-P_2
= z\bar z - 2ze^{-\i\theta}(y_0\sin\varphi + \cos\varphi)
+ \left(ze^{-\i\theta}\right)^2
\ee
The crucial phenomenon is that the coefficient of the $y_0$-linear
term is actually a common factor $z$ in the whole expression.
Furthermore, it is not identically zero in the ring and thus can be
eliminated.
This provides a {\it new} element of the boundary ring which
in this case is  automatically linear in
{\it all} the $y$-variables:
\be
%\widetilde{\tilde P}\phantom.^{-+}
%\{\theta+\varphi,\theta-\varphi\} =
%\widetilde{\tilde P}\{\phi_2^-,\phi_1^+\} =
\frac{{\cal C}\{\phi_2^-,\phi_1^+\}-P_2}{ze^{-\i\theta}}
=\ \bar z e^{\i\theta} + ze^{-\i\theta} -
2(y_0\sin\varphi + \cos\varphi) = -2{\cal L}_\angle
\label{Langle}
\ee
Indeed, substituting $\ ze^{-\i\phi} = 1+i\sigma y_0\ $
we get:
\be
\left.\frac{1}{2}
{\cal L}_\angle\Big\{(\theta+\varphi)^-,(\theta-\varphi)^+\Big\}
\right|_{z = (1+i\sigma y_0)e^{\i\phi}} =
%\Big(z = (1+i\sigma y_0)e^{\i\phi}\Big) =
\Big(\cos(\theta-\phi)-\cos\varphi\Big) +
y_0\Big(\sigma \sin(\theta-\phi) - \sin\varphi\Big)
\ee
and this expression obviously vanishes for
$\theta-\phi = \pm\varphi$ and $\sigma = \pm 1$.

Note that despite we obtained it from the complex-valued
character ${\cal C}_\angle$, this new element (\ref{Langle})
is {\it real}:
\be
{\cal L}_\angle^{-+} = y_0\sin\varphi + \cos\varphi -
y_1\cos\theta - y_2\sin\theta =
y_0\cos\alpha + \sin\alpha -
y_1\cos\theta - y_2\sin\theta
\label{Langelexp}
\ee
We do not divide the r.h.s. by $\cos\alpha$ to simplify
the formulas, however, this hides the singularity of
the limit $\alpha \rightarrow \frac{\pi}{2}$.
At other values of $\alpha$
the boundary ring ${\cal R}^{-+}_\angle$ is nicely
described by the pair $({\cal L}_\angle, P_2)$.

Existence of ${\cal L}$ is a non-trivial phenomenon.
We do not need to go far away to find a situation when it does
not exist: it is enough to switch from alternating $\sigma$
to a constant one.

\subsubsection{The case of $\ \sigma_2=\sigma_1$
\label{anglepp}}

In this case we obtain:
\be
{\cal C}\{\phi_2^+,\phi_1^+\} =
{\cal C}_|(\phi_2^+){\cal C}_|(\phi_1^+) =
\Big(y_0 - {\i}(1-e^{-{\i}\phi_1}z)\Big)
\Big(y_0 - {\i}(1-e^{-{\i}\phi_2}z)\Big) = \nn \\ =
1+y_0^2 - 2(1-ze^{-\i\theta}\cos\varphi)(1+\i y_0)
- \left(ze^{-\i\theta}\right)^2 =
P_2 + z\bar z - 2(1-ze^{-\i\theta}\cos\varphi)(1+\i y_0)
- \left(ze^{-\i\theta}\right)^2
\ee
We can again subtract $P_2$ in order to obtain an
$y_0$-linear element of ${\cal R}^{++}_\angle$:
\be
%\tilde P\{\phi_2^+,\phi_1^+\} =
{\cal C}\{\phi_2^+,\phi_1^+\}-P_2 =
z\bar z  - 2(1-ze^{-\i\theta}\cos\varphi)(1+\i y_0)
- \left(ze^{-\i\theta}\right)^2
\label{Qangle}
\ee
However, the coefficient of $y_0$ term is now not a
common factor of entire expression and can not be eliminated.
A linear element exists in ${\cal R}^{+-}_\angle$ but not in
${\cal R}^{++}_\angle$.

This last part of this conclusion has a remarkable exception:
$\cos\varphi = 0$, i.e. $\varphi=\frac{\pi}{2}$.

\subsubsection{Two parallel lines. The case of $\sigma_2=-\sigma_1$
\label{Lparpm}}

In many non-generic examples, like $Z_n$-symmetric configurations
of \cite{malda3} or $z_2\times Z_2$-symmetric rhombus of
\cite{am1} the possible building block is a pair of parallel
lines, which is a particular choice of our angle with $2\alpha=0$.
Moreover, both cases $\sigma_2=\pm\sigma_1$ are needed for
this kind of application, even if we are interested in
$n$-angle polygons with even $n$ and alternated $\sigma_a=(-)^{a-1}$:
for $n=4k-2$ the parallel sides will have opposite $\sigma$'s,
while for $n=4k$ their $\sigma$'s will be the same.

Substituting $\varphi = \frac{\pi}{2}$,
i.e. $\theta = \frac{\pi}{2}+\phi$ into (\ref{Langle}),
we obtain:
\be
{\cal L}_{||}^{-+} = y_0 - {\rm Re}(ze^{-i\theta}) =
y_0 - {\rm Im}(ze^{-i\phi}) =
y_0 - y_1\cos\theta - y_2\sin\theta = y_0 +y_1\sin\phi -y_2\cos\phi
\label{Lparmp}
\ee
and
\be
{\cal L}_{||}^{+-} = y_0 + {\rm Re}(ze^{-i\theta}) =
y_0 + {\rm Im}(ze^{-i\phi}) =
y_0 + y_1\cos\theta + y_2\sin\theta = y_0 -y_1\sin\phi +y_2\cos\phi
\label{Lpar}
\ee
and ${\cal L}_{||}^{-\sigma\sigma}=0$ implies that
\be
y_0 = \sigma(-sy_1+cy_2)
\ee

Comparing (\ref{Lsegm}) and (\ref{Lpar}), we can observe that
\be
{\cal L}_{||}^{^{-\sigma\sigma}} = {\cal L}_|^\sigma
\ee
This is manifestation of the fact, which we already
mentioned in the end of s.\ref{1segsec}.
Now we can formulate it in a better way:
it turns out that ${\cal L}_|$ is not just an element of
the boundary ring ${\cal R}_|$, it actually lies in its
sub-ring:
\be
{\cal L}_| \in {\cal R}_{||} \subset {\cal R}_|
\ee
Whenever the boundary $\Pi = \Pi_1\cup \Pi_2$ is decomposed
into two components, we have
\be
{\cal R}_{\Pi_1\cup \Pi_2} \subset {\cal R}_{\Pi_1}, \ \ \
{\cal R}_{\Pi_1\cup \Pi_2} \subset {\cal R}_{\Pi_2}
\ee
and all the elements of a polygon boundary ring naturally belong
to the bigger boundary rings of its particular segments,
angles, triangles etc.
What we encountered, however, is a kind of an opposite phenomenon:
in our attempt to build up representation of a given boundary
ring, namely ${\cal R}_|$ we actually obtained elements of
its sub-ring ${\cal R}_{||}$ instead of elements in generic position!
We shall encounter more examples of this kind below, and one
should always be careful to check what the actual nature of emerging
elements is.

\subsubsection{Two parallel lines. The case of $\sigma_2=\sigma_1$
\label{Lparpp}}

As mentioned at the very end of s.\ref{anglepp},
two parallel lines provide a practically important exception
from the rule that there are no $y_0$-linear elements in
${\cal R}_\angle^{++}$.
This exception, however, has a number of non-trivial properties.
At $\varphi=\frac{\pi}{2}$ and $\theta=\phi+\frac{\pi}{2}$
eq.(\ref{Qangle}) gives:
\be
{\cal C}^{++}_{||}-P_2 = z\bar z -2(1+\i y_0) +z^2e^{-2\i\phi}
\ee
The real and imaginary parts of this complex expression are:
\be
{\rm Re}\Big({\cal C}^{++}_{||}\Big)-P_2 =
z\bar z -2 + (y_1^2-y_2^2)\cos(2\phi) + 2y_1y_2\sin(2\phi)=\nn\\
= -2\Big(1-(y_1\cos\phi + y_2\sin\phi)^2\Big)
\ \stackrel{(\ref{Ppolsdef})}{=}\ -2P_{||}(y_1,y_2)
\ee
and
\be
{\rm Im}\Big({\cal C}^{++}_{||}\Big) = -2y_0
- (y_1^2-y_2^2)\sin(2\phi) + 2y_1y_2\cos(2\phi) \equiv
-2{\cal L}_{||}^{++}
\ee
For $\sigma_2=\sigma_1= -1$ the answer will differ by sign
in front of $y_0$, and we obtain the linear element in
${\cal R}_{||}^{\sigma\sigma}$ in the form:
\be
{\cal L}_{||}^{\sigma\sigma}(\phi) =
y_0  -\sigma \Big(y_1y_2\cos(2\phi) +
\frac{1}{2}(y_2^2-y_1^2)\sin(2\phi)\Big)
= y_0-\sigma y_1^\phi y_2^\phi,
\label{Lparapp}
\ee
where
\be
y_1^\phi = y_1\cos\phi - y_2\sin\phi, \nn \\
y_2^\phi = y_1\sin\phi + y_2\cos\phi
\ee
are rotated coordinates $y_1$ and $y_2$.
In particular, for two vertical lines ($\phi=0$) we obtain:
\be
{\cal L}_{||}^{++} =y_0- y_1y_2
\ee
It is now obvious that what we obtained is not just an element
from ${\cal R}_{||}$ -- it actually belongs to its sub-ring
${\cal R}_\Box$: vanishes on {\it four} sides of the unit square,
not only on the two vertical lines, which formed our $\Pi$:
\be
{\cal L}_{||}^{++} \in {\cal R}_\Box
\subset {\cal R}_{||}^{++}
\ee

Worse than that, in this case one can not find any element of the
boundary ring ${\cal R}_{||}^{++}$
which could be used as a complement of $P_2$ in
adequate description of the boundary ring: such description
is available only without $P_2$, for example ${\cal C}_{||}^{++}$
in this case is a pair $\{1-y_1^2,\ y_0-y_1y_2\}$.
This in turn means that our approach to NG solutions would not
work in this case: and indeed two parallel lines with
coincident $\sigma$'s form an impossible $\Pi$, such {\it diangle}
formed by two null lines is simply non-existing
(while a similar diangle with $\sigma_2=-\sigma_1$ does exist,
and is an $n=2$ version of the $Z_n$-symmetric configurations
of \cite{malda3} with (\ref{Lpar}) providing (together with the
usual $r^2=P_2$) an exact solution to NG equations.

\subsection{Pairs of parallel lines: from square to hexagons
\label{parahexa}}

The boundary rings for a square and, more generally,
for a rhombus can be constructed from already available
building blocks in two ways: by combining two pairs of
parallel lines and by combining two non-adjacent angles.
Only the second one of these options is available for
kite and for generic skew quadrilateral, but it is a little
more complicated and we begin from analysis of the first one.

\subsubsection{Square}

We know already that ${\cal L}^{++}_{||}$
occasionally belongs to ${\cal R}_\Box$
and we do not need to do any more calculations.
However, we know this because the situation is very simple
and all answers are immediately clear "from the first look".
But what we need, is a kind of a systematic approach to
construction of boundary rings, not relying upon accidental
observations.
Therefore we proceed regularly in this trivial example and
use it to illustrate the general procedure.
This procedure implies that we take $y_0$-linear elements,
associated with our building blocks, multiply them and
try to make them $y_0$-linear again by subtracting
the always-available
polynomials $P_2$ and $P$, $\tilde P$, $\widetilde{\tilde P}$
from (\ref{Ppolsdef}).
If we are building the square from two pairs of parallel lines,
this means that write:
\be
{\cal L}^{--}_{||}\left(\frac{\pi}{2}\right)
{\cal L}^{++}_{||}\left(0\right)
\ \stackrel{(\ref{Lparapp})}{=}\
\left(y_0+(-y_1y_2)\right)(y_0-y_1y_2) = (y_0 - y_1y_2)^2
y_0^2 -2y_0y_1y_2 + y_1^2y_2^2
\ee
Next we subtract $P_2$ to eliminate the term $y_0^2$:
\be
{\cal L}^{--}_{||}\left(\frac{\pi}{2}\right)
{\cal L}^{++}_{||}\left(0\right) - P_2 =
-2y_0y_1y_2 + y_1^2y_2^2 - 1 + y_1^2 +y_2^2
\label{squareint1}
\ee
This element does not deserve the name of ${\cal L}_\Box$,
because the coefficient in front of $y_0$ is not constant.
This coefficient does not belong to ${\cal R}_\Box$ thus
in principle we could eliminate it.
Unfortunately, it is not a common factor in front of
entire expression, so we can not simply get rid of it.
What we can do, however, is to make use of
\be
P_\Box \ \stackrel{(\ref{Ppolsdef})}{=}\
(1-y_1^2)(1-y_2^2)
\ee
which is an "obvious" element of ${\cal R}_\Box$.
Adding it to (\ref{squareint1}) we obtain:
\be
{\cal L}^{--}_{||}\left(\frac{\pi}{2}\right)
{\cal L}^{++}_{||}\left(0\right) - P_2 + P_\Box =
-2y_1y_2y_0 + 2y_1^2y_2^2 =
-2y_1y_2(y_0 - y_1y_2) = -2y_1y_2{\cal L}_\Box
\ee
Now the coefficient of $y_0$ is a common factor and can
be thrown away to give
\be
{\cal L}_\Box = y_0 - y_1y_2
\ee

\subsubsection{Rhombus}

Above procedure is immediately generalized to the case of rhombus:
$$
{\cal L}^{--}_{||}\left(\frac{\pi}{4}+\varphi\right)
{\cal L}^{++}_{||}\left(\frac{\pi}{4}-\varphi\right)
\ \stackrel{(\ref{Lparapp})}{=}
$$ $$
= \Big(y_0  + y_1y_2\cos\left(\frac{\pi}{2} + 2\varphi\right) +
\frac{1}{2}(y_2^2-y_1^2)\sin\left(\frac{\pi}{2} + 2\varphi\right)
\Big)
\Big(y_0  - y_1y_2\cos\left(\frac{\pi}{2} - 2\varphi\right) -
\frac{1}{2}(y_2^2-y_1^2)\sin\left(\frac{\pi}{2} - 2\varphi\right)
\Big) =
$$ $$
= \Big(y_0  - y_1y_2\sin\left(2\varphi\right) +
\frac{1}{2}(y_2^2-y_1^2)\cos\left(2\varphi\right)\Big)
\Big(y_0  - y_1y_2\sin\left(2\varphi\right) -
\frac{1}{2}(y_2^2-y_1^2)\cos\left(2\varphi\right)
\Big) = \nn
$$
\vspace{-0.4cm}
\be
= \Big(y_0  - y_1y_2\sin\left(2\varphi\right)\Big)^2 -
\frac{1}{4}(y_2^2-y_1^2)^2\cos^2\left(2\varphi\right)
\ee
In the case o square $\varphi = \frac{\pi}{4}$
and $2\varphi=\frac{\pi}{2}$.
Subtraction of $P_2$ converts this expression into
\be
{\cal L}^{--}_{||}\left(\frac{\pi}{4}+\varphi\right)
{\cal L}^{++}_{||}\left(\frac{\pi}{4}-\varphi\right) - P_2
= -2y_0y_1y_2\sin(2\varphi) +
y_1^2y_2^2\sin^2(2\varphi) +y_1^2+y_2^2-1
-\frac{1}{4}(y_2^2-y_1^2)^2\cos^2\left(2\varphi\right)
\nn
\ee
Now we need to get rid of the terms that are not divisible by
$y_1y_2$, and again we have $P_\diamond$ to try to achieve this.
Substituting $\phi_1 = \frac{\pi}{4}-\varphi$,
$\phi_2 = \frac{\pi}{4}+\varphi$, $\phi_3=\phi_1+\pi$
and $\phi_4 = \phi_2 + \pi$ into the
first line of (\ref{Ppolsdef}), we obtain:
$$
P_\diamond = \Big(1 - (y_1\cos\phi_1 + y_2\sin\phi_1)^2\Big)
\Big(1 - (y_1\cos\phi_2 + y_2\sin\phi_2)^2\Big) =
$$ $$
= \left(1 - \frac{1}{2}\Big(y_1(c+s)+y_2(c-s)\Big)^2\right)
\left(1 - \frac{1}{2}\Big(y_1(c-s)+y_2(c+s)\Big)^2\right) =
$$ $$
= \left(1 - \frac{1}{2}\Big(y_1^2+y_2^2+2y_1y_2\cos(2\varphi)
+ (y_1^2-y_2^2)\sin(2\varphi)\Big)\right)
\left(1 - \frac{1}{2}\Big(y_1^2+y_2^2+2y_1y_2\cos(2\varphi)
- (y_1^2-y_2^2)\sin(2\varphi)\Big)\right) =
$$
\vspace{-0.4cm}
\be
= 1 -y_1^2-y_2^2 - 2y_1y_2\cos(2\varphi) +
\frac{1}{4}\left(\Big(y_1^2+y_2^2+2y_1y_2\cos(2\varphi)\Big)^2
- (y_1^2-y_2^2)^2\sin^2(2\varphi)\right)
\ee
At intermediate stage we denoted $c=\cos\varphi$ and $s=\sin\varphi$.
Now we are ready to combine:
$$
{\cal L}^{--}_{||}\left(\frac{\pi}{4}+\varphi\right)
{\cal L}^{++}_{||}\left(\frac{\pi}{4}-\varphi\right) - P_2
+ P_\diamond =
$$ $$
= -2y_0y_1y_2\sin(2\varphi) - 2y_1y_2\cos(2\varphi) +
y_1^2y_2^2\sin^2(2\varphi)
+\frac{1}{4}\left(\Big(y_1^2+y_2^2+2y_1y_2\cos(2\varphi)\Big)^2
- (y_1^2-y_2^2)^2\right) =
$$ $$
= -2y_1y_2\left(y_0\sin(2\varphi) + \cos(2\varphi)
- \frac{1}{2}y_1y_2\sin^2(2\varphi)
-\frac{1}{2}\Big(y_1+y_2\cos(2\varphi)\Big)
\Big(y_2+y_1\cos(2\varphi)\Big)\right)
= $$
\vspace{-0.3cm}
\be
= -2y_1y_2\left(y_0\sin(2\varphi) +
\cos(2\varphi)\left(1-\frac{1}{2}(y_1^2+y_2^2)\right) - y_1y_2\right)
= -2y_1y_2\sin(2\varphi){\cal L}_\diamond
\ee
All terms, which were not divisible by $y_1y_2$,
canceled and we finally obtain:
\be
{\cal L}_\diamond =
y_0  - \frac{1}{\sin(2\varphi)}y_1y_2 +
\frac{\cos(2\varphi)}{\sin(2\varphi)}
\left(1-\frac{1}{2}(y_1^2+y_2^2)\right) =
y_0 - y_1y_2\cosh\xi
+ \left(1-\frac{1}{2}(y_1^2+y_2^2)\right)\sinh\xi
\label{Lromb}
\ee
where a new parameter $\xi$ introduced, related to $\varphi$ by
\be
\cosh\xi = \frac{1}{\sin(2\varphi)} = \frac{1}{\cos(2\phi_1)},\ \ \
\sinh\xi = \frac{\cos(2\varphi)}{\sin(2\varphi)} = \tan(2\phi_1)
\ee

Thus we {\it derived} an expression for ${\cal L}_\diamond$.
It is {\it canonical} in the sense that this is the only element
of ${\cal R}_\diamond$, which is linear in $y_0$ and satisfies
(\ref{linearity1}).
Moreover, it has degree $2 = \frac{n}{2}$ in $y$-variables!
Any other element of degree $2$ in ${\cal R}_\diamond$
can be obtained by adding $P_2$ with some constant coefficient.
It is within this $1$-parametric family
\be
{\cal L}_\diamond + \mu P_2 = 0
\ee
that we expect to find the solution to Plateau problem
(since we know from section \ref{quadrila} that for $n=4$ the
solution is quadratic in $y$):
\be
{\cal S}_\diamond \sim {\cal L}_\diamond + \mu_\diamond P_2
\ee
the value $\mu_\diamond$ can not be found
by the study of the boundary ring alone:
it is defined either by NG equations or by minimization
of regularized action w.r.t. to $\mu$-variable.
Since we actually know what ${\cal S}_\diamond$ is,
we can use this answer, eq.(\ref{rhomeq}),
\be
{\cal S}_\diamond \ \stackrel{(\ref{rhomeq})}{\sim}
y_0 - y_1y_2\cosh\xi + \frac{1}{2}(1-y_0^2)\sinh\xi,
\ee
to get:
\be
\mu_\diamond = -\frac{1}{2}\sinh\xi
\label{mudia}
\ee

\subsubsection{A two-parametric family of hexagons}

If instead of two pairs of parallel lines we consider three,
what we obtain will be a hexagon.
It will be not a generic hexagon with inscribed circle,\footnote{
If conditions (\ref{ads3}) of $AdS_3$-embedding are not imposed,
hexagons form a $3n-8=10$-parametric family:
$3n$ coordinates $(y_1,y_2,y_3)$ of $n=6$ vertices
minus $3$ parallel transports, minus $3$ rotations,
minus one rescaling and minus
one constraint $\sum_a \sigma_a l_a = 0$ which guarantees that
$\Pi$ formed from null-segments closes in $y_0$ direction.
If only space-flatness condition $y_3=0$ is imposed, the space
of relevant hexagons reduces to $2n-5=7$ dimensions.
Inscribed-circle condition (it makes sense only if $y_3=0$)
imposes $n-4$ extra constraints
and brings the dimension down to $n-1=5$:
$n$ angles $\phi_a$ minus one common rotation.
}
which form a family with $n-1=5$ parameters
(),
but a $2$-parametric sub-family, which, however,
contains the $Z_6$-symmetric hexagon, considered in
\cite{malda3}.

We assume that the first (and thus also the forth) side
of the hexagon is parallel to the $y_2$-axis, $\phi_1=0$,
$\phi_4=\pi$ -- this fixes rotation freedom.
Remaining two parameters are $\phi_2 = \phi$ and
$\phi_3=\pi-\phi'$.
We denote their sines and cosines by
$c=\cos\phi=\cos\phi_2$, $s=\sin\phi= \sin\phi_2$,
$c'=\cos\phi' = -\cos\phi_3$, $s'=\sin\phi' = \sin\phi_3$.
This time we should use ${\cal L}_{||}^{-+}$ and ${\cal L}_{||}^{+-}$
rather than ${\cal L}_{||}^{++}$    as the building
blocks.
$$
{\cal L}_{||}^{-+}(\phi_3){\cal L}_{||}^{+-}(\phi_2)
{\cal L}_{||}^{-+}(\phi_1)
\ \stackrel{(\ref{Lparmp})\&(\ref{Lpar})}{=}\
(y_0+s'y_1+c'y_2)(-y_0+sy_1-cy_2)(y_0-y_2) =
$$ $$
-y_0^3+ y_0^2\Big((s-s')y_1+(1-c-c')y_2\Big) +
$$
\vspace{-0.3cm}
\be
+ y_0\Big(ss'y_1^2+\big(\sin(\phi-\phi')+(s'-s)\big)y_1y_2
+(c+c'-cc')y_2^2\Big)
+\Big(-ss'y_1^2y_2-\sin(\phi-\phi')y_1y_2^2+cc'y_2^3\Big)
\ee
This time we get an element of ${\cal R}_{hexa}$, which is
cubic in $y$-variables, in particular it is cubic in $y_0$
In order to obtain an $y_0$-linear expression we need to
subtract $P_2$, multiplied by a polynomial which is not
just a constant, but contains also a first power of $y_0$.
Note, however, that since we are multiplying
${\cal L}_{||}^{-+}$ instead of ${\cal L}_{||}^{++}$,
the product has power $n/2=3$ in {\it all} of the $y$-variables,
and thus we can not make use of polynomials
(\ref{Ppolsdef}) in order to further simplify it:
all these polynomials are of degree $n=6>3$.
\be
{\cal L}_{hexa} =
{\cal L}_{||}^{-+}(\phi_3){\cal L}_{||}^{+-}(\phi_2)
{\cal L}_{||}^{-+}(\phi_1) + \Big(y_0 +(s-s')y_1+(c+c'-1)y_2\Big)P_2
= \nn \\
= y_0\Big(1-(1-ss')y_1^2+\big(\sin(\phi-\phi')-(s-s')\big)y_1y_2
-(1-c)(1-c')y_2^2\Big) +
\label{hexacc'}
\ee
\vspace{-0.3cm}
$$
+\Big((s-s')y_1^3 +(1-c-c'-ss')y_1^2y_2
+ \big(s-s'-\sin(\phi-\phi')\big)y_1y_2^2 +
(1-c)(1-c')y_2^2 %+ \nn \\
+(s'-s)y_1+(c+c'-1)y_2\Big)
$$
Note that this time ${\cal L}_{hexa}$ is linear only in $y_0$,
it satisfies (\ref{linearity1}),
but the coefficient in front of $y_0$ is non-trivial function
of $y_1$ and $y_2$ which can {\it not} be eliminated.

This expression is considerably simplified if we restrict to a
$Z_2\times Z_2$-symmetric one-parametric family of hexagons
with $\phi'=\phi$. Then
\be
{\cal L}_{hexa} = y_0\Big(1-c^2y_1^2-(1-c)^2y_2^2\Big)
+ y_2\Big(-c(2-c)y_1^2 + (1-c)^2y_2^2 +(2c-1)\Big)
\label{hexac}
\ee
In the case of $Z_6$-symmetry, when $\phi'=\phi=\frac{\pi}{3}$
and $c=\frac{1}{2}$,
it further simplifies to
\be
{\cal L}_{hexa} = y_0\left(1-\frac{1}{4}(y_1^2+y_2^2)\right)
-\frac{1}{4}y_2(3y_1^2-y_2^2)
\ee
This expression is familiar from \cite{malda3}, and now
we derived it applying a systematical, constructive and
generalizable method.

For hexagons the full family of $y$-cubic ($n/2=3$) elements in
${\cal R}_{hexa}$ is $4$-parametric:
\be
\Big\{ {\cal L}_{hexa} + (\vmu\vy) P_2 + \nu P_2\Big\}
\ee
We know from \cite{malda3} that exact solution to Plateau
problem does {\it not} lie entirely in this space,
but
\be
\vmu_{hexa} = 0, \ \ \  \nu_{hexa} = 0, \ \ \
{\cal S}_{hexa} \approx {\cal L}_{hexa}
+ (\vmu\vy)P_2 + \nu P_2
\ee
provides a nice first approximation, which can be further
improved by methods of s.\ref{comprec} -- with $\mu$ promoted
to a power series.

\subsection{Combining angles
\label{comban}}

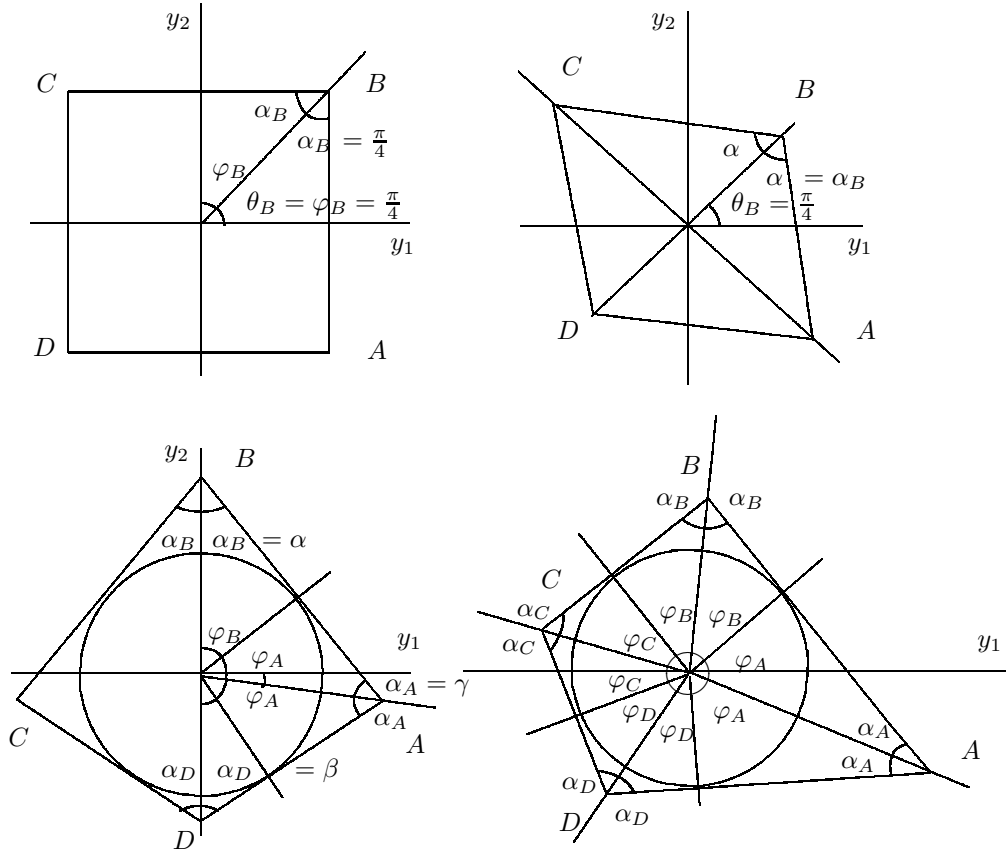
\begin{figure}
\begin{center}
%%
%%
%%

%TeXCAD Picture [polygs3.tex]. Options:
%\grade{\on}
%\emlines{\off}
%\epic{\off}
%\beziermacro{\on}
%\reduce{\on}
%\snapping{\off}
%\pvinsert{% Your \input, \def, etc. here}
%\quality{8.000}
%\graddiff{0.005}
%\snapasp{1}
%\zoom{5.6569}
\unitlength 1mm % = 2.845pt
\linethickness{0.4pt}
\ifx\plotpoint\undefined\newsavebox{\plotpoint}\fi
% GNUPLOT compatibility
\begin{picture}(136.48,122.117)(0,0)
%\circle(94.495,23.342){31.427}
\put(110.209,23.342){\line(0,1){.7745}}
\put(110.19,24.117){\line(0,1){.7726}}
\put(110.132,24.89){\line(0,1){.7689}}
\multiput(110.037,25.658)(-.03326,.19081){4}
{\line(0,1){.19081}}
\multiput(109.904,26.422)(-.028416,.125957){6}
{\line(0,1){.125957}}
\multiput(109.734,27.177)(-.029648,.106631){7}
{\line(0,1){.106631}}
\multiput(109.526,27.924)(-.030509,.09191){8}
{\line(0,1){.09191}}
\multiput(109.282,28.659)(-.031113,.080262){9}
{\line(0,1){.080262}}
\multiput(109.002,29.381)(-.031528,.070768){10}
{\line(0,1){.070768}}
\multiput(108.687,30.089)(-.031798,.062844){11}
{\line(0,1){.062844}}
\multiput(108.337,30.78)(-.031952,.0561){12}
{\line(0,1){.0561}}
\multiput(107.953,31.454)(-.0320107,.050268){13}
{\line(0,1){.050268}}
\multiput(107.537,32.107)(-.0319888,.0451556){14}
{\line(0,1){.0451556}}
\multiput(107.09,32.739)(-.0318971,.0406225){15}
{\line(0,1){.0406225}}
\multiput(106.611,33.349)(-.0317443,.0365634){16}
{\line(0,1){.0365634}}
\multiput(106.103,33.934)(-.0335078,.0349543){16}
{\line(0,1){.0349543}}
\multiput(105.567,34.493)(-.0351899,.0332603){16}
{\line(-1,0){.0351899}}
\multiput(105.004,35.025)(-.0392389,.0335845){15}
{\line(-1,0){.0392389}}
\multiput(104.415,35.529)(-.0408466,.0316097){15}
{\line(-1,0){.0408466}}
\multiput(103.803,36.003)(-.0453802,.0316693){14}
{\line(-1,0){.0453802}}
\multiput(103.167,36.446)(-.0504926,.0316552){13}
{\line(-1,0){.0504926}}
\multiput(102.511,36.858)(-.056324,.031555){12}
{\line(-1,0){.056324}}
\multiput(101.835,37.237)(-.063066,.031354){11}
{\line(-1,0){.063066}}
\multiput(101.141,37.581)(-.070989,.031028){10}
{\line(-1,0){.070989}}
\multiput(100.431,37.892)(-.08048,.030546){9}
{\line(-1,0){.08048}}
\multiput(99.707,38.167)(-.092123,.02986){8}
{\line(-1,0){.092123}}
\multiput(98.97,38.406)(-.124644,.033711){6}
{\line(-1,0){.124644}}
\multiput(98.222,38.608)(-.151385,.033032){5}
{\line(-1,0){.151385}}
\multiput(97.465,38.773)(-.19104,.03191){4}
{\line(-1,0){.19104}}
\put(96.701,38.901){\line(-1,0){.7695}}
\put(95.932,38.99){\line(-1,0){.773}}
\put(95.159,39.042){\line(-1,0){.7746}}
\put(94.384,39.056){\line(-1,0){.7743}}
\put(93.61,39.031){\line(-1,0){.7722}}
\put(92.838,38.969){\line(-1,0){.7682}}
\multiput(92.069,38.868)(-.152454,-.027684){5}
{\line(-1,0){.152454}}
\multiput(91.307,38.729)(-.125753,-.029304){6}
{\line(-1,0){.125753}}
\multiput(90.553,38.554)(-.106419,-.0304){7}
{\line(-1,0){.106419}}
\multiput(89.808,38.341)(-.091693,-.031157){8}
{\line(-1,0){.091693}}
\multiput(89.074,38.092)(-.080041,-.031679){9}
{\line(-1,0){.080041}}
\multiput(88.354,37.806)(-.070544,-.032027){10}
{\line(-1,0){.070544}}
\multiput(87.648,37.486)(-.062618,-.032241){11}
{\line(-1,0){.062618}}
\multiput(86.96,37.132)(-.055873,-.032347){12}
{\line(-1,0){.055873}}
\multiput(86.289,36.743)(-.0500408,-.0323647){13}
{\line(-1,0){.0500408}}
\multiput(85.639,36.323)(-.0449287,-.0323066){14}
{\line(-1,0){.0449287}}
\multiput(85.01,35.87)(-.0403963,-.032183){15}
{\line(-1,0){.0403963}}
\multiput(84.404,35.388)(-.0363385,-.0320015){16}
{\line(-1,0){.0363385}}
\multiput(83.822,34.876)(-.0326748,-.0317682){17}
{\line(-1,0){.0326748}}
\multiput(83.267,34.336)(-.0330112,-.0354238){16}
{\line(0,-1){.0354238}}
\multiput(82.739,33.769)(-.0333067,-.039475){15}
{\line(0,-1){.039475}}
\multiput(82.239,33.177)(-.0335578,-.0440021){14}
{\line(0,-1){.0440021}}
\multiput(81.769,32.561)(-.0313482,-.0456026){14}
{\line(0,-1){.0456026}}
\multiput(81.33,31.922)(-.0312981,-.0507147){13}
{\line(0,-1){.0507147}}
\multiput(80.923,31.263)(-.031157,-.056545){12}
{\line(0,-1){.056545}}
\multiput(80.549,30.584)(-.030908,-.063286){11}
{\line(0,-1){.063286}}
\multiput(80.209,29.888)(-.030526,-.071206){10}
{\line(0,-1){.071206}}
\multiput(79.904,29.176)(-.033724,-.09078){8}
{\line(0,-1){.09078}}
\multiput(79.634,28.45)(-.033382,-.105522){7}
{\line(0,-1){.105522}}
\multiput(79.401,27.711)(-.03283,-.124879){6}
{\line(0,-1){.124879}}
\multiput(79.204,26.962)(-.031962,-.151614){5}
{\line(0,-1){.151614}}
\multiput(79.044,26.204)(-.03056,-.19126){4}
{\line(0,-1){.19126}}
\put(78.922,25.439){\line(0,-1){.7701}}
\put(78.837,24.669){\line(0,-1){1.548}}
\put(78.783,23.121){\line(0,-1){.7742}}
\put(78.813,22.347){\line(0,-1){.7717}}
\put(78.881,21.575){\line(0,-1){.7674}}
\multiput(78.987,20.807)(.02876,-.152254){5}{\line(0,-1){.152254}}
\multiput(79.131,20.046)(.030191,-.125543){6}{\line(0,-1){.125543}}
\multiput(79.312,19.293)(.03115,-.106202){7}{\line(0,-1){.106202}}
\multiput(79.53,18.549)(.031803,-.091471){8}{\line(0,-1){.091471}}
\multiput(79.785,17.818)(.032243,-.079815){9}{\line(0,-1){.079815}}
\multiput(80.075,17.099)(.032524,-.070316){10}{\line(0,-1){.070316}}
\multiput(80.4,16.396)(.032682,-.062389){11}{\line(0,-1){.062389}}
\multiput(80.759,15.71)(.032741,-.055643){12}{\line(0,-1){.055643}}
\multiput(81.152,15.042)(.032717,-.0498111){13}
{\line(0,-1){.0498111}}
\multiput(81.578,14.395)(.0326229,-.0446996){14}
{\line(0,-1){.0446996}}
\multiput(82.034,13.769)(.0324673,-.0401682){15}
{\line(0,-1){.0401682}}
\multiput(82.521,13.166)(.0322572,-.0361117){16}
{\line(0,-1){.0361117}}
\multiput(83.037,12.589)(.031998,-.0324498){17}
{\line(0,-1){.0324498}}
\multiput(83.581,12.037)(.0356559,-.0327603){16}
{\line(1,0){.0356559}}
\multiput(84.152,11.513)(.039709,-.0330273){15}
{\line(1,0){.039709}}
\multiput(84.748,11.017)(.0442378,-.0332464){14}
{\line(1,0){.0442378}}
\multiput(85.367,10.552)(.0493475,-.0334122){13}
{\line(1,0){.0493475}}
\multiput(86.008,10.118)(.055179,-.033518){12}
{\line(1,0){.055179}}
\multiput(86.671,9.715)(.061924,-.033553){11}
{\line(1,0){.061924}}
\multiput(87.352,9.346)(.069853,-.033507){10}
{\line(1,0){.069853}}
\multiput(88.05,9.011)(.079355,-.033359){9}{\line(1,0){.079355}}
\multiput(88.764,8.711)(.091016,-.033083){8}{\line(1,0){.091016}}
\multiput(89.493,8.446)(.105755,-.032636){7}{\line(1,0){.105755}}
\multiput(90.233,8.218)(.125108,-.031948){6}{\line(1,0){.125108}}
\multiput(90.984,8.026)(.151836,-.030892){5}{\line(1,0){.151836}}
\put(91.743,7.872){\line(1,0){.7659}}
\put(92.509,7.755){\line(1,0){.7707}}
\put(93.279,7.676){\line(1,0){.7737}}
\put(94.053,7.635){\line(1,0){.7747}}
\put(94.828,7.632){\line(1,0){.7739}}
\put(95.602,7.668){\line(1,0){.7712}}
\put(96.373,7.741){\line(1,0){.7667}}
\multiput(97.139,7.853)(.152048,.029833){5}{\line(1,0){.152048}}
\multiput(97.9,8.002)(.125327,.031076){6}{\line(1,0){.125327}}
\multiput(98.652,8.188)(.10598,.031899){7}{\line(1,0){.10598}}
\multiput(99.394,8.412)(.091244,.032448){8}{\line(1,0){.091244}}
\multiput(100.123,8.671)(.079585,.032805){9}{\line(1,0){.079585}}
\multiput(100.84,8.967)(.070085,.033019){10}{\line(1,0){.070085}}
\multiput(101.541,9.297)(.062156,.033121){11}{\line(1,0){.062156}}
\multiput(102.224,9.661)(.055411,.033133){12}{\line(1,0){.055411}}
\multiput(102.889,10.059)(.049579,.0330677){13}{\line(1,0){.049579}}
\multiput(103.534,10.489)(.0444683,.0329375){14}{\line(1,0){.0444683}}
\multiput(104.156,10.95)(.0399381,.03275){15}{\line(1,0){.0399381}}
\multiput(104.755,11.441)(.0358832,.0325112){16}{\line(1,0){.0358832}}
\multiput(105.33,11.961)(.0322232,.0322262){17}{\line(0,1){.0322262}}
\multiput(105.877,12.509)(.0325079,.0358862){16}{\line(0,1){.0358862}}
\multiput(106.397,13.083)(.0327463,.0399411){15}{\line(0,1){.0399411}}
\multiput(106.889,13.682)(.0329334,.0444713){14}{\line(0,1){.0444713}}
\multiput(107.35,14.305)(.0330631,.0495821){13}{\line(0,1){.0495821}}
\multiput(107.78,14.949)(.033127,.055414){12}{\line(0,1){.055414}}
\multiput(108.177,15.614)(.033116,.062159){11}{\line(0,1){.062159}}
\multiput(108.541,16.298)(.033013,.070088){10}{\line(0,1){.070088}}
\multiput(108.871,16.999)(.032798,.079588){9}{\line(0,1){.079588}}
\multiput(109.167,17.715)(.03244,.091247){8}{\line(0,1){.091247}}
\multiput(109.426,18.445)(.031889,.105983){7}{\line(0,1){.105983}}
\multiput(109.649,19.187)(.031064,.12533){6}{\line(0,1){.12533}}
\multiput(109.836,19.939)(.029819,.15205){5}{\line(0,1){.15205}}
\put(109.985,20.699){\line(0,1){.7667}}
\put(110.096,21.466){\line(0,1){.7712}}
\put(110.17,22.237){\line(0,1){1.1051}}
%\end
%\emline(121.452,15.387)(96.925,45.882)
\multiput(121.452,15.387)(-.03373796424,.04194635488){727}
{\line(0,1){.04194635488}}
%\end
%\emline(96.925,45.882)(74.827,28.425)
\multiput(96.925,45.882)(-.0426592664,-.0337017375){518}
{\line(-1,0){.0426592664}}
%\end
%\emline(74.827,28.425)(83.445,6.547)
\multiput(74.827,28.425)(.0336621094,-.0854589844){256}
{\line(0,-1){.0854589844}}
%\end
%\emline(83.445,6.547)(126.535,9.42)
\multiput(83.445,6.547)(.50104651,.03340116){86}
{\line(1,0){.50104651}}
%\end
%\emline(126.535,9.42)(121.452,15.387)
\multiput(126.535,9.42)(-.03365894,.039519868){151}
{\line(0,1){.039519868}}
%\end
%\emline(94.495,23.122)(98.03,56.93)
\multiput(94.495,23.122)(.033666667,.32197619){105}
{\line(0,1){.32197619}}
%\end
\put(64.442,22.9){\line(1,0){72.038}}
%\emline(94.495,22.68)(66.21,30.855)
\multiput(94.495,22.68)(-.116399177,.033641975){243}
{\line(-1,0){.116399177}}
%\end
%\emline(94.272,22.457)(79.027,.14)
\multiput(94.272,22.457)(-.0337278761,-.049375){452}
{\line(0,-1){.049375}}
%\end
%\emline(94.715,22.68)(131.617,7.432)
\multiput(94.715,22.68)(.0816426991,-.0337334071){452}
{\line(1,0){.0816426991}}
%\end
%\emline(94.495,22.68)(112.172,37.927)
\multiput(94.495,22.68)(.0391095133,.0337334071){452}
{\line(1,0){.0391095133}}
%\end
%\emline(94.272,22.9)(79.69,41.24)
\multiput(94.272,22.9)(-.0336778291,.0423556582){433}
{\line(0,1){.0423556582}}
%\end
%\emline(94.272,22.457)(72.84,14.502)
\multiput(94.272,22.457)(-.090815678,-.033707627){236}
{\line(-1,0){.090815678}}
%\end
%\emline(94.495,22.237)(95.82,5.002)
\multiput(94.495,22.237)(.033125,-.430875){40}
{\line(0,-1){.430875}}
%\end
\put(94.272,22.68){\circle{5.812}}
\qbezier(93.61,43.007)(96.482,41.02)(99.355,42.567)
\qbezier(121.232,9.2)(120.68,12.515)(123.22,13.177)
\qbezier(75.932,25.552)(77.7,28.092)(77.7,30.192)
\qbezier(82.34,9.642)(85.767,9.2)(86.982,6.99)
\put(132.722,26.1){$y_1$}
\put(130.512,11.19){$A$}
\put(93.17,49.207){$B$}
\put(74.722,34){$C$}
\put(77.11,1.687){$D$}
\put(116.812,14.5){$\alpha_A$}
\put(114.602,10){$\alpha_A$}
\put(97.587,16.712){$\varphi_A$}
\put(100.46,23.462){$\varphi_A$}
\put(99.692,45){$\alpha_B$}
\put(90,45){$\alpha_B$}
\put(71.5,30.307){$\alpha_C$}
\put(69.585,25.667){$\alpha_C$}
\put(77.5,7.295){$\alpha_D$}
\put(84.5,3){$\alpha_D$}
\put(97,29.425){$\varphi_B$}
\put(90.5,30){$\varphi_B$}
\put(85.645,26.215){$\varphi_C$}
\put(83.655,20.795){$\varphi_C$}
\put(85.5,16.817){$\varphi_D$}
\put(90.5,14.167){$\varphi_D$}
\put(11.852,65.327){\line(0,1){34.692}}
\put(11.852,100.02){\line(1,0){34.693}}
\put(46.545,100.02){\line(0,-1){34.693}}
\put(46.545,65.327){\line(-1,0){34.693}}
\put(6.77,82.562){\line(1,0){50.383}}
\put(29.53,62.232){\line(0,1){49.498}}
%\emline(29.53,82.342)(51.405,105.322)
\multiput(29.53,82.342)(.0337057011,.0354083205){649}
{\line(0,1){.0354083205}}
%\end
\qbezier(29.53,85.215)(32.402,84.882)(32.622,82.342)
\qbezier(42.125,100.02)(43.007,95.6)(46.545,96.482)
\put(24.89,108.857){$y_2$}
\put(54.72,79){$y_1$}
\put(51.627,64.442){$A$}
\put(51.405,100.02){$B$}
\put(7.652,100.02){$C$}
\put(7.21,64.885){$D$}
\put(35.557,84.252){$\theta_B=\varphi_B=\frac{\pi}{4}$}
\put(36.6,96.705){$\alpha_B$}
\put(48.3,92.505){$=\frac{\pi}{4}$}
\put(42.3,92.505){$\alpha_B$}
\put(31,89){$\varphi_B$}
\put(71.955,82.12){\line(1,0){45.52}}
%\emline(14.502,122.117)(14.725,121.895)
\multiput(14.502,122.117)(.031786,-.031786){7}
{\line(1,0){.031786}}
%\end
\put(94.272,61.127){\line(0,1){49.497}}
%\emline(81.457,70.187)(108.415,95.82)
\multiput(81.457,70.187)(.03547039474,.03372697368){760}
{\line(1,0){.03547039474}}
%\end
%\emline(71.735,102.67)(114.382,64)
\multiput(71.735,102.67)(.03718177855,-.03371403662){1147}
{\line(1,0){.03718177855}}
%\end
%\emline(110.847,66.872)(106.87,94.052)
\multiput(110.847,66.872)(-.033707627,.230338983){118}
{\line(0,1){.230338983}}
%\end
%\emline(106.87,94.052)(76.375,98.252)
\multiput(106.87,94.052)(-.24396,.0336){125}
{\line(-1,0){.24396}}
%\end
%\emline(76.375,98.252)(81.677,70.41)
\multiput(76.375,98.252)(.033560127,-.176218354){158}
{\line(0,-1){.176218354}}
%\end
%\emline(81.677,70.41)(110.625,67.095)
\multiput(81.677,70.41)(.29239899,-.033484848){99}
{\line(1,0){.29239899}}
%\end
\qbezier(103.112,94.495)(103.665,90.96)(107.31,90.96)
\qbezier(97.147,84.772)(98.25,83.887)(98.472,82.12)
\put(115.487,79){$y_1$}
\put(89.412,108.857){$y_2$}
\put(116.592,67.095){$A$}
\put(108.415,99.135){$B$}
\put(77.48,102.23){$C$}
\put(76.817,67.095){$D$}
\put(98.915,91.402){$\alpha$}
\put(104.66,87.645){$\alpha$}
\put(109.66,87.645){$=\alpha_B$}
\put(100,84.11){$\theta_B = \frac{\pi}{4}$}
\put(29.53,52.51){\line(0,-1){51.485}}
\put(4.337,22.68){\line(1,0){53.032}}
%\circle(29.53,22.457){32.265}
\put(45.662,22.457){\line(0,1){.79}}
\put(45.643,23.247){\line(0,1){.7881}}
\put(45.585,24.036){\line(0,1){.7843}}
\multiput(45.489,24.82)(-.0337,.19465){4}{\line(0,1){.19465}}
\multiput(45.354,25.598)(-.028797,.128511){6}{\line(0,1){.128511}}
\multiput(45.181,26.369)(-.030047,.108812){7}{\line(0,1){.108812}}
\multiput(44.971,27.131)(-.030922,.093809){8}{\line(0,1){.093809}}
\multiput(44.723,27.882)(-.031536,.081939){9}{\line(0,1){.081939}}
\multiput(44.439,28.619)(-.031959,.072267){10}{\line(0,1){.072267}}
\multiput(44.12,29.342)(-.032236,.064196){11}{\line(0,1){.064196}}
\multiput(43.765,30.048)(-.032396,.057329){12}{\line(0,1){.057329}}
\multiput(43.377,30.736)(-.0324592,.0513912){13}{\line(0,1){.0513912}}
\multiput(42.955,31.404)(-.0324412,.0461872){14}{\line(0,1){.0461872}}
\multiput(42.5,32.051)(-.032353,.0415737){15}{\line(0,1){.0415737}}
\multiput(42.015,32.674)(-.032203,.0374434){16}{\line(0,1){.0374434}}
\multiput(41.5,33.273)(-.031998,.0337145){17}{\line(0,1){.0337145}}
\multiput(40.956,33.846)(-.0336105,.0321072){17}{\line(-1,0){.0336105}}
\multiput(40.385,34.392)(-.0373387,.0323244){16}{\line(-1,0){.0373387}}
\multiput(39.787,34.909)(-.0414685,.0324877){15}{\line(-1,0){.0414685}}
\multiput(39.165,35.397)(-.0460817,.0325909){14}{\line(-1,0){.0460817}}
\multiput(38.52,35.853)(-.0512855,.0326258){13}{\line(-1,0){.0512855}}
\multiput(37.853,36.277)(-.057223,.032582){12}{\line(-1,0){.057223}}
\multiput(37.167,36.668)(-.064091,.032444){11}{\line(-1,0){.064091}}
\multiput(36.462,37.025)(-.072163,.032194){10}{\line(-1,0){.072163}}
\multiput(35.74,37.347)(-.081837,.031802){9}{\line(-1,0){.081837}}
\multiput(35.003,37.633)(-.093708,.031226){8}{\line(-1,0){.093708}}
\multiput(34.254,37.883)(-.108713,.0304){7}{\line(-1,0){.108713}}
\multiput(33.493,38.096)(-.128417,.029213){6}{\line(-1,0){.128417}}
\multiput(32.722,38.271)(-.155632,.027468){5}{\line(-1,0){.155632}}
\put(31.944,38.408){\line(-1,0){.784}}
\put(31.16,38.507){\line(-1,0){.7879}}
\put(30.372,38.568){\line(-1,0){1.5799}}
\put(28.792,38.573){\line(-1,0){.7882}}
\put(28.004,38.518){\line(-1,0){.7846}}
\multiput(27.22,38.424)(-.19476,-.03307){4}{\line(-1,0){.19476}}
\multiput(26.44,38.291)(-.128604,-.028379){6}{\line(-1,0){.128604}}
\multiput(25.669,38.121)(-.108908,-.029694){7}{\line(-1,0){.108908}}
\multiput(24.907,37.913)(-.093908,-.030617){8}{\line(-1,0){.093908}}
\multiput(24.155,37.668)(-.082041,-.03127){9}{\line(-1,0){.082041}}
\multiput(23.417,37.387)(-.072371,-.031725){10}{\line(-1,0){.072371}}
\multiput(22.693,37.07)(-.0643,-.032028){11}{\line(-1,0){.0643}}
\multiput(21.986,36.717)(-.057434,-.03221){12}{\line(-1,0){.057434}}
\multiput(21.297,36.331)(-.0514962,-.0322922){13}{\line(-1,0){.0514962}}
\multiput(20.627,35.911)(-.0462923,-.0322911){14}{\line(-1,0){.0462923}}
\multiput(19.979,35.459)(-.0416785,-.0322179){15}{\line(-1,0){.0416785}}
\multiput(19.354,34.976)(-.0375478,-.0320813){16}{\line(-1,0){.0375478}}
\multiput(18.753,34.462)(-.0338182,-.0318884){17}{\line(-1,0){.0338182}}
\multiput(18.178,33.92)(-.0322161,-.0335061){17}{\line(0,-1){.0335061}}
\multiput(17.631,33.351)(-.0324454,-.0372336){16}{\line(0,-1){.0372336}}
\multiput(17.111,32.755)(-.0326221,-.0413629){15}{\line(0,-1){.0413629}}
\multiput(16.622,32.135)(-.0327403,-.0459757){14}{\line(0,-1){.0459757}}
\multiput(16.164,31.491)(-.0327921,-.0511794){13}{\line(0,-1){.0511794}}
\multiput(15.737,30.826)(-.032767,-.057117){12}{\line(0,-1){.057117}}
\multiput(15.344,30.14)(-.032652,-.063985){11}{\line(0,-1){.063985}}
\multiput(14.985,29.436)(-.032428,-.072058){10}{\line(0,-1){.072058}}
\multiput(14.661,28.716)(-.032067,-.081733){9}{\line(0,-1){.081733}}
\multiput(14.372,27.98)(-.03153,-.093606){8}{\line(0,-1){.093606}}
\multiput(14.12,27.231)(-.030752,-.108614){7}{\line(0,-1){.108614}}
\multiput(13.905,26.471)(-.02963,-.128321){6}{\line(0,-1){.128321}}
\multiput(13.727,25.701)(-.027973,-.155542){5}{\line(0,-1){.155542}}
\put(13.587,24.923){\line(0,-1){.7836}}
\put(13.485,24.14){\line(0,-1){.7877}}
\put(13.422,23.352){\line(0,-1){.7898}}
\put(13.398,22.562){\line(0,-1){.7901}}
\put(13.412,21.772){\line(0,-1){.7884}}
\put(13.465,20.984){\line(0,-1){.7849}}
\multiput(13.556,20.199)(.03244,-.19487){4}{\line(0,-1){.19487}}
\multiput(13.686,19.419)(.033554,-.154434){5}{\line(0,-1){.154434}}
\multiput(13.854,18.647)(.02934,-.109004){7}{\line(0,-1){.109004}}
\multiput(14.059,17.884)(.030312,-.094007){8}{\line(0,-1){.094007}}
\multiput(14.302,17.132)(.031003,-.082142){9}{\line(0,-1){.082142}}
\multiput(14.581,16.393)(.03149,-.072473){10}{\line(0,-1){.072473}}
\multiput(14.896,15.668)(.031819,-.064404){11}{\line(0,-1){.064404}}
\multiput(15.246,14.96)(.032023,-.057538){12}{\line(0,-1){.057538}}
\multiput(15.63,14.269)(.0321249,-.0516008){13}{\line(0,-1){.0516008}}
\multiput(16.048,13.598)(.0321407,-.0463968){14}{\line(0,-1){.0463968}}
\multiput(16.498,12.949)(.0320825,-.0417829){15}{\line(0,-1){.0417829}}
\multiput(16.979,12.322)(.0319593,-.0376517){16}{\line(0,-1){.0376517}}
\multiput(17.49,11.72)(.0317785,-.0339215){17}{\line(0,-1){.0339215}}
\multiput(18.03,11.143)(.0334014,-.0323247){17}{\line(1,0){.0334014}}
\multiput(18.598,10.594)(.0371281,-.032566){16}{\line(1,0){.0371281}}
\multiput(19.192,10.072)(.0412568,-.0327562){15}{\line(1,0){.0412568}}
\multiput(19.811,9.581)(.0458692,-.0328893){14}{\line(1,0){.0458692}}
\multiput(20.453,9.121)(.0510727,-.032958){13}{\line(1,0){.0510727}}
\multiput(21.117,8.692)(.057011,-.032952){12}{\line(1,0){.057011}}
\multiput(21.801,8.297)(.063879,-.03286){11}{\line(1,0){.063879}}
\multiput(22.504,7.935)(.071953,-.032661){10}{\line(1,0){.071953}}
\multiput(23.224,7.609)(.081629,-.032332){9}{\line(1,0){.081629}}
\multiput(23.958,7.318)(.093503,-.031833){8}{\line(1,0){.093503}}
\multiput(24.706,7.063)(.108514,-.031105){7}{\line(1,0){.108514}}
\multiput(25.466,6.845)(.128225,-.030046){6}{\line(1,0){.128225}}
\multiput(26.235,6.665)(.15545,-.028478){5}{\line(1,0){.15545}}
\put(27.012,6.523){\line(1,0){.7833}}
\put(27.796,6.418){\line(1,0){.7875}}
\put(28.583,6.353){\line(1,0){.7897}}
\put(29.373,6.326){\line(1,0){.7901}}
\put(30.163,6.337){\line(1,0){.7886}}
\put(30.952,6.388){\line(1,0){.7852}}
\multiput(31.737,6.477)(.19497,.03181){4}{\line(1,0){.19497}}
\multiput(32.517,6.604)(.154542,.033053){5}{\line(1,0){.154542}}
\multiput(33.289,6.769)(.109099,.028986){7}{\line(1,0){.109099}}
\multiput(34.053,6.972)(.094105,.030007){8}{\line(1,0){.094105}}
\multiput(34.806,7.212)(.082243,.030737){9}{\line(1,0){.082243}}
\multiput(35.546,7.489)(.072575,.031254){10}{\line(1,0){.072575}}
\multiput(36.272,7.801)(.064507,.03161){11}{\line(1,0){.064507}}
\multiput(36.981,8.149)(.057642,.031836){12}{\line(1,0){.057642}}
\multiput(37.673,8.531)(.0517047,.0319573){13}{\line(1,0){.0517047}}
\multiput(38.345,8.946)(.0465009,.03199){14}{\line(1,0){.0465009}}
\multiput(38.996,9.394)(.0418868,.0319467){15}{\line(1,0){.0418868}}
\multiput(39.625,9.874)(.0377552,.031837){16}{\line(1,0){.0377552}}
\multiput(40.229,10.383)(.036151,.0336475){16}{\line(1,0){.036151}}
\multiput(40.807,10.921)(.0324329,.0332963){17}{\line(0,1){.0332963}}
\multiput(41.358,11.487)(.0326864,.0370223){16}{\line(0,1){.0370223}}
\multiput(41.881,12.08)(.0328899,.0411503){15}{\line(0,1){.0411503}}
\multiput(42.375,12.697)(.033038,.0457622){14}{\line(0,1){.0457622}}
\multiput(42.837,13.338)(.0331235,.0509655){13}{\line(0,1){.0509655}}
\multiput(43.268,14)(.033137,.056903){12}{\line(0,1){.056903}}
\multiput(43.666,14.683)(.033067,.063772){11}{\line(0,1){.063772}}
\multiput(44.029,15.384)(.032895,.071846){10}{\line(0,1){.071846}}
\multiput(44.358,16.103)(.032597,.081523){9}{\line(0,1){.081523}}
\multiput(44.652,16.837)(.032137,.093399){8}{\line(0,1){.093399}}
\multiput(44.909,17.584)(.031457,.108412){7}{\line(0,1){.108412}}
\multiput(45.129,18.343)(.030462,.128126){6}{\line(0,1){.128126}}
\multiput(45.312,19.111)(.028982,.155357){5}{\line(0,1){.155357}}
\put(45.457,19.888){\line(0,1){.7829}}
\put(45.563,20.671){\line(0,1){.7872}}
\put(45.632,21.458){\line(0,1){.999}}
%\end
%\emline(29.53,3.012)(53.615,18.922)
\multiput(29.53,3.012)(.0510275424,.0337076271){472}
{\line(1,0){.0510275424}}
%\end
%\emline(53.615,18.922)(29.53,48.755)
\multiput(53.615,18.922)(-.033732493,.04178221289){714}
{\line(0,1){.04178221289}}
%\end
%\emline(29.53,48.755)(5.002,19.145)
\multiput(29.53,48.755)(-.03373796424,-.04072902338){727}
{\line(0,-1){.04072902338}}
%\end
%\emline(5.002,19.145)(29.307,3.012)
\multiput(5.002,19.145)(.0507411273,-.0336795407){479}
{\line(1,0){.0507411273}}
%\end
\qbezier(26.215,44.777)(29.64,43.45)(32.622,44.777)
\qbezier(26.657,4.34)(29.417,5.665)(31.74,4.34)
%\emline(29.53,22.68)(46.765,36.157)
\multiput(29.53,22.68)(.0430875,.03369375){400}
{\line(1,0){.0430875}}
%\end
%\emline(29.53,22.237)(60.685,18.04)
\multiput(29.53,22.237)(.24924,-.03358){125}{\line(1,0){.24924}}
%\end
%\emline(29.53,22.237)(40.357,6.107)
\multiput(29.53,22.237)(.0337305296,-.0502492212){321}
{\line(0,-1){.0502492212}}
%\end
\qbezier(51.185,21.575)(48.865,19.365)(50.522,17.155)
\put(56.72,12.062){$A$}
\put(55.605,26.435){$y_1$}
\put(24.667,51.405){$y_2$}
\put(33.947,50.08){$B$}
\put(3.897,12.957){$C$}
\put(25.772,-.743){$D$}
\put(24.215,39.5){$\alpha_B$}
\put(31,39.5){$\alpha_B$}
\put(37.622,39.5){$=\alpha$}
\put(24.215,8.875){$\alpha_D$}
\put(31.622,8.875){$\alpha_D$}
\put(42,8.875){$ =\beta$}
\put(30.402,27.192){$\varphi_B$}
\qbezier(29.53,25.995)(33.175,25.882)(32.845,21.795)
\put(36.157,24.005){$\varphi_A$}
\put(35.495,18.702){$\varphi_A$}
\put(52.5,15.712){$\alpha_A$}
%\put(51.2,12.212){$\alpha_A$}
\put(54,20.3){$\alpha_A=\gamma$}
%\put(46.765,22.68){$\alpha_A$}
\qbezier(32.88,21.567)(32.085,18.473)(29.522,18.562)
\qbezier(37.83,22.451)(38.272,21.125)(37.653,20.86)
%\vector[middle](51.796,34.295)(39.421,22.097)
%%\put(45.608,28.196){\vector(-1,-1){4.15}}
%%\multiput(51.796,34.295)(-.0341833645,-.0336950308){362}
%%{\line(-1,0){.0341833645}}
%\end
%%\put(52.326,37.477){$\tilde\theta_A = - \theta_A$}
\end{picture}
%%
%%
%%
%%
%%
%%\input{./pics/polygs3.tex}
%{\includegraphics[width=200pt,height=200pt]
%{./pics/rho.jpg}}
%\input{./pics/FigZnpol.tex}
\caption{{\footnotesize
The four embedded families: square, rhombus, kite and
generic skew quadrilateral with inscribed circle,
considered in s.\ref{comban}.
Shown are various angle variables used in the text.
Vertices are labeled
counterclockwise by alphabetically ordered capital letters,
directions
to corresponding vertices are denoted through $\theta$,
directions of normals -- by $\phi$
(not shown in {\it this} picture),-- angles
between these normals and vertex directions -- by $\varphi$,--
finally, the angles of polygons are $2\alpha$.
Obvious relations are:
$\alpha_a+\varphi_a=\frac{\pi}{2}$,
$\theta_{a+1}-\theta_a = \varphi_{a+1}+\varphi_a$
Relations involving $\phi$'s depends on the labeling of
polygon sides.
If vector (external momentum) ${\bf p}_a$ points from vertex
$a$ to vertex $a+1$, i.e. the vertex $a$ is at
the intersection of sides $a$ and $a-1$, then
$\theta_{a}-\varphi_a = \phi_{a-1}$ and
$\theta_a+\varphi_a = \phi_a$.
}}
\label{figpolygs}
\end{center}
\end{figure}

Instead of combining parallel lines, we can combine angles.
This enlarges the set of possible configurations and is
simply a necessary thing to do for description of generic
asymmetric configurations, like $3$-parametric family
of skew quadrilaterals and its $2$-parametric sub-family of
{\it kites} at $n=4$.
Rhombi and square are further restrictions of this family
to $1$- and $0$-parametric sub-sets.
Consideration of multiple angles is straightforward,
however a new phenomenon arises: particular element of
the boundary ring which we obtain can depend on the choice
of angles in the polygon, but canonical elements like
${\cal L}$ will, of course, coincide.
A variety of angle variables appearing in calculations
is shown in combined Fig.\ref{figpolygs}.

\subsubsection{Square}

As usual, we begin from the simplest case: the square.
This time we want to obtain ${\cal L}_\Box \in {\cal R}_\Box$
from two boundary rings ${\cal R}_\angle^{-+}$, associated
with two {\it opposite} right angles, say, at vertices
$B$ and $D$.
Following our standard procedure, we multiply the
canonical ${\cal L}$ elements of these two rings,
then subtract $P_2$ in order to eliminate the $y_0^2$-term
and afterwards look at the coefficient in front of $y_0$:
if it is not constant we add more "obvious" elements
(\ref{Ppolsdef}) to make this coefficient into a common
factor and then throw it away.
Actually, the last step will appear unnecessary in
the study of a pair of angles (this is {\it a priori} obvious
because the degree of appearing polynomials will be lower
than $n$, and polynomials (\ref{Ppolsdef}) can not mix with
them).

Throughout this subsection we use the following notation:
\be
{\cal L}_\angle^\sigma (\theta|\alpha)
\ \stackrel{(\ref{Langelexp})}{=}\
\sigma y_0\cos\alpha + \sin\alpha -
y_1\cos\theta - y_2\sin\theta
\ee
We remind that $\theta$ denotes direction to the vertex
of the angle, while its size is $2\alpha$.

In the case of square $2\alpha=\frac{\pi}{2}$ and we locate
two opposite angles at $\theta_B = \frac{\pi}{4}$ and
$\theta_D = \frac{5\pi}{4}$.
Then
\be
{\cal L}_\angle^+\left(\frac{\pi}{4}\Big|\frac{\pi}{4}\right)
{\cal L}_\angle^+\left(\frac{5\pi}{4}\Big|\frac{\pi}{4}\right)
= \frac{1}{2}\Big(y_0+1-y_1-y_2\Big)
\Big(y_0+1+y_1+y_2\Big) = \nn \\
= \frac{1}{2}\Big((y_0+1)^2 - (y_1+y_2)^2\Big) =
\frac{1}{2}\Big(P_2 + 2(y_0-y_1y_2)\Big) =
\frac{1}{2}P_2 + {\cal L}_\Box
\ee
${\cal L}_\Box = y_0-y_1y_2$ is our familiar expression,
both the ${\cal L}$-element of ${\cal R}_\Box$ and exact
solution $S_\Box$ to the $AdS$ Plateau problem.

\subsubsection{Rhombus}

In the case of rhombus we keep $\theta$'s the same,
$\theta_B = \frac{\pi}{4}$ and
$\theta_D = \frac{5\pi}{4}$, but angles at the vertices
are now not restricted to be $\frac{\pi}{4}$. Then
\be
{\cal L}_\angle^+\left(\frac{\pi}{4}\Big|\alpha\right)
{\cal L}_\angle^+\left(\frac{5\pi}{4}\Big|\alpha\right)
= \left(y_0\cos\alpha+\sin\alpha-\frac{1}{\sqrt{2}}
(y_1+y_2)\right)
\left(y_0\cos\alpha+\sin\alpha+\frac{1}{\sqrt{2}}
(y_1+y_2)\right) = \nn \\
= (y_0\cos\alpha+\sin\alpha)^2-\frac{1}{2}(y_1+y_2)^2 =
\frac{1}{2}P_2 + \left(\frac{1}{2}(y_0^2-1)\cos(2\alpha)
+y_0\sin(2\alpha)-y_1y_2\right) =\nn \\
= P_2 \cos^2\alpha + \left\{y_0\sin(2\alpha) -y_1y_2 -
\left(1-\frac{1}{2}(y_1^2+y_2^2)\right)\cos(2\alpha)\right\}
= P_2\cos^2\alpha + {\cal L}_\diamond \sin(2\alpha)
\ee
For comparison with the other formulas for ${\cal L}_\diamond$,
like (\ref{Lromb}),
one should keep in mind that $\alpha = \frac{\pi}{2}-\varphi$,
so that $\sin(2\alpha) = \sin(2\varphi)$ and
$\cos(2\alpha) = -\cos(2\varphi)$.

Finally, if we use in this formula another angle of the rhombus
$\alpha' = \pi-\alpha$ instead of $\alpha$, then $\sin(2\alpha)$
changes sign. However, simultaneously one should change $\sigma$
to $-\sigma$, since the starting side of the rhombus in above
derivation has also changed. Changing sign of $\sigma$
is equivalent to changing sing of $y_0$, thus the product
$y_0\sin(2\alpha)= \sigma y_0\sin(2\alpha) =
(-\sigma) y_0\sin(2(\pi-\alpha))$ does not change and
${\cal L}_\diamond$ remains the same -- as it should, since
it is a canonically defined element of the boundary ring
${\cal R}_\diamond$.

\subsubsection{Kite}

In the case of kite we can consider two essentially
inequivalent choices of opposite angles:
$\alpha=\alpha_B$ and $\beta=\alpha_D$  or
$\gamma=\alpha_A$ and $\gamma=\alpha_C=\alpha_A=\frac{\pi}{2} -
\frac{\alpha+\beta}{2}$.
The corresponding angles $\theta$ will also be different:
either $\theta_B = \frac{\pi}{2}$ and $\theta_D = \frac{3\pi}{2}$
or $\theta_A = \frac{\alpha-\beta}{2}$
and $\theta_C = \pi - \theta_A$.

A product of two $y$-linear elements ${\cal L}_\angle$ is
usually quadratic in $y$ and we denote it ${\cal Q}$.
Thus in the case of kite we are interested in two different
quantities ${\cal Q} \in {\cal R}_{kite}$:
$$
{\cal Q}_{BD} =
{\cal L}^+_\angle\Big(\theta_D\Big|\alpha_D\Big)
{\cal L}_\angle^+\Big(\theta_B\Big|\alpha_B\Big)
={\cal L}^+_\angle\Big(\frac{3\pi}{2}\Big|\beta\Big)
{\cal L}_\angle^+\Big(\frac{\pi}{2}\Big|\alpha\Big)
\ \stackrel{(\ref{Langelexp})}{=}\
\Big(y_0\cos\beta + \sin\beta + y_2\Big)
%- y_1\cos\theta_D-y_2\sin\theta_D\Big)
\Big(y_0\cos\alpha + \sin\alpha - y_2\Big)
%- y_1\cos\theta_B-y_2\sin\theta_B\Big)
= $$
\vspace{-0.4cm}
\be
= y_0^2\cos\alpha\cos\beta
+ y_0\Big(\sin(\alpha+\beta) +y_2(\cos\alpha-\cos\beta)\Big)
 + (\sin\beta + y_2)(\sin\alpha-y_2)
\label{QBDkite}
\ee
and
$$
{\cal Q}_{AC} =
{\cal L}^-_\angle\Big(\theta_C\Big|\alpha_C\Big)
{\cal L}_\angle^-\Big(\theta_A\Big|\alpha_A\Big)
={\cal L}^-_\angle\Big(\pi-\frac{\alpha-\beta}{2}\Big|\gamma\Big)
{\cal L}_\angle^-\Big(\frac{\alpha-\beta}{2}\Big|\gamma\Big)
\ \stackrel{(\ref{Langelexp})}{=}\
$$ $$ =
\Big(-y_0\cos\gamma + \sin\gamma + y_1\cos\frac{\alpha-\beta}{2}-
y_2\sin\frac{\alpha-\beta}{2}\Big)
\Big(-y_0\cos\gamma + \sin\gamma - y_1\cos\frac{\alpha-\beta}{2}-
y_2\sin\frac{\alpha-\beta}{2}\Big)
= $$ $$ =
\left(y_0\sin\frac{\alpha+\beta}{2} + y_2\sin\frac{\alpha-\beta}{2}
- \cos\frac{\alpha+\beta}{2}\right)^2 -
y_1^2\cos^2\frac{\alpha-\beta}{2} =
$$ $$
= y_0^2\sin^2\frac{\alpha+\beta}{2} + \cos^2\frac{\alpha+\beta}{2}
- y_0\sin(\alpha+\beta)
+ 2y_0y_2\sin\frac{\alpha+\beta}{2}\sin\frac{\alpha-\beta}{2}
-2y_2\cos\frac{\alpha+\beta}{2}\sin\frac{\alpha-\beta}{2} -
$$
\vspace{-0.3cm}
\be
- y_1^2\cos^2\frac{\alpha-\beta}{2}
+ y_2^2\sin^2\frac{\alpha-\beta}{2} \
=\ P_2\cos^2\frac{\alpha-\beta}{2} + {\cal Q}_{BD}
\ee
Thus the two ways of construction provides us with two different
elements of the boundary ring.
They both belong to the family ${\cal Q}_{BD}+\nu P_2$,
consisting of all the elements of ${\cal R}_{kite}$ of degree $2$.
Expression (\ref{QBDkite}) is already familiar to us:
it appeared in (\ref{kitesol2}) and we also know from there
how exact solution to Plateau problem is embedded into this family:
\be
{\cal S}_{kite} \ \stackrel{(\ref{kitesol2})}{\sim}\
{\cal Q}_{BD} -\frac{1}{2}P_2\cos(\alpha-\beta)
\ee
In order to convert ${\cal Q}_{BD}$ into a $y_0$-linear
expression ${\cal L}_{kite}$ we need to subtract
$P_2\cos\alpha\cos\beta$.
However in the resulting
\be
\sin(\alpha+\beta){\cal L}_{kite} =
{\cal Q}_{BD} - P_2\cos\alpha\cos\beta =
y_0\Big(\sin(\alpha+\beta) +y_2(\cos\beta-\cos\alpha)\Big)
% + (\sin\beta + y_2)(\sin\alpha-y_2)
-\cos(\alpha+\beta)+\nn \\
+y_2(\sin\alpha-\sin\beta)
+ y_1^2\cos\alpha\cos\beta - y_2^2(1-\cos\alpha\cos\beta)
\ee
the coefficient in front of $y_0$ is non-trivial function
of $y_2$ and it {\it can not} be eliminated.
Still such ${\cal L}_{kite}$ satisfies the condition
(\ref{linearity1}).
if we parameterize the family of quadratic elements in
${\cal R}_{kite}$ canonically:
$\Big\{ {\cal L}_{kite} + \mu P_2\Big\}$ then exact
solution ${\cal S}_{kite}$ is associated with
\be
\mu_{kite} =
\frac{1}{2}\frac{\cos(\alpha+\beta)}{\sin(\alpha+\beta)}
\ee
In the particular case of $\alpha=\beta$ kite becomes
rhombus and we reproduce (\ref{mudia}):
\be
\left.\mu_{kite}\right|_{\alpha=\beta} =
\frac{1}{2}\frac{\cos(2\alpha)}{\sin(2\alpha)} =
-\frac{1}{2}\frac{\cos(2\varphi)}{\sin(2\varphi)}
= \mu_\diamond
\ee

\subsubsection{Generic skew quadrilateral}

As basic variables, parameterizing the skew quadrilateral
(possessing an inscribed circle) we take the four angles
$\varphi_A$, $\varphi_B$, $\varphi_C$, $\varphi_D$.
Actually these are three independent variables, since
$\varphi_A+\varphi_B+\varphi_C+\varphi_D=\pi$.
The angles $2\alpha_A$, $2\alpha_B$, $2\alpha_C$ and $2\alpha_D$
of the quadrilateral are easily expressed through these
$\varphi$'s:
\be
\alpha_a = \frac{\pi}{2} - \varphi_a
\ee
The normals directions $\phi_a$ and those of the vertices
$\theta_a$ are also expressed through $\varphi_a$, provided one
fixes the freedom of overall rotation in the $(y_1,y_2)$ plane.
In this subsection we do this by putting $\phi_1=0$,
so that the side $AB$ is parallel to ordinate axis,
see Fig,\ref{figpolygs}.
Then
\be
\phi_1 = 0, \ \ \ \phi_2 = 2\varphi_B, \ \ \
\phi_3 = 2\varphi_B+\varphi_C, \ \ \
\phi_4 = -2\varphi_A-\varphi_D
\nn \\
\theta_A = -\varphi_A, \ \ \ \theta_B = \varphi_B, \ \ \
\theta_C = 2\varphi_B + \varphi_C, \ \ \
\theta_D = 2\varphi_D-\varphi_A
\ee

Like kite, the boundary ring for generic skew quadrilateral
can be obtained from rings for two different pairs of
angles: $B$ and $D$ or $A$ and $C$.
$$
Q_{BD} =
{\cal L}_\angle^+(\theta_D|\alpha_D)
{\cal L}_\angle^+(\theta_B|\alpha_B)
\ \stackrel{(\ref{Langelexp})}{=}
$$ $$
=\Big(y_0\cos\alpha_D + \sin\alpha_D
- y_1\cos\theta_D-y_2\sin\theta_D\Big)
\Big(y_0\cos\alpha_B + \sin\alpha_B
- y_1\cos\theta_B-y_2\sin\theta_B\Big)
= $$ $$
= y_0^2\cos\alpha_B\cos\alpha_D
+y_1^2\cos\theta_B\cos\theta_D
+ y_2^2\sin\theta_B\sin\theta_D
-y_1y_2\sin(\theta_B+\theta_D)
+ y_0\sin(\alpha_B + \alpha_D)
+ \sin\alpha_B\sin\alpha_D
- $$ $$
-y_0y_1\Big(\cos\alpha_B\cos\theta_D + \cos\alpha_D\cos\theta_B\Big)
-y_0y_2\Big(\cos\alpha_B\sin\theta_D + \cos\alpha_D\sin\theta_B\Big)
- $$
\vspace{-0.4cm}
\be
-y_1\Big(\sin\alpha_B\cos\theta_D + \sin\alpha_D\cos\theta_B\Big)
-y_2\Big(\sin\alpha_B\sin\theta_D + \sin\alpha_D\sin\theta_B\Big)
\ee
Similarly we can define
\be
Q_{AC} =
{\cal L}_\angle^-(\theta_C|\alpha_C)
{\cal L}_\angle^-(\theta_A|\alpha_A)
\ee
It is given by the same formula with $(B,D)$ changed for $(A,C)$
and the sign of $y_0$ reversed (because the starting segment
is now different and therefore ${\cal L}^-_\angle$
is used instead of ${\cal L}^+_\angle$).
Both these quantities can be used to find the $y_0$-linear
element ${\cal L}$:
\be
Q_{BD} - P_2\cos\alpha_B\cos\alpha_D
= \sin(\alpha_B+\alpha_D){\cal L}_{quadri}, \nn \\
Q_{AC} - P_2\cos\alpha_A\cos\alpha_C
= \sin(\alpha_A+\alpha_C){\cal L}_{quadri}
\label{quadriel}
\ee
The fact that ${\cal L}$ is the same in both cases is
a direct, but somewhat tedious consistency check.
Both expressions can be considered as explicit expression
for ${\cal L}_{quadri}$ -- but written in terms of two
different sets of independent parameters:
$(\alpha_B,\alpha_D,\theta_B,\theta_D)$ in one case and
$(\alpha_A,\alpha_C,\theta_A,\theta_C)$ in the other.

This ${\cal L}$ is exactly the ${\cal L}_{quadri}$
which appeared in eq.(\ref{geskqu}), which describes its
relation to exact solution of $AdS$ Plateau problem for
generic skew quadrilateral.

\subsection{Summary}

We now give a short summary of our consideration of the boundary
rings.

\subsubsection{Boundary ring and Plateau problem}

Suggested strategy is to represent the ring ${\cal R}_\Pi$
by {\it canonical} element ${\cal L}_\Pi$, which is linear in $y_0$
and satisfies the condition (\ref{linearity1}):
\be
{\cal L}_\Pi = y_0\Big(1+ O(y_1,y_2)\Big)
+ {\cal K}_\Pi(y_1,y_2)
\label{linearity}
\ee
For $\bar\Pi$ possessing an inscribed circle and thus a degree-two
element $P_2\in {\cal R}_\Pi$, such element can be constructed
from the product of complex-valued generators
${\cal C}_| \in {\cal R}_|$  of individual segments and
eliminating higher powers of $y_0$ by subtracting $P_2$ with
various coefficients.
In this way, however, we obtain a polynomial of degree $n$
in $y_1$ and $y_2$ which is not unique, since one can always
combine it with the "obvious"
${\cal P}_\Pi \in {\cal R}_\Pi$, see eq.(\ref{Ppolsdef}),
which also has degree $n$.
Worse than that, this polynomial can not serve as ${\cal L}_\Pi$
because it does not necessarily satisfy (\ref{linearity}).
In the case when $y_0$ in $\Pi$ flips (changes direction) at
every vertex, one can always adjust the combination with $P_\Pi$
in such a way that a common multiplier of degree $n/2$ factors out,
and after throwing it away (what is possible because this
expression is not identical zero in ${\cal R}_\Pi$)
we finally obtain the ${\cal L}_\Pi$, which turns out
to be of degree $n/2$ in $y_1$ and $y_2$.
This ${\cal L}_\Pi$ can be also constructed straightforwardly
from building blocks ${\cal L}_\angle^{\mp\pm}$,
associated with $n/2$ non-adjacent angles of $\Pi$ instead of
its $n$ sides.
Since ${\cal L}_\angle^{\mp\pm}$ is itself linear in all
$y$-variables, the product of such building blocks provides
an element of degree $n/2$ and modulo $P_2$ it is linear in $y_0$,
as requested. It turns out that it automatically
(after appropriate rescaling) satisfied (\ref{linearity}).

Thus canonical element ${\cal L}_\Pi$

$\bullet$ is linear in $y_0$,
\be
{\cal L}_\Pi = y_0Q_\Pi(y_1,y_2) - {\cal K}_\Pi(y_1,y_2);
\ee

$\bullet$ satisfies (\ref{linearity}), i.e.
\be
Q_\Pi(y_1,y_2) = 1 + O(y_1,y_2);
\ee

$\bullet$ is of degree $n/2$ in $y_1$ and $y_2$, more
precisely ${\cal K}_\Pi$ is of degree $n/2$ and $Q_\Pi$
is of degree $n/2-1$.

Such element is unique, up to overall rotation of the
$(y_1,y_2)$ plane.
Unfortunately, there is no distinguished way to fix
this freedom and historically it was done in different ways
in different particular cases.
Among existing options are:
$\theta_B=\frac{\pi}{4}$ (square and rhombus in \cite{am1}),
$\theta_B = \frac{\pi}{2}$ (kite, a natural choice),
$\phi_1=0$ (square and other $Z_n$-symmetric configurations
of \cite{malda3}, generic quadrilateral and skew hexagons
in this paper).
Vertex $B$ is the one where $y_0$ is takes its maximal
positive value.
Still another option is to require that the coefficient
in front of $y_1^{n/2}$ -- the maximal power of $y_1$ --
vanishes.
Rotational freedom should be taken into account in comparison
of different formulas in this paper.

Entire family of elements of degree
$n/2$ in ${\cal R}_\Pi$ is spanned by polynomials of
degree $n-2$ of three variables $y_0,y_1,y_2$:
\be
\Big\{ {\cal L}_\Pi + \mu({\bf y})P_2\Big\}
\ee
The suggestion is to look for the first approximation to
solution of the $AdS$ Plateau problem within this set --
finding the optimal {\it point}
$\mu_\Pi$ in this {\it moduli space} (made of polynomials),
either directly from NG equations or from minimization of
the regularized action over $\mu$ {\it a la} \cite{mmt1}.
Then this approximate solution can be further perturbed,
as described in \cite{malda3} and s.\ref{comprec} above.

\subsubsection{List of the simplest ${\cal L}_\Pi$}

We now list briefly the simplest examples of
${\cal L}$, obtained in the previous subsections
what provides a general look on the problem.

{\bf Single segment:}
\be
{\cal C}_|^{\pm}(\phi) =  1\pm\i y_0 - ze^{-i\phi}
\ee
is the complex generator, consisting of two real ones:
\be
{\rm Re}({\cal L}_|^C) = 1 -cy_1-sy_2
\ \stackrel{(\ref{Ppolsdef})}{=}\ P_|(y_1,y_2), \ \ \ \
c=\cos\phi,\ s=\sin\phi
\ee
and
\be
{\cal L}_|^\pm = {\rm Re}({\cal L}_|^{C\pm}) = \pm y_0 + sy_1-cy_2
\label{1segL}
\ee
$P_\Pi$ does not contain $y_0$ and is independent
of the sign $\sigma$.
${\cal L}_|^\pm$ is actually an element of a special sub-ring
in ${\cal R}_|$,
\be
{\cal L}_|^\pm \in {\cal R}_{||}^{\mp\pm} \subset {\cal R}_|^\pm,
\ee
and does not adequately represent ${\cal R}_|$ itself.
Angle $\phi$ specifies the direction of a {\it normal} to the
segment, direction of the segment itself is $\phi+\frac{\pi}{2}$.

\bigskip
\noindent
{\bf Two segments}, forming an angle of the size $2\alpha$
with flipping $\sigma_2=-\sigma_1=1$:
\be
{\cal L}_\angle(\theta|\alpha) = {\cal L}_\angle^{-+}
= y_0\cos\alpha + \sin\alpha - y_1\cos\theta-y_2\sin\theta
%-{\rm Re} (ze^{-i\theta})
\ee
$\theta$ defines the direction to the angle's vertex.
It is related to the single-segment quantities by
\be
ze^{-i\theta} {\cal L}_\angle(\theta|\alpha) =
{\cal C}_|^+(\theta-\varphi) {\cal C}_|^-(\theta+\varphi) - P_2
\ee
Here the normal directions are $\phi_1 = \theta - \varphi$
and $\phi_2 = \theta + \varphi$, so that
$\varphi = \frac{\pi}{2}-\alpha$.
In particular, for {\bf two parallel segments} we have:
\be
{\cal L}_{||}^{\mp\pm} = y_0\mp (y_1\sin\phi + y_2\cos\phi)
\ee
Such combination appears in description of symmetric $n$-angle
polygons with $n=4k-2$, including $n=2$ (see s.2.1 of \cite{malda3})
and $n=6$ (hexagon).
For $n=4k$ another combination of $\sigma$'s is needed, then:
\be
{\cal L}_{||}^{\pm\pm} = y_0 \mp \left(y_1y_2 \cos(2\phi) +
\frac{1}{2}(y_2^2-y_1^2)\sin(2\phi)\right)
\ee
Square and rhombus belong to this class of examples.

\bigskip
\noindent
{\bf Four segments} can be described as a combination of two
non-adjacent angles with alternating $\sigma$:
$$
{\cal L}_{quadri} = \frac{1}{\sin(\alpha_1+\alpha_3)}
\Big(
{\cal L}_\angle^{-+}(\theta_3|\alpha_3)
{\cal L}_\angle^{-+}(\theta_1|\alpha_1)
-P_2\cos\alpha_1\cos\alpha_2\Big) =
$$ $$
= y_0\left(1
-\frac{\cos\alpha_1\cos\theta_3 + \cos\alpha_3\cos\theta_1}
{\sin(\alpha_1+\alpha_3)}\,y_1
-\frac{\cos\alpha_1\sin\theta_3 + \cos\alpha_3\sin\theta_1}
{\sin(\alpha_1+\alpha_3)}\,y_2\right) +
$$ $$
+ \frac{\cos\theta_1\cos\theta_3 + \cos\alpha_1\cos\alpha_3}
{\sin(\alpha_1+\alpha_3)}\,y_1^2
+ \frac{\sin\theta_1\sin\theta_3 + \cos\alpha_1\cos\alpha_3}
{\sin(\alpha_1+\alpha_3)}\,y_2^2
-\frac{\sin(\theta_1+\theta_3)}
{\sin(\alpha_1+\alpha_3)}\,y_1y_2
- $$
\vspace{-0.4cm}
\be
-\frac{\sin\alpha_1\cos\theta_3 + \sin\alpha_3\cos\theta_1}
{\sin(\alpha_1+\alpha_3)}\,y_1
-\frac{\sin\alpha_1\sin\theta_3 + \sin\alpha_3\sin\theta_1}
{\sin(\alpha_1+\alpha_3)}\,y_2
-\frac{\cos(\alpha_1+\alpha_3)}{\sin(\alpha_1+\alpha_3)}
\ee
For particular sub-families this expression simplifies:\\
{\bf Kite}, $\theta_3-\theta_1=\pi$:
(we also put $\theta_1=\frac{\pi}{2}$, i.e.
$y_{1,2}=y_{1,2}^{\theta_1-{\pi}/{2}}$)
\be
{\cal L}_{kite} =
y_0\left(1+
\frac{\sin\frac{\alpha_1-\alpha_3}{2}}
{\cos\frac{\alpha_1+\alpha_3}{2}}\,y_2\right)
+ \frac{\sin\frac{\alpha_1-\alpha_3}{2}}
{\sin\frac{\alpha_1+\alpha_3}{2}}\,y_2 +
\frac{\cos\alpha_1\cos\alpha_3}{\sin(\alpha_1+\alpha_3)}\,y_1^2
- \frac{1-\cos\alpha_1\cos\alpha_3}{\sin(\alpha_1+\alpha_3)}\,y_2^2
-\frac{\cos(\alpha_1+\alpha_3)}{\sin(\alpha_1+\alpha_3)}
\ee
{\bf Rhombus}, i.e. kite with $\alpha_3=\alpha_1=\alpha$:
%($y_{1,2}=y_{1,2}^{\theta_1-{\pi}/{4}}$)
\be
{\cal L}_\diamond = y_0
+ \frac{1+\cos(2\alpha)}{\sin(2\alpha)}(y_1^2+y_2^2) -
\frac{1}{\sin(2\alpha)}y_2^2 - \frac{\cos(2\alpha)}{2\sin(2\alpha)}
\ee
Rotation by $\pi/4$, $(y_1,y_2) \rightarrow
\frac{1}{\sqrt{2}}(y_1-y_2,y_1+y_2)$,
and substitution $2\alpha=\frac{\pi}{2}-2\phi$
convert this into
$$
{\cal L}_\diamond
= y_0 - \frac{1}{\cos(2\phi)}y_1y_2 -
\frac{\sin(2\phi)}{\cos(2\phi)}\left(1-\frac{1}{2}y^2\right)
= y_0 - y_1y_2\cosh\xi - \left(1-\frac{1}{2}y^2\right)\sinh\xi
= $$
\vspace{-0.3cm}
\be
= y_0-\frac{1+b^2}{1-b^2}y_1y_2 -
\frac{2b}{1-b^2}\left(1-\frac{1}{2}y^2\right)
\ee
with $y^2=y_1^2+y_2^2$, $\cosh\xi = \frac{1}{\cos(2\phi)}$,
$\sinh\xi = \frac{\sin(2\phi)}{\cos(2\phi)}$ and
$b = \tan\phi$.\\
{\bf Square}, i.e. rhombus with $2\alpha = \frac{\pi}{2}$:
%In particular for the {\bf square} ($\xi = 0$)
\be
{\cal L}_{\Box} = y_0 - y_1y_2
\ee

These examples are concisely represented in the following table.
Its first part contains examples which are symmetric
under $z \rightarrow -z$ accompanied by either $y_0 \rightarrow
y_0$ or $y_0 \rightarrow - y_0$.
Examples in the second part of the table do not have this symmetry.
An element ${\cal A}-{\cal B}$ of ${\cal R}_\Pi$
is often written as ${\cal A}={\cal B}$.

\bigskip

\centerline{$
\begin{array}{|c|c|c|c|}
\hline
&&&\\
\Pi ={\rm set\ of}  &{\rm set\ of}
&{\cal L}_\Pi\in{\cal R}_\Pi
&{\cal S}_\Pi\in \overline{{\cal R}_\Pi}, \ \ \ \
{\cal S}_\Pi={\cal L}_\Pi - \mu_\Pi P_2 = 0\\
n\ {\rm segments} &\sigma'{\rm s}
&({\rm linear\ in}\ y_0,\ {\rm of\ degree}\ \frac{n}{2}\
{\rm in}\ y_1,y_2)
&{\rm is\ exact\ solution\ of}\ AdS\ {\rm Plateau\ problem}\\
&&&\\
\hline\hline
&&&\\
&&\pm y_0 = -sy_1+cy_2&\\
{\rm single\ segment} &  &
=-y_1\sin\phi + y_2\cos\phi
&  \\
| & \pm & {\rm actually\ belongs\ to\
{\cal R}_{||}^{\mp\pm}\subset {\cal R}_{|}^{\pm}}& -\\
&&{\cal C}_|=1\pm y_0 - ze^{-i\phi} &\\
&&&\\
\hline
&&&\\
2\ {\rm parallel\ segms}&++
&y_0 = y_1y_2
& -  \\
||&n=4k&  {\rm actually\ belongs\ to\
{\cal R}_\Box\subset {\cal R}_{||}^{++}}&\\
&&&\\
\hline
&&&\\
{\rm square}&-+-+&y_0=y_1y_2& \mu_\Box = 0:\\
\Box &+-+-&-y_0 = y_1y_2& {\cal S}_\Box={\cal L}_\Box \\
&&&\\
\hline
&&&\\
{\rm rhombus}&-+-+&
y_0 = y_1y_2\cosh\xi + (1-\frac{1}{2}y^2)\sinh\xi
&\mu_\diamond = -\frac{1}{2}\tan(2\phi): \ \ \
{\cal S}_\diamond = {\cal L}_\diamond -\frac{1}{2}\sinh\xi P_2
\\
\diamond&&\cosh\xi=\frac{1}{\cos(2\phi)},
\ \ \sinh\xi = \frac{\sin(2\phi)}{\cos(2\phi)}
&\sim y_1y_2-\frac{1}{2}(1-y_0^2)\sin(2\phi) -y_0\cos(2\phi)\\
&&&\\
\hline
&&&\\
2\ {\rm parallel\ segms}&\mp\pm&\pm y_0 = {\rm Im}(ze^{-i\phi})
& \mu_{||}=0: \\
||&n=4k-2&=-sy_1+cy_2&  {\cal S}_{||} = {\cal L}_{||}^{-+} \\
&&&\\
\hline
&&&\\
{\rm hexagon}&-+-+-+&y_0(1-\frac{1}{4}y^2)
=\frac{1}{4}y_2(3y_1^2-y_2^2)
& \mu_{hexa}=y_0B(y_1,y_2)\ \cite{malda3},\ \ \
{\cal S}_{hexa}\approx {\cal L}_{hexa}\\
&&&\\
\hline
&&&\\
Z_n-{\rm symmetric}&{\rm alternated}&y_0Q_n(y^2)
=\frac{1}{2^{n/2-1}}{\rm Im}(z^{n/2})
& {\cal S}_{Z_n}\approx {\cal L}_{Z_n}, \ \ \
\mu_{Z_n}=y_0B_n(y_1,y_2)\\
{\rm polygon},\ \ n\ {\rm even}&&
& \ {\rm with\ non-polynomial}\ B_n,\ {\rm see}\ \cite{malda3}\\
&&&\\
\hline\hline
&&&\\
{\rm angle\ of\ size }\ 2\alpha &-+&y_0\cos\alpha+\sin\alpha
= {\rm Re}(ze^{-i\theta})
&\mu_\angle = 0: \ \ \ {\cal S}_\angle = {\cal L}_\angle\\
\angle && =y_1\cos\theta + y_2\sin\theta
& {\rm NG\ eqs\ are\ singular\ in\ this\ case},\ L_{NG}=0\\
&&& \\
\hline
&&&\\
{\rm kite}&-+-+
&y_0\left(1 +
y_2\frac{\cos\alpha-\cos\beta}{\sin(\alpha+\beta)}\right)
+ y_2\frac{\sin\alpha-\sin\beta}{\sin(\alpha+\beta)}
& \mu_{kite} = -\frac{1}{2}\cot(\alpha+\beta):
\\
&&+(y_1^2+y_2^2)\frac{\cos\alpha\cos\beta}{\sin(\alpha+\beta)}
-\frac{1}{\sin(\alpha+\beta)}y_2^2
& {\cal S}_{kite}={\cal L}_{kite}
+\frac{\cos(\alpha+\beta)}{2\sin(\alpha+\beta)}P_2 \sim \\
&&-\frac{\cos(\alpha+\beta)}{\sin(\alpha+\beta)}
&\sim  y_1^2\cos(\alpha-\beta)
-y_2^2\big(2-\cos(\alpha-\beta)\big)\\
&& (y_1,y_2) \ {\rm here\ are\ rotated\ by}\ \frac{\pi}{4}
%\ {\rm w.r.t.}
&+(y_0^2-1)\cos(\alpha+\beta) +2y_2(\sin\alpha-\sin\beta)\\
&& {\rm w.r.t. the\ rhombus\ and\ square}
&+2y_0\sin(\alpha+\beta)+2y_0y_2(\cos\alpha-\cos\beta)\\
&&&\\
\hline
&&&\\
{\rm generic\ skew}&-+-+&{\rm see\ eqs.}
(\ref{quadrisol})\ {\rm and}\ (\ref{quadriel})
&\mu_{quadri} =
-\frac{t_At_C-(t_A+t_C)t_B+(2t_At_C-1)t_B^2}
{2(t_A+t_C)(1+t_B^2)}\\
{\rm quadrilateral}&&{\rm for\ two\ different\ parametrizations}
&{\cal S}_{quadri} = {\cal L}_{quadri} + \mu_{quadri}P_2\\
&&&\\
\hline
\end{array}
$}

\subsubsection{Solutions to $AdS$ Plateau problem}

We are still not in position to describe exact solutions
in general situation, even under assumptions (\ref{ads3}).
Still, according to \cite{malda3}, a reasonable approximation
can be found within the families of the boundary-ring elements
of degree $n/2$:
\be
{\cal S}_\Pi \approx {\cal L}_\Pi - \mu_\Pi({\bf y})\cdot P_2
\label{SvsLmu}
\ee
The optimal choice of the polynomial $\mu_\Pi$ of degree
$n/2-2$ can be dictated by two kinds of argument:

$\bullet$ by NG equations, that -- according to s.\ref{comprec} --
imply that ${\cal S}_\Pi$ should be a properly
perturbed harmonic function,

$\bullet$ by minimization of regularized area, evaluated as
a {\it height function} on the space of coefficients of $\mu$,
as suggested in \cite{mmt1} and \cite{mmt2}.

Exact solutions, available at $n=4$ fit into this scheme
{\it exactly}: always belong to the family (\ref{SvsLmu}),
however, unlike in the $Z_n$-symmetric case considered in
\cite{malda3}, the relevant $\mu_\Pi \neq 0$.
This means that the third way to specify $\mu_\Pi$ --

$\bullet$ by some algebraic criterium\\
-- still remains to be found:
hypothesis ${\cal S}_\Pi\ \stackrel{?}{\approx}\ {\cal L}_\Pi$
does {\it not} work for asymmetric $\Pi$.

One can easily play with $3d$ plots of above functions
to see how nice these approximations are and how strong
is dependence on the deviations of $\mu({\bf y})$ from the
optimal values.
Unfortunately, today such plots can not be adequately represented
in a {\it paper}, even on computer screen, since they necessarily
use additional software, allowing to {\it rotate} $3d$ images.
However, after the functions ${\cal L}$ are explicitly constructed
in this paper, it takes two minutes to write a two-line "program"
in MAPLE or Mathematica to make these plots and start investigating
them.
As explained in \cite{malda3}, it is more informative to
plot $r(\vec y) = \sqrt{\Big(y_0(y_1,y_2)\Big)^2+1-y_1^2-y_2^2}$
than $y_0(y_1,y_2)$ itself.
As long as $\mu({\bf y})$ is taken to be independent of $y_0$
the equation ${\cal L} + \mu P_2 = 0$ is quadratic for $y_0$
and can be analytically resolved -- this simplifies the
computer program even further and makes it working fast on
not-very-modern laptops.\footnote{For the sake of convenience
we suggest a version of such MAPLE program here:\\ \\
L:= ?? : \ \ \ \ \ \hfill
\#  for example, for the square  L:=$\ y0\ -\ y1*y2$:\\
mu:=?? : \ \ \ \ \  \hfill
\# function of y1 and y2 with NUMERICAL coefficients should be
substituted \\ \\
P2:=\ $y0^2+1-y1^2-y2^2$:\\
s:=2 \ \ \ \hfill \# this parameter can be adjusted to focus on the
domain bounded by our polygon \\ \\
Y:=\ solve(\ L + mu*P2,\ y0\ )[1]: \ \ \
\hfill \# sometime one needs to change
"[1]" for "[2]" to choose appropriate root of quadratic equation\\
\phantom{.}  \hfill \# ATTENTION: if mu=0
then there is only one root and "[1]" should be omitted!\\
plot3d( sqrt($Y^2+1-y1^2-y2^2$),\ y1:=-s..s,\ y2:=-s..s,\
axes=boxed,\ grid=[100,100] ):\\ \\
The first two lines contain input:
explicit expression for ${\cal L}_\Pi$ from  this paper
or \cite{malda3} and one's favorite parametrization of the trial
constant/polynomial/function $\mu({\bf y})$.
The last two lines are the plotting program itself.
It can be better to substitute trigonometric functions of
angles by their rational expressions through tangents of the
one-half angle, otherwise MAPLE should be taught trigonometric
identities.
Before one reaches asymmetric hexagons at $n=6$ one can begin
from substituting numbers for $\mu$.
For $n\geq 6$ polynomials of $y_1$ and $y_2$ of degree $n/2-2$
are a nice starting point.
If $\mu$ is non-trivial function of $y_0$ one can need to
switch to {\it pointplot} commands which takes computers
more time to work with.
}

If one wants to go beyond  {\it approximate} methods,
then for $n>4$ the restriction that $\mu_\Pi$ is
a polynomial of degree $n/2-2$, should be lifted.
In \cite{malda3} and s.\ref{comprec} it is shown how one
can proceed with {\it formal series} for $\mu({\bf y})$.
It would be most interesting to identify a narrow class
of functions, which $\mu_\Pi({\bf y})$ actually belongs to.
As shown in s.\ref{comprec}, hypergeometric
functions can be a better choice than polynomials to address
this problem.

\section{Appendix. A list of notational agreements}

Notations in this paper are somewhat sophisticated, thus
it make sense to list them in a separate appendix.

\subsection{Polygons and angles}

The most difficult are angular variables, associated with
our polygons. All of them refer to planar polygons
$\bar\Pi$, obtained by projection of $\Pi$ onto the
plane $(y_1,y_2)$.
Polygon $\bar\Pi$ have $n$ sides and $n$ vertices,
which we enumerate counterclockwise, assuming that
vertex $\# a$ is the intersection of the sides
$\#(a-1)$ and $\# a$. In other words, the side $a$
(i.e. external momentum ${\bf p}_a$) originates at
vertex $\# a$ and ends at vertex $\# (a+1)$.
The $y_0$ variable is either growing or decreasing
when we move along this side, this choice is labeled
by discrete parameter $\sigma_a = \pm 1$, associated with
each side of $\bar\Pi$.

The origin of coordinate system in $(y_1,y_2)$ plane
is located at the center of the circle, inscribed into
$\bar\Pi$.
In this paper we assume that such circle exists, see
discussion around eq.(\ref{ads3}.
This, of course, unjustly restricts the choice of $\Pi$,
but considerably simplifies the formalism.
The general scale is fixed by requiring that the
circle radius is unity.
Rotational symmetry is not fixed in any universal way,
it is done in different ways in different examples,
because it is done so in
%in order to allow comparison with
existing literature.

Direction of sides of $\bar\Pi$ are defined through
directions of {\it normals} to these sides, which are
labeled by angles $\phi_a$. This means that direction of the
side itself is $\frac{\pi}{2}+\phi_a$.
Directions towards the vertices are labeled by the angles
$\theta_a$.
The difference between $\theta$ and $\phi$ variables is
denoted by $\varphi$.
With above-described convention about comparative
enumeration of sides and vertices
\be
\theta_a-\varphi_a = \phi_{a-1}, \ \ \
\theta_a+\varphi_a = \phi_a
\ee
Both these formulas contain the same $varphi_a$ -- this
is a corollary of inscribed-circle condition.

In some examples vertices are also labeled by alphabetically
ordered capital letters $1,2,3,4=A,B,C,D$.
Angles of the polygon are denoted $2\alpha_a$,
$\alpha_a = \frac{\pi}{2}-\varphi_a$ is one {\it half}
of the polygon angle.

\subsection{Boundary rings and exact solutions}

"Linear in $y_0$" means that the expression has
the form $Ay_0 + B$ with $B$ not necessarily vanishing.
This is the usual form of our canonical element
${\cal L}_\Pi = y_0Q_\Pi(y_1,y_2) - {\cal K}_\Pi(y_1,y_2)$.

Multiplicative character is a number-valued homomorphism
of the ring multiplication.
When we consider a union, $\Pi = \Pi_1\cup \Pi_2$,
the boundary rings are multiplied and so do characters:
a family of functions $C_\Pi(y_0,y_1,y_2)$ is
a multiplicative character if
$C_\Pi = C_{Pi_1}C_{\Pi_2}$. Examples of multiplicative
characters are "obvious" elements (\ref{Ppolsdef})
of the polygon boundary rings and also the complex-valued
${\cal C}_\Pi$ from s.\ref{frosito}.

Calligraphic letters denote elements of the boundary rings,
as well as the rings themselves.
However there are exceptions,
not {\it all} elements of the ring are denoted
by calligraphic letters and some objects, though denoted by
calligraphic letters, do not belong to the ring.
Among the elements of the ring ${\cal R}_\Pi$
are: complex characters ${\cal C}_\Pi$, canonical
$y_0$-linear elements ${\cal L}_\Pi$, most of solutions
${\cal S}_\Pi$ to $AdS$ Plateau problem mentioned in this
paper. However, real-valued characters (\ref{Ppolsdef})
are also elements ${\cal R}_\Pi$, still they are denoted
by ordinary capital letters $P$.
This is because ${\cal P}_\Pi$ was used in \cite{malda3}
and in s.\ref{goal} to denote a "nice" element of
${\cal R}_\Pi$ -- a notion that we still did not manage
to extend beyond $Z_n$-symmetric case in the present paper.
For $Z_n$ -symmetric $\Pi$ this ${\cal P}_\Pi = {\cal L}_\Pi$
and simultaneously ${\cal S}_\Pi\approx {\cal P}_\Pi$, but
this in general ${\cal S}_\Pi \neq {\cal L}_\Pi$.
The difference is measured by $\mu_\Pi$, which is constant
for exact solutions considered in this paper, and this
constant is non-vanishing in asymmetric situations
(starting from rhombus).
In general, for $n>4$, $\mu_\Pi$ is not a constant and, perhaps,
not even a polynomial, this means that in general ${\cal S}_\Pi$
is not quite an element of the polynomial boundary ring
${\cal R}_\Pi$, it rather belongs to some completion
$\overline{{\cal R}_\Pi}$, which can hopefully be made
smaller than just the formal series made from elements of
${\cal R}_\Pi$.
The prototype of $\mu_\Pi$ is called ${\cal B}$ in s.\ref{goal},
despite denoted by calligraphic letter, it is {\it not}
and element of the boundary ring or of its completion:
$P_2{\cal B}$ is. The same is true about  ${\cal K}_\Pi$:
it does not belong to the ring, ${\cal L}_\Pi = y_0Q_\Pi-
{\cal K}_\Pi$ does.

\section*{Acknowledgements}

We appreciate collaboration and discussions with A.Mironov
on the main topics of his paper.
H.Itoyama acknowledges the hospitality of ITEP during his visit
to Moscow at the beginning of this work.
A.Morozov is indebted for hospitality to Osaka
City University and for support of JSPS.
The work of H.I. is partly supported by Grant-in-Aid for
Scientific Research 18540285 from the Ministry of Education, Science
and Culture, Japan and the XXI Century COE program "Constitution of
wide-angle mathematical basis focused on knots", the work of
A.M. is partly supported by Russian Federal Nuclear Energy Agency,
by the joint grant 06-01-92059-CE,  by NWO project 047.011.2004.026,
by INTAS grant 05-1000008-7865, by ANR-05-BLAN-0029-01 project and
by the Russian President's Grant of Support for the Scientific
Schools NSh-8004.2006.2, and by RFBR grant 07-02-00645.

\end{document}